\documentclass[a4paper, traditabstract,longauth]{aa}

\def\setsymbol#1#2{\expandafter\def\csname #1\endcsname{#2}}
\def\getsymbol#1{\csname #1\endcsname}

\def\Planck{\textit{Planck}}

\def\HeJT{$^4$He-JT}

\newbox\tablebox    \newdimen\tablewidth
\def\leaderfil{\leaders\hbox to 5pt{\hss.\hss}\hfil}
\def\endPlancktable{\tablewidth=\columnwidth 
    $$\hss\copy\tablebox\hss$$
    \vskip-\lastskip\vskip -2pt}
\def\endPlancktablewide{\tablewidth=\textwidth 
    $$\hss\copy\tablebox\hss$$
    \vskip-\lastskip\vskip -2pt}
\def\tablenote#1 #2\par{\begingroup \parindent=0.8em
    \abovedisplayshortskip=0pt\belowdisplayshortskip=0pt
    \noindent
    $$\hss\vbox{\hsize\tablewidth \hangindent=\parindent \hangafter=1 \noindent
    \hbox to \parindent{$^#1$\hss}\strut#2\strut\par}\hss$$
    \endgroup}
\def\doubleline{\vskip 3pt\hrule \vskip 1.5pt \hrule \vskip 5pt}

\def\L2{\ifmmode L_2\else $L_2$\fi}

\def\DeltaT{\ifmmode \Delta T\else $\Delta T$\fi}
\def\deltat{\ifmmode \Delta t\else $\Delta t$\fi}
\def\fknee{\ifmmode f_{\rm knee}\else $f_{\rm knee}$\fi}
\def\Fmax{\ifmmode F_{\rm max}\else $F_{\rm max}$\fi}
\def\solar{\ifmmode{\rm M}_{\mathord\odot}\else${\rm M}_{\mathord\odot}$\fi}
\def\Msolar{\ifmmode{\rm M}_{\mathord\odot}\else${\rm M}_{\mathord\odot}$\fi}
\def\Lsolar{\ifmmode{\rm L}_{\mathord\odot}\else${\rm L}_{\mathord\odot}$\fi}
\def\inv{\ifmmode^{-1}\else$^{-1}$\fi}
\def\mo{\ifmmode^{-1}\else$^{-1}$\fi}
\def\sup#1{\ifmmode ^{\rm #1}\else $^{\rm #1}$\fi}
\def\expo#1{\ifmmode \times 10^{#1}\else $\times 10^{#1}$\fi}
\def\,{\thinspace}
\def\lsim{\mathrel{\raise .4ex\hbox{\rlap{$<$}\lower 1.2ex\hbox{$\sim$}}}}
\def\gsim{\mathrel{\raise .4ex\hbox{\rlap{$>$}\lower 1.2ex\hbox{$\sim$}}}}

\def\simprop{\mathrel{\raise .4ex\hbox{\rlap{$\propto$}\lower 1.2ex\hbox{$\sim$}}}}
\def\deg{\ifmmode^\circ\else$^\circ$\fi}
\def\pdeg{\ifmmode $\setbox0=\hbox{$^{\circ}$}\rlap{\hskip.11\wd0 .}$^{\circ}
          \else \setbox0=\hbox{$^{\circ}$}\rlap{\hskip.11\wd0 .}$^{\circ}$\fi}
\def\arcs{\ifmmode {^{\scriptstyle\prime\prime}}
          \else $^{\scriptstyle\prime\prime}$\fi}
\def\arcm{\ifmmode {^{\scriptstyle\prime}}
          \else $^{\scriptstyle\prime}$\fi}
\newdimen\sa  \newdimen\sb
\def\parcs{\sa=.07em \sb=.03em
     \ifmmode \hbox{\rlap{.}}^{\scriptstyle\prime\kern -\sb\prime}\hbox{\kern -\sa}
     \else \rlap{.}$^{\scriptstyle\prime\kern -\sb\prime}$\kern -\sa\fi}
\def\parcm{\sa=.08em \sb=.03em
     \ifmmode \hbox{\rlap{.}\kern\sa}^{\scriptstyle\prime}\hbox{\kern-\sb}
     \else \rlap{.}\kern\sa$^{\scriptstyle\prime}$\kern-\sb\fi}
\def\ra[#1 #2 #3.#4]{#1\sup{h}#2\sup{m}#3\sup{s}\llap.#4}
\def\dec[#1 #2 #3.#4]{#1\deg#2\arcm#3\arcs\llap.#4}
\def\deco[#1 #2 #3]{#1\deg#2\arcm#3\arcs}
\def\rra[#1 #2]{#1\sup{h}#2\sup{m}}

\def\dots{\relax\ifmmode \ldots\else $\ldots$\fi}
\def\WHzsr{\ifmmode $W\,Hz\mo\,sr\mo$\else W\,Hz\mo\,sr\mo\fi}
\def\mHz{\ifmmode $\,mHz$\else \,mHz\fi}
\def\GHz{\ifmmode $\,GHz$\else \,GHz\fi}
\def\mKs{\ifmmode $\,mK\,s$^{1/2}\else \,mK\,s$^{1/2}$\fi}
\def\muKs{\ifmmode \,\mu$K\,s$^{1/2}\else \,$\mu$K\,s$^{1/2}$\fi}
\def\muKRJs{\ifmmode \,\mu$K$_{\rm RJ}$\,s$^{1/2}\else \,$\mu$K$_{\rm RJ}$\,s$^{1/2}$\fi}
\def\muKHz{\ifmmode \,\mu$K\,Hz$^{-1/2}\else \,$\mu$K\,Hz$^{-1/2}$\fi}
\def\MJysr{\ifmmode \,$MJy\,sr\mo$\else \,MJy\,sr\mo\fi}
\def\MJysrmK{\ifmmode \,$MJy\,sr\mo$\,mK$_{\rm CMB}\mo\else \,MJy\,sr\mo\,mK$_{\rm CMB}\mo$\fi}
\def\microns{\ifmmode \,\mu$m$\else \,$\mu$m\fi}

\def\muK{\ifmmode \,\mu$K$\else \,$\mu$\hbox{K}\fi}
\def\microK{\ifmmode \,\mu$K$\else \,$\mu$\hbox{K}\fi}
\def\muW{\ifmmode \,\mu$W$\else \,$\mu$\hbox{W}\fi}
\def\kms{\ifmmode $\,km\,s$^{-1}\else \,km\,s$^{-1}$\fi}
\def\kmsMpc{\ifmmode $\,\kms\,Mpc\mo$\else \,\kms\,Mpc\mo\fi}
\providecommand{\sorthelp}[1]{}

\usepackage[english]{babel} 
\usepackage{inputenc} 
\usepackage[T1]{fontenc}
\usepackage[mathscr]{eucal}
\usepackage{graphicx,amsmath}
\usepackage{epstopdf}
\usepackage[draft,breaklinks, colorlinks, citecolor=blue]{hyperref}
\usepackage{ifthen}
\usepackage{newlfont}
\usepackage{textcomp}
\usepackage{times}
\usepackage{mathrsfs}
\usepackage{mathbbol}
\usepackage{natbib}
\bibpunct{(}{)}{;}{a}{}{,} 

\usepackage{txfonts}
\usepackage{amsfonts}
\usepackage{amssymb}
\usepackage{color}
\usepackage{longtable,lscape}
\usepackage{fixltx2e}
\usepackage{afterpage}
\usepackage{verbatim}
\usepackage[usenames,dvipsnames,svgnames,table]{xcolor}
\usepackage{pdfsync}

\def\GHz{\ifmmode $GHz$\else \,GHz\fi}
\def\MHz{\ifmmode $MHz$\else \,MHz\fi}
\def\Hz{\ifmmode $Hz$\else \,Hz\fi}

\def\microm{\ifmmode \,\mu$m$\else \,$\mu$\hbox{m}\fi}

\graphicspath{{./Figure/}}

\begin{document}
\title{\Planck\ 2015 results. VII. High Frequency Instrument data processing: Time-ordered information and beams}

\titlerunning{Planck 2015 results. VII. HFI TOI and beams}

\author{
\author{\small
Planck Collaboration: R.~Adam\inst{71}
\and
P.~A.~R.~Ade\inst{82}
\and
N.~Aghanim\inst{55}
\and
M.~Arnaud\inst{69}
\and
M.~Ashdown\inst{65, 5}
\and
J.~Aumont\inst{55}
\and
C.~Baccigalupi\inst{81}
\and
A.~J.~Banday\inst{90, 9}
\and
R.~B.~Barreiro\inst{60}
\and
N.~Bartolo\inst{28, 61}
% ...
}
\institute{\small
APC, AstroParticule et Cosmologie, Universit\'{e} Paris Diderot, CNRS/IN2P3, CEA/lrfu, Observatoire de Paris, Sorbonne Paris Cit\'{e}, 10, rue Alice Domon et L\'{e}onie Duquet, 75205 Paris Cedex 13, France\goodbreak
\and
African Institute for Mathematical Sciences, 6-8 Melrose Road, Muizenberg, Cape Town, South Africa\goodbreak
\and
Agenzia Spaziale Italiana Science Data Center, Via del Politecnico snc, 00133, Roma, Italy\goodbreak
\and
Aix Marseille Universit\'{e}, CNRS, LAM (Laboratoire d'Astrophysique de Marseille) UMR 7326, 13388, Marseille, France\goodbreak
\and
Astrophysics Group, Cavendish Laboratory, University of Cambridge, J J Thomson Avenue, Cambridge CB3 0HE, U.K.\goodbreak
\and
Astrophysics \& Cosmology Research Unit, School of Mathematics, Statistics \& Computer Science, University of KwaZulu-Natal, Westville Campus, Private Bag X54001, Durban 4000, South Africa\goodbreak
\and
Atacama Large Millimeter/submillimeter Array, ALMA Santiago Central Offices, Alonso de Cordova 3107, Vitacura, Casilla 763 0355, Santiago, Chile\goodbreak
\and
CITA, University of Toronto, 60 St. George St., Toronto, ON M5S 3H8, Canada\goodbreak
\and
CNRS, IRAP, 9 Av. colonel Roche, BP 44346, F-31028 Toulouse cedex 4, France\goodbreak
\and
CRANN, Trinity College, Dublin, Ireland\goodbreak
\and
California Institute of Technology, Pasadena, California, U.S.A.\goodbreak
\and
Centre for Theoretical Cosmology, DAMTP, University of Cambridge, Wilberforce Road, Cambridge CB3 0WA, U.K.\goodbreak
\and
Computational Cosmology Center, Lawrence Berkeley National Laboratory, Berkeley, California, U.S.A.\goodbreak
\and
DSM/Irfu/SPP, CEA-Saclay, F-91191 Gif-sur-Yvette Cedex, France\goodbreak
\and
DTU Space, National Space Institute, Technical University of Denmark, Elektrovej 327, DK-2800 Kgs. Lyngby, Denmark\goodbreak
\and
D\'{e}partement de Physique Th\'{e}orique, Universit\'{e} de Gen\`{e}ve, 24, Quai E. Ansermet,1211 Gen\`{e}ve 4, Switzerland\goodbreak
\and
Departamento de F\'{\i}sica, Universidad de Oviedo, Avda. Calvo Sotelo s/n, Oviedo, Spain\goodbreak
\and
Department of Astronomy and Astrophysics, University of Toronto, 50 Saint George Street, Toronto, Ontario, Canada\goodbreak
\and
Department of Astrophysics/IMAPP, Radboud University Nijmegen, P.O. Box 9010, 6500 GL Nijmegen, The Netherlands\goodbreak
\and
Department of Physics \& Astronomy, University of British Columbia, 6224 Agricultural Road, Vancouver, British Columbia, Canada\goodbreak
\and
Department of Physics and Astronomy, Dana and David Dornsife College of Letter, Arts and Sciences, University of Southern California, Los Angeles, CA 90089, U.S.A.\goodbreak
\and
Department of Physics and Astronomy, University College London, London WC1E 6BT, U.K.\goodbreak
\and
Department of Physics, Florida State University, Keen Physics Building, 77 Chieftan Way, Tallahassee, Florida, U.S.A.\goodbreak
\and
Department of Physics, Gustaf H\"{a}llstr\"{o}min katu 2a, University of Helsinki, Helsinki, Finland\goodbreak
\and
Department of Physics, Princeton University, Princeton, New Jersey, U.S.A.\goodbreak
\and
Department of Physics, University of California, Santa Barbara, California, U.S.A.\goodbreak
\and
Department of Physics, University of Illinois at Urbana-Champaign, 1110 West Green Street, Urbana, Illinois, U.S.A.\goodbreak
\and
Dipartimento di Fisica e Astronomia G. Galilei, Universit\`{a} degli Studi di Padova, via Marzolo 8, 35131 Padova, Italy\goodbreak
\and
Dipartimento di Fisica e Scienze della Terra, Universit\`{a} di Ferrara, Via Saragat 1, 44122 Ferrara, Italy\goodbreak
\and
Dipartimento di Fisica, Universit\`{a} La Sapienza, P. le A. Moro 2, Roma, Italy\goodbreak
\and
Dipartimento di Fisica, Universit\`{a} degli Studi di Milano, Via Celoria, 16, Milano, Italy\goodbreak
\and
Dipartimento di Matematica, Universit\`{a} di Roma Tor Vergata, Via della Ricerca Scientifica, 1, Roma, Italy\goodbreak
\and
Discovery Center, Niels Bohr Institute, Blegdamsvej 17, Copenhagen, Denmark\goodbreak
\and
European Southern Observatory, ESO Vitacura, Alonso de Cordova 3107, Vitacura, Casilla 19001, Santiago, Chile\goodbreak
\and
European Space Agency, ESAC, Planck Science Office, Camino bajo del Castillo, s/n, Urbanizaci\'{o}n Villafranca del Castillo, Villanueva de la Ca\~{n}ada, Madrid, Spain\goodbreak
\and
European Space Agency, ESTEC, Keplerlaan 1, 2201 AZ Noordwijk, The Netherlands\goodbreak
\and
Facolt\`{a} di Ingegneria, Universit\`{a} degli Studi e-Campus, Via Isimbardi 10, Novedrate (CO), 22060, Italy\goodbreak
\and
Gran Sasso Science Institute, INFN, viale F. Crispi 7, 67100 L'Aquila, Italy\goodbreak
\and
HGSFP and University of Heidelberg, Theoretical Physics Department, Philosophenweg 16, 69120, Heidelberg, Germany\goodbreak
\and
Helsinki Institute of Physics, Gustaf H\"{a}llstr\"{o}min katu 2, University of Helsinki, Helsinki, Finland\goodbreak
\and
INAF - Osservatorio Astronomico di Padova, Vicolo dell'Osservatorio 5, Padova, Italy\goodbreak
\and
INAF - Osservatorio Astronomico di Roma, via di Frascati 33, Monte Porzio Catone, Italy\goodbreak
\and
INAF - Osservatorio Astronomico di Trieste, Via G.B. Tiepolo 11, Trieste, Italy\goodbreak
\and
INAF/IASF Bologna, Via Gobetti 101, Bologna, Italy\goodbreak
\and
INAF/IASF Milano, Via E. Bassini 15, Milano, Italy\goodbreak
\and
INFN, Sezione di Bologna, Via Irnerio 46, I-40126, Bologna, Italy\goodbreak
\and
INFN, Sezione di Roma 1, Universit\`{a} di Roma Sapienza, Piazzale Aldo Moro 2, 00185, Roma, Italy\goodbreak
\and
INFN, Sezione di Roma 2, Universit\`{a} di Roma Tor Vergata, Via della Ricerca Scientifica, 1, Roma, Italy\goodbreak
\and
IPAG: Institut de Plan\'{e}tologie et d'Astrophysique de Grenoble, Universit\'{e} Grenoble Alpes, IPAG, F-38000 Grenoble, France, CNRS, IPAG, F-38000 Grenoble, France\goodbreak
\and
IUCAA, Post Bag 4, Ganeshkhind, Pune University Campus, Pune 411 007, India\goodbreak
\and
Imperial College London, Astrophysics group, Blackett Laboratory, Prince Consort Road, London, SW7 2AZ, U.K.\goodbreak
\and
Infrared Processing and Analysis Center, California Institute of Technology, Pasadena, CA 91125, U.S.A.\goodbreak
\and
Institut N\'{e}el, CNRS, Universit\'{e} Joseph Fourier Grenoble I, 25 rue des Martyrs, Grenoble, France\goodbreak
\and
Institut Universitaire de France, 103, bd Saint-Michel, 75005, Paris, France\goodbreak
\and
Institut d'Astrophysique Spatiale, CNRS (UMR8617) Universit\'{e} Paris-Sud 11, B\^{a}timent 121, Orsay, France\goodbreak
\and
Institut d'Astrophysique de Paris, CNRS (UMR7095), 98 bis Boulevard Arago, F-75014, Paris, France\goodbreak
\and
Institut f\"ur Theoretische Teilchenphysik und Kosmologie, RWTH Aachen University, D-52056 Aachen, Germany\goodbreak
\and
Institute of Astronomy, University of Cambridge, Madingley Road, Cambridge CB3 0HA, U.K.\goodbreak
\and
Institute of Theoretical Astrophysics, University of Oslo, Blindern, Oslo, Norway\goodbreak
\and
Instituto de F\'{\i}sica de Cantabria (CSIC-Universidad de Cantabria), Avda. de los Castros s/n, Santander, Spain\goodbreak
\and
Istituto Nazionale di Fisica Nucleare, Sezione di Padova, via Marzolo 8, I-35131 Padova, Italy\goodbreak
\and
Jet Propulsion Laboratory, California Institute of Technology, 4800 Oak Grove Drive, Pasadena, California, U.S.A.\goodbreak
\and
Jodrell Bank Centre for Astrophysics, Alan Turing Building, School of Physics and Astronomy, The University of Manchester, Oxford Road, Manchester, M13 9PL, U.K.\goodbreak
\and
Kavli Institute for Cosmological Physics, University of Chicago, Chicago, IL 60637, USA\goodbreak
\and
Kavli Institute for Cosmology Cambridge, Madingley Road, Cambridge, CB3 0HA, U.K.\goodbreak
\and
LAL, Universit\'{e} Paris-Sud, CNRS/IN2P3, Orsay, France\goodbreak
\and
LERMA, CNRS, Observatoire de Paris, 61 Avenue de l'Observatoire, Paris, France\goodbreak
\and
LESIA, Observatoire de Paris, CNRS, UPMC, Universit\'{e} Paris-Diderot, 5 Place J. Janssen, 92195 Meudon, France\goodbreak
\and
Laboratoire AIM, IRFU/Service d'Astrophysique - CEA/DSM - CNRS - Universit\'{e} Paris Diderot, B\^{a}t. 709, CEA-Saclay, F-91191 Gif-sur-Yvette Cedex, France\goodbreak
\and
Laboratoire Traitement et Communication de l'Information, CNRS (UMR 5141) and T\'{e}l\'{e}com ParisTech, 46 rue Barrault F-75634 Paris Cedex 13, France\goodbreak
\and
Laboratoire de Physique Subatomique et Cosmologie, Universit\'{e} Grenoble-Alpes, CNRS/IN2P3, 53, rue des Martyrs, 38026 Grenoble Cedex, France\goodbreak
\and
Laboratoire de Physique Th\'{e}orique, Universit\'{e} Paris-Sud 11 \& CNRS, B\^{a}timent 210, 91405 Orsay, France\goodbreak
\and
Lawrence Berkeley National Laboratory, Berkeley, California, U.S.A.\goodbreak
\and
Lebedev Physical Institute of the Russian Academy of Sciences, Astro Space Centre, 84/32 Profsoyuznaya st., Moscow, GSP-7, 117997, Russia\goodbreak
\and
Max-Planck-Institut f\"{u}r Astrophysik, Karl-Schwarzschild-Str. 1, 85741 Garching, Germany\goodbreak
\and
McGill Physics, Ernest Rutherford Physics Building, McGill University, 3600 rue University, Montr\'{e}al, QC, H3A 2T8, Canada\goodbreak
\and
National University of Ireland, Department of Experimental Physics, Maynooth, Co. Kildare, Ireland\goodbreak
\and
Nicolaus Copernicus Astronomical Center, Bartycka 18, 00-716 Warsaw, Poland\goodbreak
\and
Niels Bohr Institute, Blegdamsvej 17, Copenhagen, Denmark\goodbreak
\and
Optical Science Laboratory, University College London, Gower Street, London, U.K.\goodbreak
\and
SISSA, Astrophysics Sector, via Bonomea 265, 34136, Trieste, Italy\goodbreak
\and
School of Physics and Astronomy, Cardiff University, Queens Buildings, The Parade, Cardiff, CF24 3AA, U.K.\goodbreak
\and
School of Physics and Astronomy, University of Nottingham, Nottingham NG7 2RD, U.K.\goodbreak
\and
Sorbonne Universit\'{e}-UPMC, UMR7095, Institut d'Astrophysique de Paris, 98 bis Boulevard Arago, F-75014, Paris, France\goodbreak
\and
Space Sciences Laboratory, University of California, Berkeley, California, U.S.A.\goodbreak
\and
Special Astrophysical Observatory, Russian Academy of Sciences, Nizhnij Arkhyz, Zelenchukskiy region, Karachai-Cherkessian Republic, 369167, Russia\goodbreak
\and
Sub-Department of Astrophysics, University of Oxford, Keble Road, Oxford OX1 3RH, U.K.\goodbreak
\and
Theory Division, PH-TH, CERN, CH-1211, Geneva 23, Switzerland\goodbreak
\and
UPMC Univ Paris 06, UMR7095, 98 bis Boulevard Arago, F-75014, Paris, France\goodbreak
\and
Universit\'{e} de Toulouse, UPS-OMP, IRAP, F-31028 Toulouse cedex 4, France\goodbreak
% ...
}
}
\authorrunning{Planck Collaboration}

\date{Submitted to A\&A February 8, 2015; Accepted July 15, 2015}

\abstract{The \Planck\ High Frequency Instrument (HFI) has observed the full
  sky at six frequencies (100, 143, 217, 353, 545, and 857\GHz) in intensity
  and at four frequencies in linear polarization (100, 143, 217, and
  353\GHz). In order to obtain sky maps, the time-ordered information (TOI)
  containing the detector and pointing samples must be processed and the
  angular response must be assessed. The full mission TOI is included
  in the \Planck\ 2015 release.
  This paper describes the HFI TOI and beam processing for the 2015 release. 
 HFI calibration and map making are described in a companion
  paper. The main pipeline has been modified since the last release (2013
  nominal mission in intensity only), by including a correction for the
  nonlinearity of the warm readout and by improving the model of the bolometer time
  response.  The beam processing is an essential tool that derives the angular
  response used in  all the \Planck\ science papers and we report an
  improvement in the effective beam window function uncertainty of
  more than a
  factor of 10 relative to the 2013
  release. Noise
  correlations introduced by pipeline filtering
  function are assessed using dedicated simulations. 
  Angular cross-power spectra using data sets that are decorrelated in time are
  immune to the main systematic effects. }

\maketitle

\clearpage

\section{Introduction: a summary of the HFI pipeline}
\label{sec:introduction}
 
This paper, one of a set associated with the 2015 \Planck\footnote{\Planck\ (\url{http://www.esa.int/Planck}) is a project of the European Space Agency  (ESA) with instruments provided by two scientific consortia funded by ESA member states and led by Principal Investigators from France and Italy, telescope reflectors provided through a collaboration between ESA and a scientific consortium led and funded by Denmark, and additional contributions from NASA (USA).} data release, is
the first of two that describe the processing of the data from
the High Frequency Instrument (HFI). The HFI is one of the two instruments on
board \Planck, the European Space Agency's mission dedicated to precision
measurements of the cosmic microwave background (CMB). The HFI uses cold
optics (at 4\,K, 1.6\,K, and 100\,mK), filters, and 52 bolometers cooled to
100\,mK. Coupled to the \Planck\ telescope, it enables us to map the continuum
emission of the sky in intensity and polarization at frequencies of 100, 143, 217, and 353\,GHz, and in intensity at 545 and
857\,GHz.  Paper~1 (this paper)
describes the processing of the data at the time-ordered level and the
measurement of the beam. Paper~2~\citep{planck2014-a09} describes the HFI
photometric calibration and map making.
 
\begin{figure}[ht!]
\begin{center}
\includegraphics[width=\columnwidth]{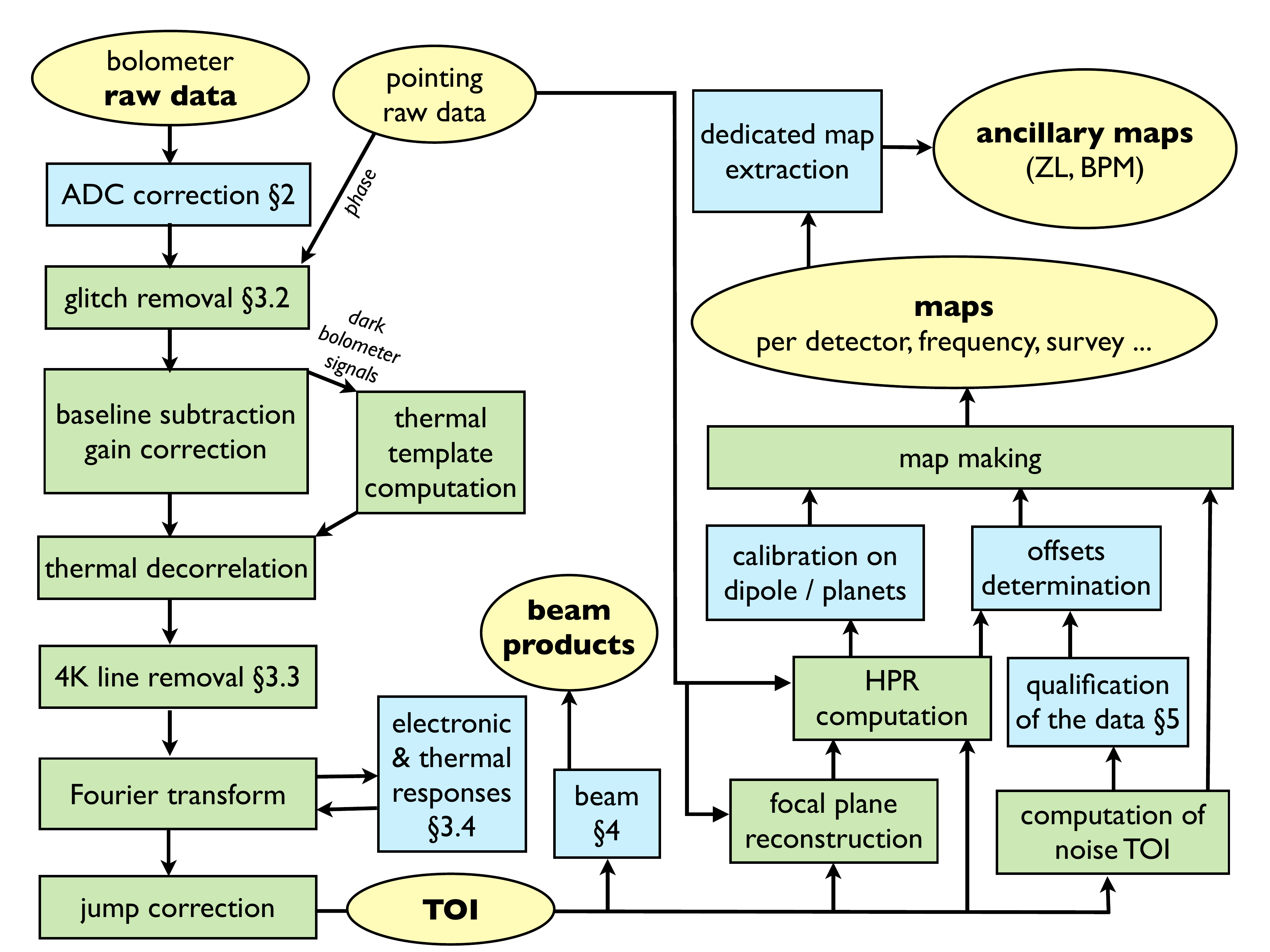}
\caption{\label{fig:DPCscheme} Schematic of the HFI
  pipeline.
  The left part of the schematic involves TOI and beams (this paper),
  while the upper-right part represents the map making steps
  (Paper~2).  Ancillary maps are composed of zodiacal light templates
  (ZL) and polarization band-pass mismatch (BPM) maps. ADC =
  analogue-to-digital converter (see Sect.~\ref{sec:adc}). HPR = {\tt HealPix}
  ring (see Paper~2).  ``Beam products'' refers to beam transfer
  functions, $B_\ell$. Blue: changes in this
  release with section numbers corresponding to this paper. Yellow: released data products.}
\end{center}
\end{figure}

The HFI data processing for this release is very similar to that used for the
2013 release~\citep{planck2013-p03}. Figure~\ref{fig:DPCscheme} provides a
summary of the main steps used in the processing, from raw data to frequency
maps both in temperature and polarization.

First, the telemetry data are converted to TOI (time ordered information). The TOI
consists of voltage measurements sampled at 180.3737\Hz\ for each of the 52
bolometers, two dark bolometers, 16 thermometers, and two devices (a resistance and
a capacitance) that comprise the HFI detector set. The TOI is then
corrected to account for nonlinearity in the analogue-to-digital conversion
(see Sect.~\ref{sec:adc}). Glitches (cosmic ray impacts on the
bolometers) are then detected, their immediate effects (data around the
maximum) are flagged, and their tails are subtracted. A baseline is computed
in order to demodulate the AC-biased TOI. A second-order polynomial correction
is applied to the demodulated TOI to linearize the bolometer response. The
minute-scale temperature fluctuations of the 100\,mK stage are subtracted from the
TOI using a combination of the TOI from the two dark bolometers. Sharp lines
in the temporal power spectrum of the TOI from the influence of the helium
Joule-Thomson (\HeJT) cooler (hereafter called 4-K lines) are removed with
interpolation in the Fourier domain (see Sect.~\ref{sec:ring_selection}). The
finite bolometer time responses are deconvolved from the TOI, also in the
Fourier domain. For this release, the time response consists of four to seven thermal
time constants for each bolometer.  Several criteria based on statistical
properties of the noise are used to reject the stable pointing periods
(hereafter called rings) that are non-stationary (see
Table~\ref{tab:OverallStat}). A subtractive jump correction is
applied; it
typically affects less than 1\,\% of the rings and the amplitude of the jumps
exceeds a tenth of the TOI rms in less than 0.1\,\% of the rings.

At this point, the TOI is cleaned but not yet calibrated. The beam is 
measured using a combination of planet observations for the main
beam and {\tt GRASP} physical optics calculations\footnote{TICRA,
  \url{http://www.ticra.com/products/software/grasp}} for the sidelobes (see
Sect.~\ref{sec:beam}). The focal plane geometry, or the relative position of
bolometers in the sky, is deduced from Mars observations. The 545 and
857\GHz\ channels are photometrically calibrated using the response to Uranus and Neptune. The
lower frequency channels (100, 143, 217, and 353\GHz, called the ``CMB''
channels) are calibrated with the orbital CMB dipole (i.e., the dipole induced
by the motion of the Lagrange point L2 around the Sun).

In this paper, we describe the changes made to the processing since the 2011 and 2013 papers
\citep{planck2011-1.7,planck2013-p03}.
Section~\ref{sec:adc} gives a view of  a step that has been added
to the beginning of the pipeline, namely the correction for ADC
(analogue-to-digital converter)
nonlinearities. This step proves to be very important for the quality of the CMB
data, especially at low multipoles. Section~\ref{sec:toiproc} deals with the addition of more long time constants in
the bolometer response  and other TOI. Section~\ref{sec:beam} presents refined beam
measurements and models. Some
consistency checks are reported in Sect.~\ref{sec:consistency}. The public HFI data
products are described in Appendix~\ref{sec:officialHFI}.

Paper~2 describes  the new scheme for the CMB dipole
calibration (the \Planck\ orbital dipole calibration is used for the first
time) and the submillimetre calibration on planets. Paper~2 also describes the
polarized map making, including the derivation of far sidelobes, and zodiacal
maps, as well as polarization correction maps due to bandpass mismatch:
a generalized least-squares destriper is used to produce maps
of the temperature and the two linear polarization Stokes parameters $Q$ and
$U$. About 3000 maps are obtained by splitting the HFI
data into different subsets by, e.g., time period or detector sets.
Consistency checks are performed in order to assess the fidelity of the
maps.

\section{ADC correction}\label{sec:adc}

\citet{planck2013-p03} reported that the HFI raw data show apparent
gain variations with time of up to 2\,\% due to nonlinearities in the HFI
readout chain. In the 2013
data release~\citep{planck2013-p03f} a correction for this systematic error was applied as an apparent gain variation
at the map making stage. The 2013 maps relied on an effective gain correction
based on the consistency constraints from the reconstructed sky maps, which
proved to be sufficient for the cosmological analysis.

For the 2015 data release we have implemented a direct ADC correction in the TOI.  In this
section, we describe the ADC effect and its correction, and its validation through
end-to-end simulations (see also Sect.~\ref{sec:e2esim}). Internal
checks of residual ADC nonlinearity are shown in 
Sect.~\ref{sec:consistency}. 

\subsection{The ADC systematic error}
The HFI bolometer electronic readout \citep{planck2014-ES}
includes a 16-bit ADC,  the
flight-qualified 
version of Maxwell Technologies model 7809LP having a very loose tolerance on the
differential nonlinearity (the maximum deviation from one least significant
bit, LSB, between two consecutive levels, over the whole range), specified to
be not worse than one LSB. The implications of this feature for HFI
performance had not been anticipated, and it  did not produce any detected effect
in ground test data. However, it proved to be a major systematic effect impacting
the flight data. A wide dynamic range at the ADC input was needed to both measure the CMB
and the sky foregrounds, and properly characterize and remove the tails of glitches from cosmic-rays.
 Operating HFI
electronics with the necessary low gains increased the effects of the ADC
scale errors on CMB data.

We have developed a method that reduces the ADC effect on the angular power spectra
by more than a factor of 10 for most bolometers. There are three main
  difficulties in making this correction.
\begin{itemize} 
\item The chip linearity defects were characterized with 
  insufficient precision before launch. We have designed and run a specific campaign to map
  the ADC nonlinearity from flight data (Sect.~\ref{sec:adcmethod}).

\item Each HFI data sample in the TOI is the sum of 40 consecutive ADC fast
  samples , corresponding to half of the modulation cycle. The full
    bandwidth of the digital signal is not transmitted to the ground. 
The effective correction of TOI samples due to ADC defects
  requires the knowledge of the shape of the fully-sampled raw data at the ADC
  input. A shape model is built from the subset of fully-sampled downloaded
  data, transmitted to the ground at a low rate (80 successive fast samples
  every $101.4\ \mathrm{s}$ for every detector), hereafter called ``the
  fully-sampled raw data.''

\item The $^4$He-JT 4-K cooler lines in the TOI
  (Sect.~\ref{sec:ring_selection}) result from a complex parasitic
  coupling. Much of this parasitic signal is a sum of 20\Hz\
  harmonics, synchronous with the readout clock, with nine slow sample
  periods fitting exactly into two parasitic periods. Capture of a
  sequence of 360 raw signal samples would have allowed a direct
  reconstruction of the full patterns of this parasitic
  signal. However, the short downloaded sequences of 80 samples always
  fall in the same 4-K phase range, allowing us to cover only 2/9 of
  the full pattern. To properly model the signal at the ADC input, one
  must decipher this parasitic signal over its full phase range. The
  present model relies both on the full sampling subset and on the
  4-K lines measured in the TOI (see Sect.~\ref{sec:ring_selection}
  and references therein).

\end{itemize} 
The signal model, including the 4-K line parasitic part, is described in
Sect.~\ref{sec:adcmodel}.

\subsection{Mapping the ADC defects}
{\label{sec:adcmethod}}
The defects of an ADC chip are fully characterized by the input levels
corresponding to the transitions between two consecutive output values (known
as digital output code, or DOC).
An ADC defect mapping is usually run on a dedicated ground test bench. We
made this measurement on two spare flight chips to
understand the typical behaviour of the circuit in the range relevant to the flight data.
These measurements revealed a 64-DOC periodic pattern, precisely
followed by most of the DOC, except at the chunk boundaries, which
allowed us to build a first, approximate defect 
model with a reduced number of parameters~\citep{planck2013-p03f}. Such
behaviour is understood from the circuit design,
where the lowest bits come from the same components over the full ADC
scale. However, because each ADC has a unique defect pattern, 
data from these ground tests could not be used directly to correct the flight data.

The parameters for the on-orbit chips were extracted for each HFI bolometer using
data samples of the thermal signal, effectively a Gaussian noise
input.  These ADC-dedicated flight data, herafter called ``warm data,'' were 
recorded during the $1.5$ years of the LFI extended mission, between
February 2012 and August 2013. During this period the bolometer temperature was
stable at about 4\,K, a temperature at which the bolometer impedance
is low, giving no input signal or parasitic pickup on the ADCs apart
from a tunable offset and Gaussian noise with rms values around $20$ equivalent mean LSBs. The
defect mapping was obtained by inverting the histograms of the accumulated
fully-sampled raw data, as explained below.

The warm data consist of large sets of $n_{\mathrm{tot}}$ counts
of 80 sample raw periods taken in stable conditions with different
bias currents and/or compensation voltages on the input stage 
tuned to mostly sample the central area of the ADC.  We
denote the signal value at the ADC input $\nu$, and $\nu_i$ is the
input level corresponding to the transition between DOC $(i-1)$ and DOC
$i$. We call the probability density function of the $k^{\mathrm{th}}$ sample
of the 80 samples $p^k(\nu)$. This is assumed to be a Gaussian
distribution with a mean $\bar{p}^k$ depending on $k$ and 
on the data set, and a variance $\sigma^2$ depending only on the data set.

Every set $k$ is a series of $\{n_i\}$ histogram bin statistics, following a
multinomial distribution. The mean values, $\{\bar{n}_i\}$, are linked
to $p^k$ and $\{\nu_i\}$ by the following set of equations:
\begin{eqnarray}
\label{eq:nibar}
\bar{n}_i = n_{\mathrm{tot}}\int_{\nu_i}^{\nu_{i+1}}p^k(\nu)d\nu.
\end{eqnarray}
A maximum likelihood technique maps the $\{\bar{n}_i\}$ to the $\{{n}_i\}$.
To extract $\{{\nu}_i\}$ from Eq.~(\ref {eq:nibar}) one needs an estimate
of $\bar{p}^k$ and $\sigma^2$. The equations are solved by recurrence,
starting from the most populated bin. Since the solution depends on
$\bar{p}^k$ and $\sigma^2$,  normally one would need to know
their true value. In fact, this system has specific  properties that
allow recovery of the correct ADC scale without any other
information. 
\begin{itemize}
\item Its
solution $\{{\nu}_i\}$ is extremely sensitive to $\bar{p}^k$.  An
incorrect input value gives unphysically diverging ADC step sizes;
this allows us to choose the $\bar{p}^k$ that gives the most stable
solution. This prescription makes the $\{{\nu}_i\}$ 
solution in the populated part of the histrogram independent from the hypothesis on $\bar{p}^k$.
\item The step size solution is proportional to $\sigma$, so the solutions for different data
sets can be intercalibrated to fit a common ADC scale, which allows the unknown
$\sigma$ to be eliminated, provided all 80 samples of a data set have the same noise value. 
\end{itemize}

\begin{figure}[ht!]
\centering
\includegraphics[width=\columnwidth]{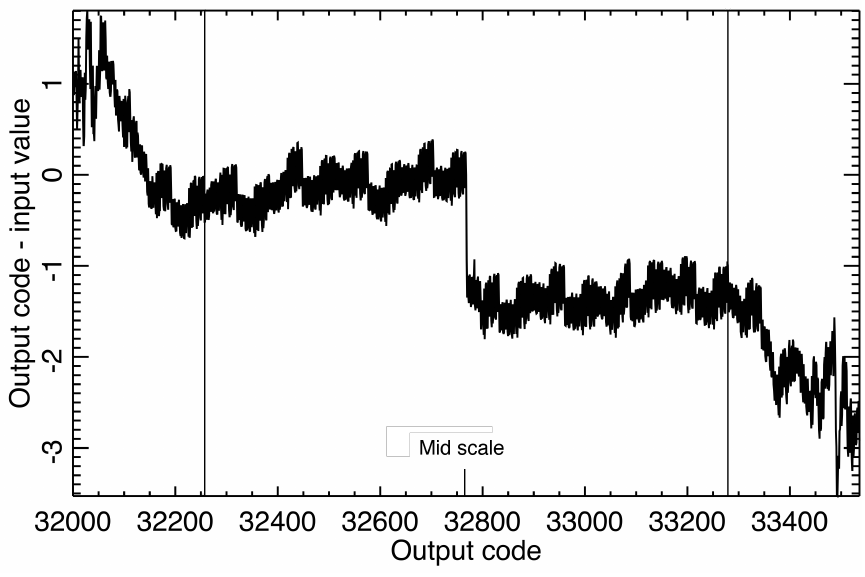}
\caption{An example of one ADC defect mapping around the mid-scale.
The useful range $\pm 512$ around mid-scale is shown with vertical lines. 
}
\label{fig:defects}
\end{figure}

  Figure~\ref{fig:defects} shows a typical example of the
maximum likelihood result, before applying the periodic feature constraints on
the solution. 
This method is based on the strong assumption that the noise
is Gaussian. Some data showing small
temperature drifts have not been used. With this method, the ADC step sizes
are nearly independently measured, so there is a random-walk type of
error on the DOC positions, which is limited by the 64 DOC periodicity
constraint. The 64 DOC pattern is obtained from the weighted average
of the values obtained in the $\pm 512\ $DOC around mid-scale, except for the first 64
DOC above mid-scale, which do not follow the same pattern.

The main step is found at mid-scale. For all channels, there are
more than $10^5$ samples per DOC histogram bin in the small $\pm$512 output
code range around mid-scale that is explored by flight data. Outside
this region, the smaller numbers of samples lead to bigger drifts in the likelihood
solution. 

Both residual distributions and simulations give an estimate of
the precision of the present defect recovery below $0.03$ LSB for any DOC over
this range. Systematic errors on large DOC distances, not taken into account
in this model, are smaller than $0.2$ LSB over a range of $512$ LSB. At this level,
we see residuals due to violation of the rms noise ($\sigma$) stability
assumption, and we are working on an improved version of the ADC
defect recovery to be included in future data releases.

\subsection{Input signal model}
{\label{sec:adcmodel}}
The signal at the input of the ADCs is the sum of several
components, including the modulation of the bolometer, noise from the
bolometer, and the electronics, and bolometer voltage changes due to the astrophysical signal. The
largest in amplitude is the modulation of the bolometer voltage
with the  combination of a triangle- and a square-wave signal, fed through bias
capacitors. Most of the modulation is balanced (i.e., the voltage amplitude is
reduced to nearly zero) by subtracting a compensating square-wave signal prior to digitization.

Spurious components also exist, such as the 4-K parasitic signal and various
electronic leakage effects. Both are well described by slowly varying periodic
patterns, assumed to be stable on a one-hour time scale.

The signal model used here is based on the linearity of
the bolometer chain. It is a steady-state approximation of the signal shape
produced by constant optical power on the bolometer. It is given by
\begin{eqnarray}
\label{eq:signal_model}
d(t) = P \times G_{\mathrm{raw}}(t) + O(t) \, ,
\end{eqnarray}
where $P$ is proportional to the sky signal, and the raw gain $G_{\mathrm{raw}}$ and the
offset $O$ are periodic functions of time.  The raw gain period is the
same length as the readout period, and the offset period is equivalent
to the 4-K cooler period.

The parameters of the model have been extracted and checked over the
whole mission from a clean subset of fully-sampled raw data (particle
glitches and planet-crossings excluded). For 
each bolometer, $G_{\mathrm{raw}}$ is given by a set of $80$ numbers,
assumed to be stable over the full mission. The offsets are given
ring by ring as sets of $360$ numbers.

\begin{figure}[ht!]
\centering
\includegraphics[width=\columnwidth]{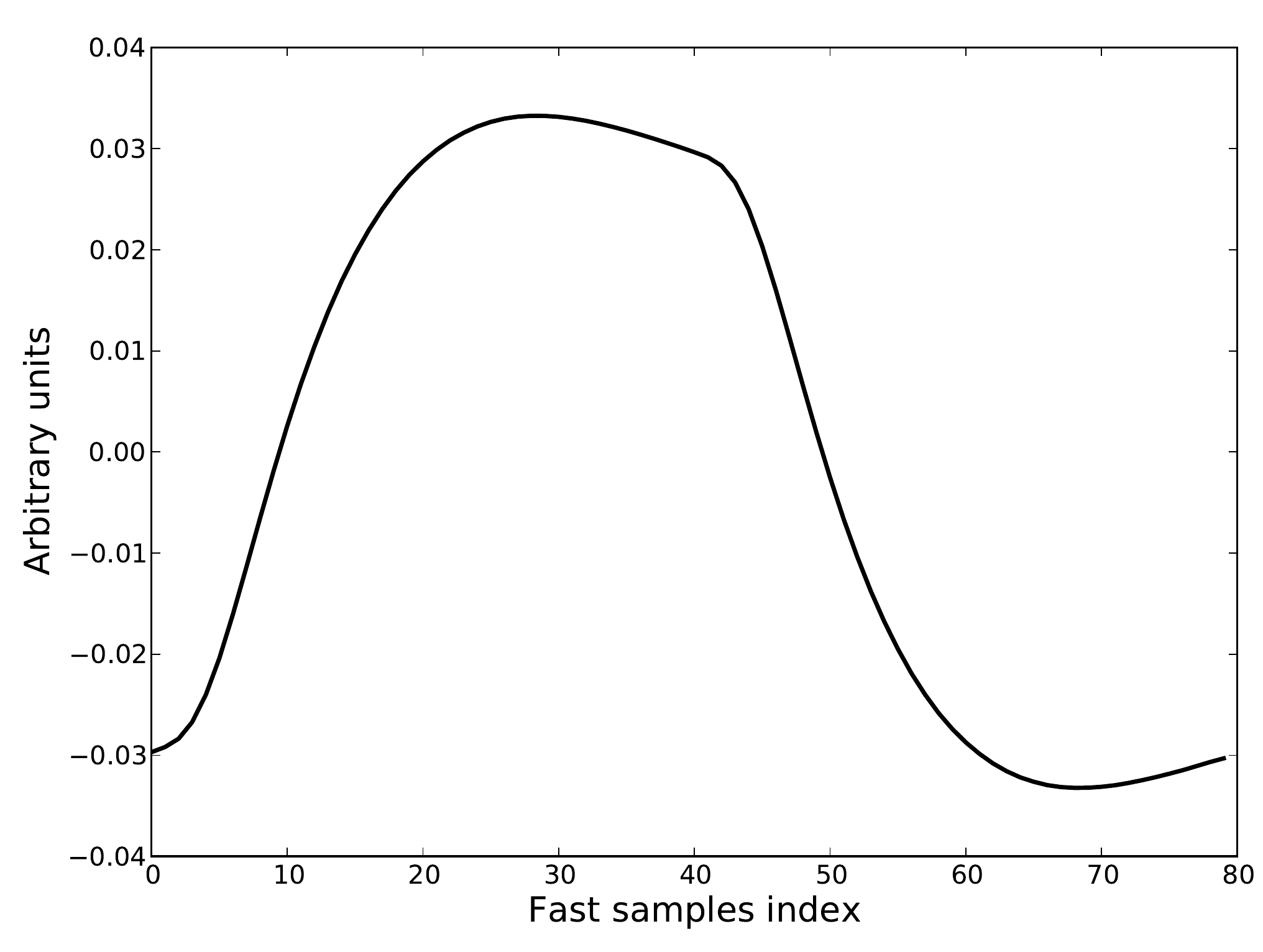}
\caption{Raw gain for the 100-1a bolometer. The horizontal scale is
  fast sample index, while the vertical  scale is arbitrary.}
\label{fig:rawgain}
\end{figure}

Figure~\ref{fig:rawgain} shows an example of the raw gain
$G_{\mathrm{raw}}$, for bolometer 100-1a.  This function represents
the fast time scale voltage across the bolometer.  The detailed shape
is due to the modulating bias current and to well-understood  
filters acting on the signal in the readout chain (for 
details, see Section~4.1 of \cite{lamarre2010}).  The signal model
(Eq.~\ref{eq:signal_model}) assumes that every pair of telemetered fast
data samples is the integral of one half of this function, linearly scaled by the
input sky signal.

\begin{figure}[ht!]
\centering
\includegraphics[width=\columnwidth]{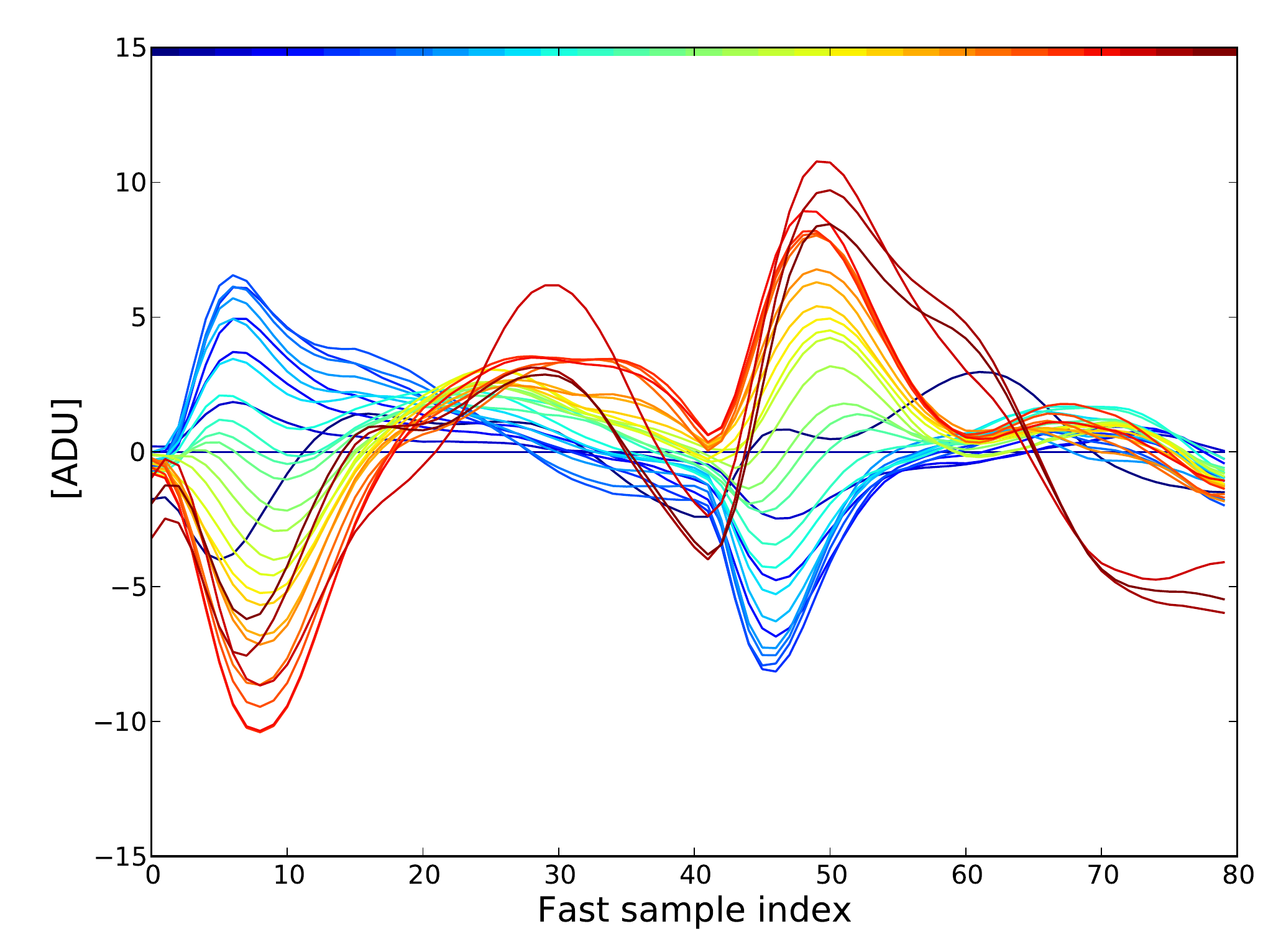}
\caption{Constant term drift for the 143-6 bolometer, relative to
  ring 1000. The colour code goes from black to red over the full 
  mission.}
\label{fig:constant}
\end{figure}

Figure~\ref{fig:constant} shows the evolution of the  constant term over the
mission, for the 143-6 bolometer, due 
to the spurious signals in the fully-sampled raw data window.

Electronics leakage is readout-synchronous and is therefore taken into
account by the present model. This is not the case for the 4-K cooler
parasitic signal. The available information in the TOI 
does not allow reconstruction of the shape of the 4-K line at the ADC input without
ambiguities. Fortunately, this parasitic signal is, in most cases, dominated
by a few harmonics that we constrain from the combined analysis of the
constant term in the fully-sampled raw data and the 4-K-folded harmonics in
the TOI.  A constrained $\chi^2$ method is used to extract
amplitudes and phases ring per ring for the 20, 160, and 200\Hz\
components. Only these three harmonics are used in the present model. The effect of this
approximation is checked by using both internal and global tests. Improved 4-K
harmonic estimators will be included in future data releases.

\subsection{ADC correction functions}
{\label{sec:adc_corr}}
The input shape model from Eq.~\ref{eq:signal_model} is used, along with
the ADC DOC patterns and the $S_{\mathrm{phase}}$ value, to compute,
ring by ring, the error induced by the ADC as a function of the summed 40 samples of the TOI.
A set of input power values $\{P\}$ is used to simulate TOI data: first,with
the input model that includes the 4-K line model and the ADC non
linearity model; 
and, second, with a parasite-free input model and a linear ADC.  We
identify real TOI data with the first result and draw from these
simulations the correction functions that are then applied to the real data.

\begin{figure}[ht!]
\centering
\includegraphics[width=\columnwidth]{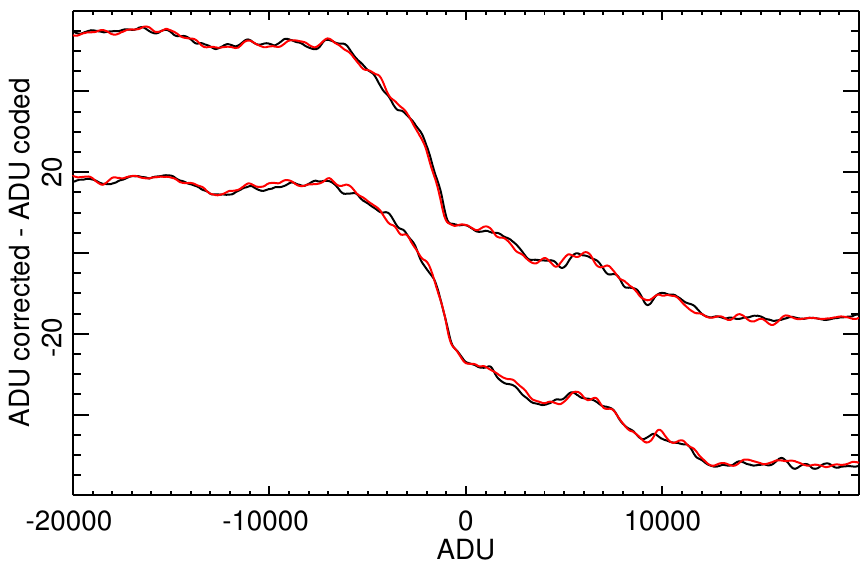}
\caption{Example ADC correction functions for both parities (upper and
  lower curve sets) and two different phases (in black and red) relative to the \HeJT\ cooler cycle. }
\label{fig:corrfuncs}
\end{figure}

The combination of the two
readout parities\footnote{The science data samples alternate between
  positive and negative parity corresponding to the positive and
  negative part of the readout modulation cycle.} with different possible phases relative to the 4-K
parasitic signal produces $18$ different correction functions that are
determined for each data ring and applied to every TOI sample.
Figure~\ref{fig:corrfuncs} shows an example.
\begin{figure}[ht!]
\centering
\includegraphics[width=\columnwidth]
{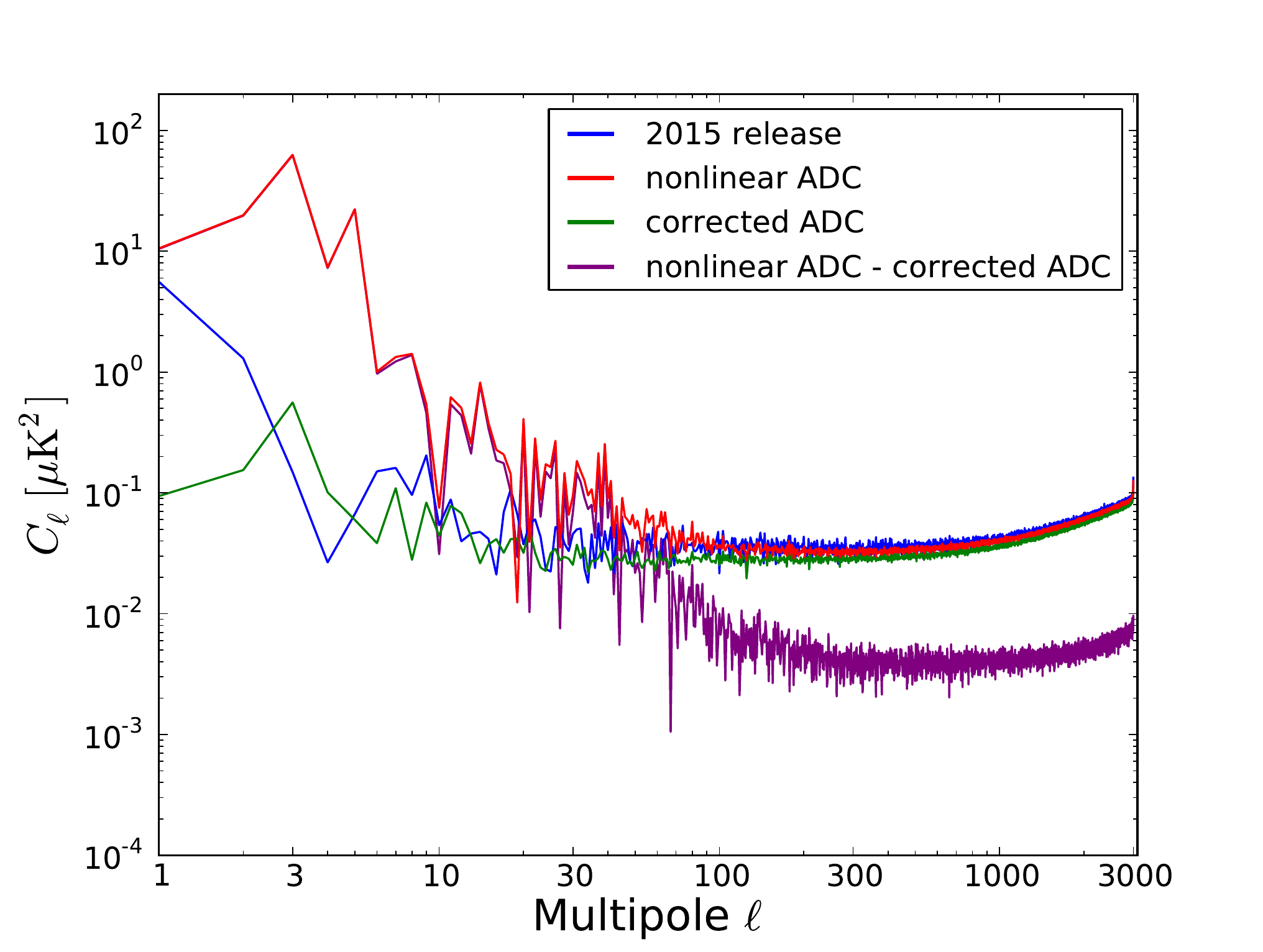}
\caption{Survey-difference angular power spectra for the 100-4b bolometer. The 2015 
  release data (blue curve) are compared with simulated data
  containing ADC nonlinearities. Green and red curves show
  simulations produced, with and without the ADC correction, respectively.}
\label{fig:simuADC}
\end{figure}

\subsection{Simulation of the ADC effect}
{\label{sec:adc_sim}}
The measured ADC defects are included in the HFI end-to-end
simulations 
(Sect.~\ref{sec:e2esim}), run at the fully-sampled raw data
level. This allows us to assess some consequences of the ADC
nonlinearity on the data. For instance, the gain variations are
well-reproduced.  Figure~\ref{fig:simuADC} shows an example of survey-difference
 angular power spectra. The systematic effect coming from the ADC behaves as $1/\ell^2$ at
low $\ell$, and is well suppressed by the ADC correction. This is an important
check on the consistency of the ADC processing.

\section{TOI processing}\label{sec:toiproc}
After the ADC correction, the software modules used to process the
2015 TOI are identical to those used for 2013, up to the stage of clean
TOIs ready for selection, calibration, and map making (see
Fig.~\ref{fig:DPCscheme}).  However, some of the modules'  input 
  parameters have been fine-tuned to better
  control residual systematic errors present in the 2013
  data. We describe these tunings here, and later assess their
  accuracy in Sect.~\ref{sec:consistency}.
 
 \subsection{Pointing and focal-plane reconstruction}\label{sec:pointing}
 
 Satellite attitude reconstruction is the same for both \Planck\ instruments
 in the 2015 release and is described in the mission overview paper
 \citep{planck2014-a01}.  The major improvement  since 2013 is
 the use of solar distance and radiometer electronics box assembly (REBA)
 thermometry as pointing error templates that are fitted and corrected.

 As in \citet{planck2013-p03}, we determine the location of the HFI detectors
 relative to the satellite boresight using bright planet observations.
 Specifically, we use the first observation of Mars to define the nominal location of
 each detector. These locations do not correspond exactly to the physical
 geometry of the focal plane, since they include any relative shift induced by
 imperfect deconvolution of the time-response of the detectors, as described in
 Sect.~\ref{focal_plane_shift}. After this initial geometrical calibration,
 the reconstruction is monitored by subsequent planet observations. We find
 that the detector positions are stable to $< 10$\arcs\ with an rms measurement
 error of about 1\arcs. 

\subsection{Cosmic ray deglitching}\label{sec:deglitching}

The deglitching method, described  in~\cite{planck2013-p03e},
consists of flagging the main part (with S/N $> 3.3$) of the response to
each cosmic ray hit and subtracting a tail computed from a template for the
remaining part. The flagged part is not used in the maps. The method and
parameters are unchanged since 2013.  As described  by
\citet{planck2013-p03e} and \citet{catalano2014}, three main populations of glitches have
been identified: (i) short glitches (with a peaked amplitude
distribution), due to the direct impact of a cosmic particle on the grid or
the thermometer; (ii) long glitches, the dominant population, due to the impact of
a cosmic particle on the silicon die, or support structure of the bolometer's
absorber; (iii) and slow glitches, with a tail similar to the long ones and
showing no fast part. The physical origin of this last population is not
yet understood.

The polarization-sensitive bolometers (PSBs) are paired, with two bolometers (called $a$ and
$b$) sharing the same housing~\citep{Jones2003}. The dies are thus
superimposed and most of the long glitches seen in one detector are also seen
in the other. A flag is computed from the sum of the $a$ and $b$
signal-subtracted TOI after each has been deglitched individually. This new
flag is then included in the total flag used for both the $a$ and $b$
bolometers.

For strong signals, the deglitcher threshold is auto-adjusted to cope with
source noise, due to the small pointing drift during a ring.
Thus, more glitches are left in data in the vicinity of bright sources (such as
the Galactic centre) than elsewhere. To mitigate this effect near bright
planets, we flag and interpolate over the signal at the planet location prior
to the TOI processing. While a simple linear interpolation was applied in the
first release \citep{planck2013-p06}, an estimate of the background signal
based on the sky map is now used to replace these samples.  For
the 2015 
release, this is done for Jupiter in all HFI frequency bands, for Saturn at
$\nu\ge 217\,\mathrm{GHz}$, and for Mars at $\nu\ge 353\,\mathrm{GHz}$.

Nevertheless, for beam and calibration studies (see Sect.~\ref{sec:beam} and
Paper~2), the TOI of all planet crossings, including the planet signals, is
needed at all frequencies. Hence, a separate data reduction is done in parallel
for those pointing periods and bolometers. For this special production, the
quality of the deglitching has been improved with respect to the 2013 data
analysis (see Appendix~\ref{sec:scanningbeampipeline}).

\subsection{$^4$He-JT cooler pickup and ring selection}
\label{sec:ring_selection}

\Planck\ scans a given ring on the sky for
roughly 45\,min before moving on to the next ring \citep{planck2013-p01}. The data between these
rings, taken while the spacecraft spin-axis is moving, are discarded as
``unstable.'' The data taken during the intervening ``stable'' periods are
subjected to a number of statistical tests to decide whether they should be flagged
as unusable \citep{planck2013-p03}. This procedure continues to be
adopted  for the present data release. Here we describe an additional selection process
introduced to mitigate the effect of the 4-K lines on the data.

The \HeJT\ cooler is the only moving part on the spacecraft. It is driven at
40\Hz, synchronously with the HFI data acquisition. Electromagnetic and
microphonic interference from the cooler reaches the readout boxes and wires in
the warm service module of the spacecraft and appears in the HFI data as a set
of very narrow lines at multiples of 10\Hz\ and at
17\Hz~\citep{planck2013-p03}. The subtraction scheme used for the 2013
release, used here as well, is based on measuring the Fourier coefficients
of these lines and interpolating them for the rings that have one of the
harmonics of their spin frequency very close to a line frequency -- a
so-called ``resonant'' ring. However, it was noticed that the amplitude of the
lines increased in the last two surveys. Therefore, as a precautionary step,
resonant rings with an expected line amplitude above a certain threshold are
now discarded.

\begin{figure}[ht!]
\begin{center}
\includegraphics[width=\columnwidth]{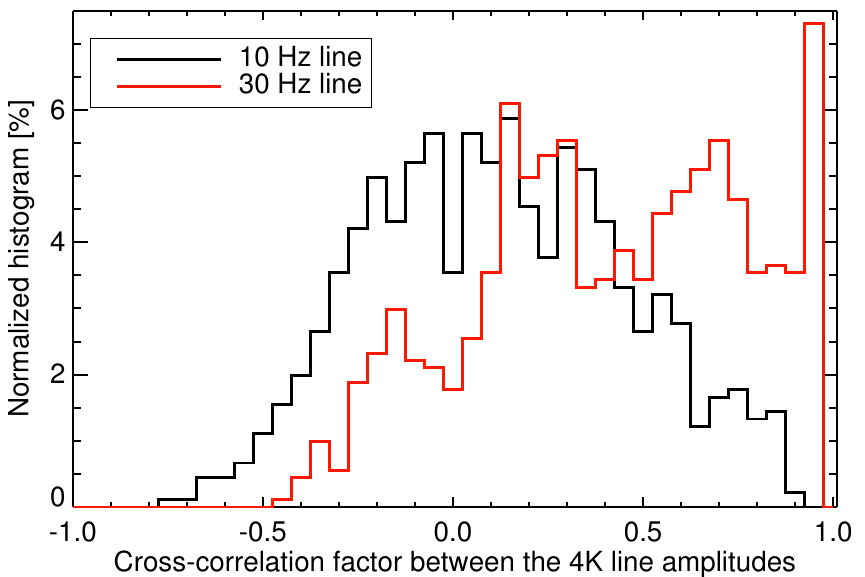}
\caption{\label{fig:Cooler30HzCorr} Normalized histograms of the correlation
  coefficients of the 10 and 30\Hz\ 4-K line amplitudes. The amplitude is
  computed per ring and per bolometer from the two coefficients (sine and
  cosine) of a given line. For each pair of distinct bolometers (from 100 to
  353\GHz), a correlation coefficient is computed between the two amplitudes
  during the mission. The black normalized histogram shows the 10\Hz\ line
  correlation coefficients of the 903 ($=43\times42/2$) pairs. The red curve
  shows the 30\Hz\ line histogram. The 30\Hz\ line is clearly correlated
  between different bolometers. This is the only line that shows a
  significant correlation.}
\end{center}
\end{figure}

\begin{figure}[ht!]
\begin{center}
\includegraphics[width=\columnwidth]{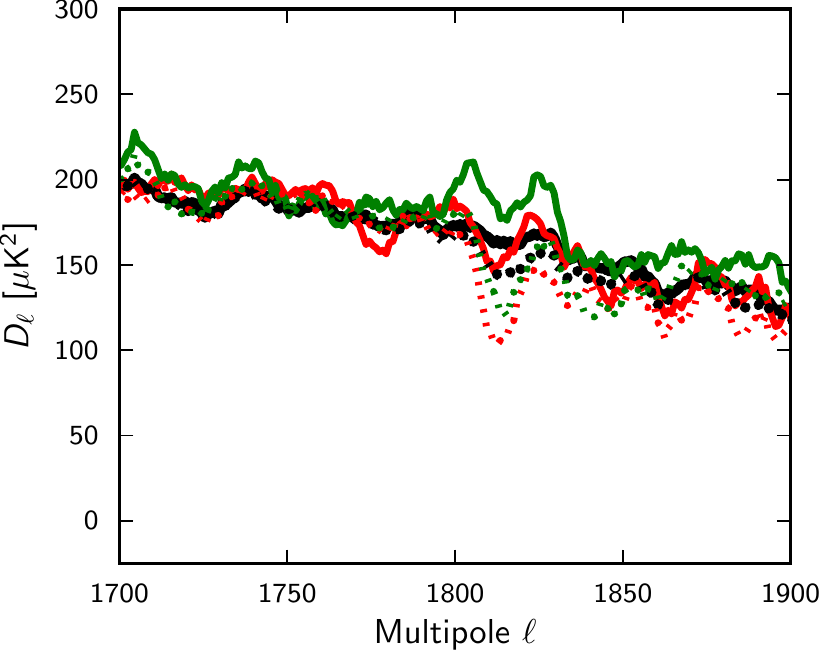}
\caption{\label{fig:Cell4K1800} Temperature cross power-spectrum of the
  217\GHz\ detector sets 1 and 2 for the full mission (black) and yearly cuts
  (Year~1 in red and Year~2 in green), comparing the 2013 (dashed lines) and
  2015 (continuous lines) data release.}
\end{center}
\end{figure}

In contrast to the other lines, the 30\Hz\ line signal is
correlated across bolometers (see Fig.~\ref{fig:Cooler30HzCorr}). It is
therefore likely that the 4-K line removal procedure leaves correlated
residuals on the 30\Hz\ line. The consequence of this correlation is that the
cross-power spectra between different detectors can show excess noise at
multipoles around $\ell\simeq1800$\footnote{The spacecraft spin period of one minute implies a
  correspondence between a TOI frequency of 30\Hz\ and $\ell\simeq1800$.} (see
the discussion in Section~1 of \citealt{planck2013-p11} and in Section~7 of
\citealt{planck2013-p08}).
 An example, computed with the {\tt anafast}
code in the {\tt HEALPix} package \citep{gorski2005}, is shown in
Fig.~\ref{fig:Cell4K1800}.  To mitigate 
this effect, we discard all 30\Hz\ resonant rings for the 16 bolometers
between 100 and 353\GHz\ for which the median average of the 30\Hz\ line
amplitude is above $10\,\mathrm{aW}$. Doing so, the $\ell=1800$ feature disappears. This issue is also addressed in Section~3.1
of \cite{planck2014-a15}.

\begin{figure*}[ht!]
\begin{center}
\includegraphics{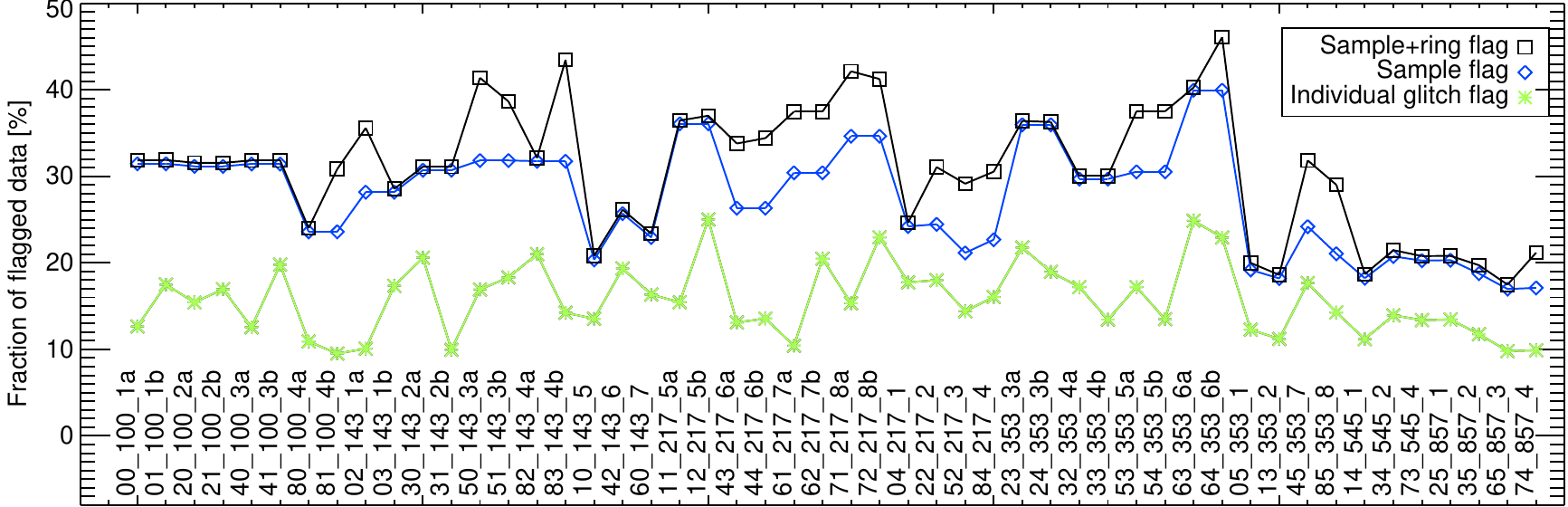}
\caption{\label{fig:DiscardedData} Fraction of discarded data per bolometer
  (squares with black line). The fraction of data discarded from glitch
  flagging alone is shown with stars and the green line. The blue line
  with diamonds indicates the average fraction of discarded samples in valid
  rings. The two bolometers showing permanent random telegraphic
  signal (RTS), i.e. 143-8 and 545-3, are not shown, because they are
  not used in the data processing. }
\end{center}
\end{figure*}

Figure~\ref{fig:DiscardedData} shows the fraction of discarded samples for
each detector over the full mission. It gathers the flags at the sample level,
which are mainly due to glitches and the depointing between rings. It also
shows the flags at the ring level, which are mostly due to the 4-K lines, but
are also due to solar flares, big manoeuvres, and end-of-life calibration
sequences, which are common to all detectors.  The main difference from the nominal
mission, presented in the 2013 papers, appears in the
fifth survey, which is somewhat disjointed,
due to solar flares arising with the increased solar activity, and to special
calibration sequences. The full cold \Planck-HFI mission lasted 885~days,
excluding the calibration and performance verification (CPV) period of 1.5 months.
 During this time, HFI data losses amount to
31\,\%, the majority of which comes from glitch flagging, as shown in
Table~\ref{tab:OverallStat}. The fraction of samples flagged due to
solar system objects (SSO), jumps, and saturation~\citep{planck2013-p03f}
is below 0.1\,\%, and hence negligible.

\begin{table*}[ht!]
\begin{center}
  \caption{ Overall budget of discarded data samples for the 50 valid
    bolometers. Two bolometers, one at 143\GHz\ and one at 545\GHz, cannot be
    used, due to a permanent random telegraphic signal (RTS) and are excluded
    from the statistics. A global average is given for glitches. Here,
    depointing denotes only the standard manoeuvres from one ring to another. Big
    manoeuvres are included in the common discarded rings (but not the 4-K line
    selection process). The RTS affects six bolometers episodically.  The 4-K
    line selection process affects 20 bolometers (see
    Sect.\ref{sec:ring_selection}). The range of values obtained for different
    bolometers is given in the last column. Note that percentages do not add
    up to the total, since depointing, common rings, RTS and 4-K flagged samples
    are already flagged at the 20\,\% level due to glitches.
\label{tab:OverallStat}
}

\vskip -8mm
\setbox\tablebox=\vbox{
\newdimen\digitwidth
\setbox0=\hbox{\rm 0}
\digitwidth=\wd0
\catcode`*=\active
\def*{\kern\digitwidth}
\newdimen\signwidth
\setbox0=\hbox{+}
\signwidth=\wd0
\catcode`!=\active
\def!{\kern\signwidth}
\newdimen\pointwidth
\setbox0=\hbox{.}
\pointwidth=\wd0
\catcode`@=\active
\def@{\kern\pointwidth}
\halign{\hbox to 2.2in{#\leaderfil}\tabskip 1.0em&
        \hfil#\hfil&
        \hfil#\hfil\tabskip 0pt\cr
\noalign{\vskip 3pt\hrule\vskip 1.5pt\hrule\vskip 5pt}
\omit\hfil Origin\hfil&  Mean fraction loss [\%] & Range [\%]\cr
\noalign{\vskip 4pt\hrule\vskip 6pt}
Glitch                 & 20 & 9--32\cr
\noalign{\vskip 4pt}
Depointing             & 8  & 8--8\cr
\noalign{\vskip 4pt}
Common discarded rings & 2  & 2--2\cr
\noalign{\vskip 4pt}
RTS                    & 0  & 0--4\cr
\noalign{\vskip 4pt}
4-K                   & 4  & 0--16\cr
\noalign{\vskip 4pt}
\bf{Total}             & \bf{31} & 17--46\cr
\noalign{\vskip 4pt}
\noalign{\vskip 5pt\hrule\vskip 4pt}
}}
\endPlancktable
\end{center}
\end{table*}

\subsection{Detector time response}\label{sec:tau}

As noted in \citet{planck2013-p03c} and \citet{planck2011-1.5}, the detector time response
is a key calibration parameter for HFI.  It describes the relation between the
optical signal incident on the detectors and the output of the readout
electronics.  This relation is characterized by a gain, and a time shift,
dependent on the temporal frequency of the incoming optical signal.  As in
previous releases, it is described by a linear complex transfer function in
the frequency domain, which we call the time transfer function.  This
time transfer function must be used to deconvolve the data in order to
correct the frequency-dependent time shift, which otherwise significantly
distorts the sky signal.
The deconvolution also restores the frequency dependent gain.  It is worth
noting that: (i) the deconvolution significantly reduces the long tail of the
scanning beam; (ii) it also symmetrizes the time response, which allows us to
combine surveys obtained by scanning in opposite directions; and (iii) given that
the gain decreases with frequency, the deconvolution boosts the noise at high
frequency, as can be seen in Figs.~\ref{fig:noise_spectra}
and~\ref{fig:PSDmodel}.  In order to avoid unacceptably high noise in the
highest temporal frequencies, a phaseless low-pass filter is applied, with
the same recipe as in Section~2.5 of~\cite{planck2013-p03c}.  This process
results in a slightly rising noise in the high frequency part of the noise
power spectrum, in particular for the slowest 100\GHz\ detectors.  This noise
property is ignored in the map making process, which assumes white noise and
low frequency noise. We note, however, that the 100\GHz\ bolometers data are not
used for CMB analysis at smaller angular scales, due to their wider main
beams.

For this release, the time transfer function is based on the same model as the
previous release. As a function of the angular frequency $\omega$, it is
defined by
\begin{equation}
\mathrm {TF}(\omega) = F(\omega) H'(\omega;S_\mathrm{phase},\tau_\mathrm{stray}),
\end{equation}
where $F(\omega)$ is the term associated with the bolometer response, and
$H'(\omega;S_\mathrm{phase},\tau_\mathrm{stray})$ is the analytic model of the
electronics transfer function, whose detailed equations and parameters are
given in Appendix~A 
of~\cite{planck2013-p03c}.
The electronics term depends only on two parameters: the phase shift between
the AC bias current and the readout sampling, encoded by the parameter
$S_\mathrm{phase}$; and the time constant $\tau_\mathrm{stray}$ associated with
the stray capacitance of the cables connecting the bolometers to the bias
capacitors (the stray capacitance enters in the $h_0$ term in the series of
filters reported in table~A.1 of \citealt{planck2013-p03c}).

The bolometer time transfer function is described by the sum of five 
single-pole low-pass functions, each with a time constant $\tau_i$ and an 
associated amplitude $a_i$: 
\begin{equation}
F(\omega) = \sum_{i=1,5} \frac{a_i}{1 + i\omega\tau_i}.
\end{equation}
In this version, five time constants are used rather than four, 
and the values of the parameters have  been measured
from a different combination of data than used previously.
The extra low-pass function and the consequent
parameter updates are motivated by the discovery of a time delay between the
measured CMB dipole and the expected one. This time delay, or phase shift, is
interpreted as the effect of an extra low-pass function, not accounted for in
the previous version of the deconvolution process.  As we will see,
some level of time delay in the dipole remains 
even after deconvolution of the five low-pass functions. This small residual can
generate problems in gain fitting and dipole subtraction, and is treated more
efficiently at the map making level (see Paper~2).

\begin{figure*}[t!]
\begin{center}
\includegraphics{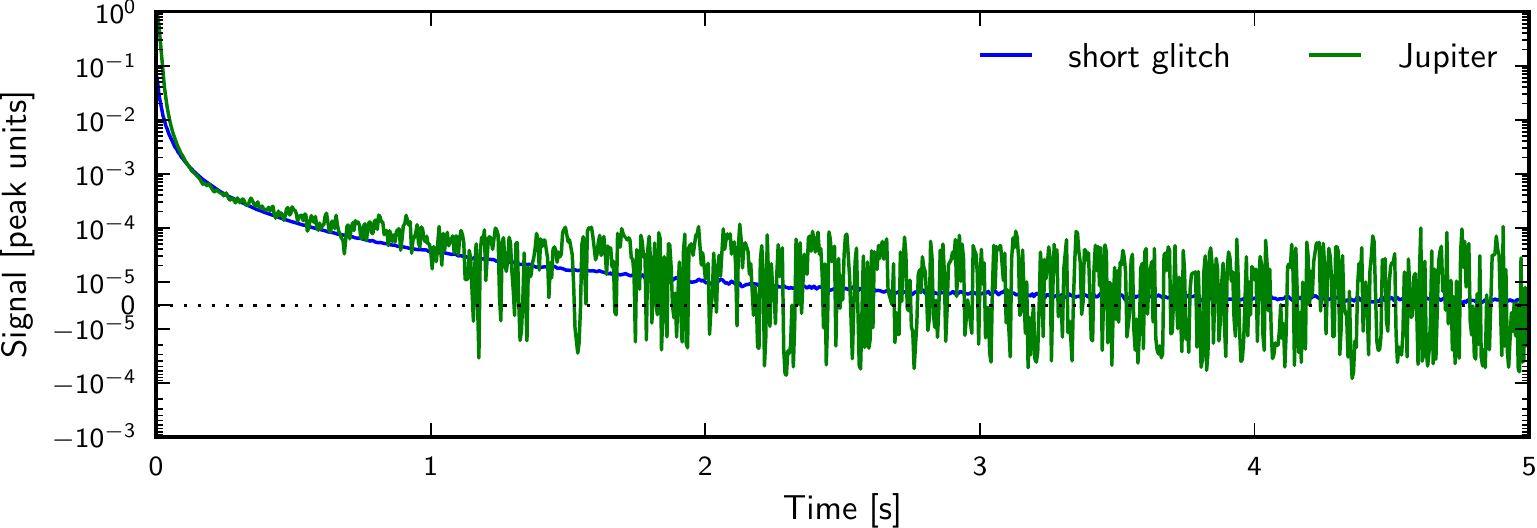}
 \caption{\label{fig:glitchvsjupiter} Impulse response of bolometer
   143-6 to short glitches and to Jupiter. The plotted axis is
 linear within the range $\pm 10^{-5}$ and logarithmic elsewhere.}
\end{center}
\end{figure*}

In addition to the dipole shift, new information comes from the stacking of
glitches induced by high energy particles hits.
As described in~\cite{planck2013-p03e}, short glitches are due to direct
interaction of particles with the bolometer grid or thermistor. The time
response of short events is then representative of the response to photons. Short
glitches show a long tail, if enough events are stacked together.  It was
observed that the signals obtained by stacking short glitches and by stacking
Jupiter scans decay to zero with the same time constants (see
Fig.~\ref{fig:glitchvsjupiter}).  Since the short glitch stacking has the
benefit of a much higher signal-to-noise ratio, it provides the most sensitive
measurement of the longest time constants.  Although the
physical process of energy injection into the detector is different for microwave photons
and energetic particles, the heat dissipation is expected to be the same.  For this reason,
it was decided to take only the time constants from the glitches, and not the
associated amplitudes.  The impact of the incomplete correction of the transfer
function is discussed in Sect.~\ref{sec:time_response_uncertainties}.

In summary, the values of the bolometer time transfer function parameters,
$a_i$ and $\tau_i$, are measured with the following logic:

\begin{itemize}

\item the two fastest time constants, $\tau_1$ and $\tau_2$, and the
  associated amplitudes $a_1$ and $a_2$, are unchanged with respect to the
  previous version, i.e., they are estimated from planet observations;
\item the two longest time constants, $\tau_4$ and $\tau_5$, 
are estimated from short glitch stacking, together with the
$a_5/a_4$ ratio;
\item $\tau_3$, $a_3$, and $a_4$ are fitted from Jupiter scans, keeping
  $\tau_1$, $a_1$, $\tau_2$, and $a_2$ fixed, while the value of $a_5$ is set to keep
  the same ratio $a_5/a_4$ as in the glitch data;
\item $a_5$ is fitted from the CMB dipole time shift.  
  It should be noted that
  the same dipole time shift can be obtained with different combinations of
  $a_5$ and $\tau_5$ and for this reason, $\tau_5$ is recovered from the short
  glitches, and only $a_5$ from the dipole time shift.  
\end{itemize}

This process was used for all channels from 100 to 217\GHz.  For the 353\GHz\
detectors, different processing was needed to avoid strong non-optical
asymmetries in the recovered scanning beam (see
Appendix~\ref{sec:appendix_time_response} for details).  For the submillimetre
channels, 545 and 857\GHz, the time transfer function is identical
to that of the 2013 release.  The values of the time response parameters are reported in
Table~\ref{table:time_response_pars}.  
The model and parameters will be improved by continuing this activity
for future releases.

\subsubsection{Time response errors}\label{sec:time_response_uncertainties}

The beam model is built from time-ordered data deconvolved by the time
transfer function. For this reason, and considering the constant rotation rate
of \Planck, the measured scanning beam absorbs, to a large extent, mismatches between
the adopted time transfer function and the true one~\citep{planck2013-p03c}.
In this sense, errors and uncertainties in the time transfer function should
not be propagated into an overall window function uncertainty, since the time
response acts as an error-free regularization function.  Biases and
uncertainties are taken into account in the beam error budget.

The beam window function correction of HFI's angular power
  spectrum will not work on time scales 
longer than than roughly $0.5\,\mathrm{s}$, given the size of the scanning
beam map and the constant scan rate. Simulations with varying time response
parameters show that errors at frequencies above 2\Hz\ are absorbed in the
scanning beam map and propagated correctly to the effective beam window
function to better than the scanning beam errors.  Errors on time scales
longer than $0.5\,\mathrm{s}$ are not absorbed by the scanning beam map, but
propagate into the shifted dipole measurement and the relative calibration
error budget (see Paper~2).

Additionally, the time response error on time scales longer than
$0.5\,\mathrm{s}$ can be checked by comparing the relative
amplitude of the first acoustic peak of CMB anisotropies between
frequency bands.   The main calibration of HFI is performed with the
CMB dipole (appearing in the TOI at 0.016\,Hz), while the first
acoustic peak at $\ell \approx 200$ appears at 6\,Hz.  
Table~3 in \cite{planck2014-a01} shows that between 
 100 and 217\,GHz the agreement is better than 0.3\,\%.

\subsubsection{Focal plane phase shift from fast Mars scan }\label{focal_plane_shift}

In December 2011, \Planck\ underwent a series of HFI end-of-life tests.  Among
these, a speed-up test was performed increasing the spin rate to 1.4~rpm
from the nominal value of 1\,rpm.  The test was executed on 7--16 December 2011, 
and included an observation of Mars. Right after
the test, a second Mars observation followed, at nominal speed.
The main result of this test was the ability to set the real position of the
detectors in the focal plane, which, when scanning at constant speed, is completely
degenerate with a time shift between bolometer data and pointing data. This
time shift was supposed to be zero. During the
test, it was found that the deconvolved bolometer signals peaked at a
different sky position for the nominal scans and for the fast scans.  This
discrepancy was solved by introducing a time-shift between bolometer readout and
pointing data. This time-shift is not the same for all detectors, and ranges
from 9.5 to 12.5~ms, with a corresponding position shift ranging from 3\parcm3 to
4\parcm5 in the scanning direction. Due to time constraints, it
was not possible to observe Mars twice with the full focal plane, so  it was decided to
favour the CMB channels and the planet was observed with all the 100, 143,
217, and 353\GHz\ detectors.  For the other detectors, we used the average
time-shift of all the measured detectors.  Notably, this detector-by-detector
shift resulted in a better agreement of the position in the focal plane of the
two PSBs in the same horn.

Working with fast spin-rate deconvolved data is
complicated by aliasing effects. For this reason the fitting procedure 
followed a forward sense approach, by modelling the signal with a beam centred
in the nominal position, then convolving the fast and nominal timelines with
the time transfer function, and fitting the correct beam centres by comparing
data and model for both fast and nominal spin-rates.  For the same reason, a
direct comparison of deconvolved timeline has not proven to provide 
better constraints on the time-response parameters.

\section{Planets and main beam description}
\label{sec:beam}
 
We follow the  nomenclature of \cite{planck2013-p03c}, where the ``scanning beam'' is
defined as the coupled response of the optical system, the deconvolved time
response function, and the software low-pass filter applied to the data.
The ``effective bea''m represents the averaging of signal due to
the scanning of the telescope and mapmaking, and varies from pixel to
pixel across the sky.  

Here we redefine the ``main beam'' to be the scanning beam out to 100$\arcm$ from
the beam axis.  The sidelobe structure at this radius is dominated by
diffraction at the mirror edges and falls as $\propto\theta^{-3}$, where
$\theta$ is the angle to the main beam axis.  The main beam is used to compute
the effective beam and the effective beam window function, which describes the
filtering of sky signals.

\begin{figure*}[ht!]
\begin{center}
\includegraphics{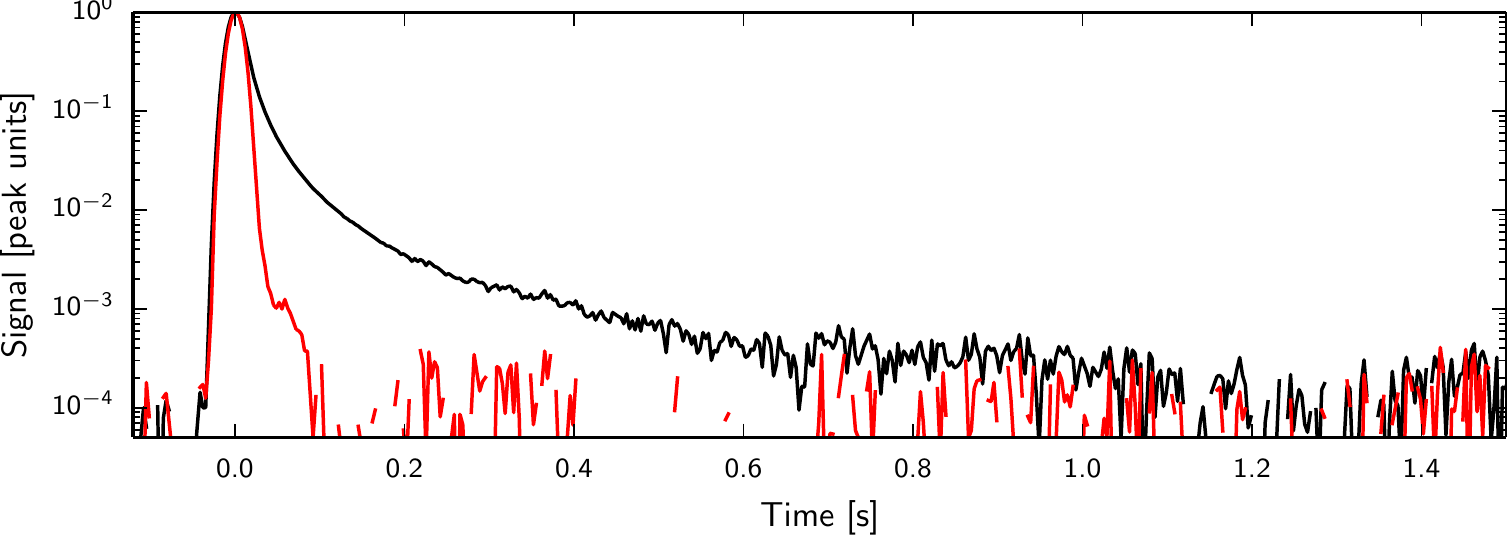}
 \caption{\label{fig:deconvjupiter} A scan across Jupiter with
   bolometer 143-1b, pre-deconvolution in black, post-deconvolution in
 red.}
\end{center}
\end{figure*}

The smearing of the main beam cannot be significantly reduced without boosting
the high-frequency noise. The regularization function (a low-pass
filter) chosen has approximately
the same width as the instrumental transfer function. The
deconvolution significantly reduces the long tail of the scanning beam
(see Fig.~\ref{fig:deconvjupiter}).  The deconvolution also produces a
more symmetric time response, so residual 
``streaking'' appears both ahead and behind the main beam, though ahead
of the beam it is at a level of less than $10^{-4}$ of the peak response. 

The ``far sidelobes'' are defined as the response from $\theta > 5^\circ$,
 roughly the minimum in the optical response.  The response begins to
rise as a function of angle $\theta$ beyond this due to spillover.  The far sidelobes
are handled separately from the beam effects (see
Sect.~\ref{sec:FSLwindowfunction} for justification and Paper~2 for details).

Observations of planets are used to estimate the main beams and to calibrate the
545 and the 857\GHz\ channels. The main beam is needed to correct for the
filtering of the CMB sky by the instrument, details of which can be found in
\cite{planck2013-p03c}. There were several key changes in the reconstruction
of the main beam since the 2013 data release, described in detail in
Appendix~\ref{sec:scanningbeampipeline}.
\begin{itemize}
\item The TOI from Saturn and Jupiter observations are merged
  prior to B-spline decomposition, taking into account residual pointing errors and
  variable seasonal brightness.  This is
  achieved by determining a scaling factor and a pointing offset by fitting
  the TOI to a template from a previous estimate of the scanning
  beams. We iterate the planet data treatment, updating the template with the
  reconstructed scanning beam. The process converges in five iterations to  an
  accurcay of better than $0.1\,\%$ in the effective beam window function.
\item Steep gradients in the signal close to the planet reduce the
  completeness of the standard glitch detection and subtraction procedure, so
  the planet timelines are deglitched a second time.
\item The beam pipeline destripes the planet data, estimating a single
  baseline between $3^\circ$ and $5^\circ$ before the peak for each scanning
  circle. Baseline values are smoothed with a sliding window of 40 circles. The
  entire scanning circle is removed from the beam reconstruction if the
  statistic in the timeline region used to estimate the baseline is far from
  Gaussian.
\item The main beam is now recontructed on a square grid that extends
  to a radius of 100\arcm\ from the centroid, as opposed to 40\arcm\ in
  the 2013 data release.  The cutoff of 100\arcm\  was chosen so 
  that a diffraction model of the beam at large angles 
  from the centroid predicts that less than $5 \times
  10^{-5}$ of the total solid angle is missing.
\item The scanning beam is constructed by combining data from Saturn
  observations, Jupiter observations, and physical optics models using {\tt
    GRASP} software.
\item No apodization is applied to the scanning beam map. 
\end{itemize}

The update of the time response deconvolved data has slightly changed the
scanning beam, the effective beam solid angles, and the effective beam window
functions.

\subsection{Hybrid beam model}

\begin{figure}
  \begin{center}
    \includegraphics[width=1\columnwidth]{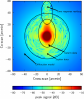} 
    \caption{\label{fig:hybridbeamdiagram}Scanning beam map for
      detector 143-6 with a rough illustration of the regions that are
      handled with a different data selection or binning.}
  \end{center}
\end{figure}

\begin{figure*}[!ht]
\centerline{\includegraphics[width=1.0\textwidth]{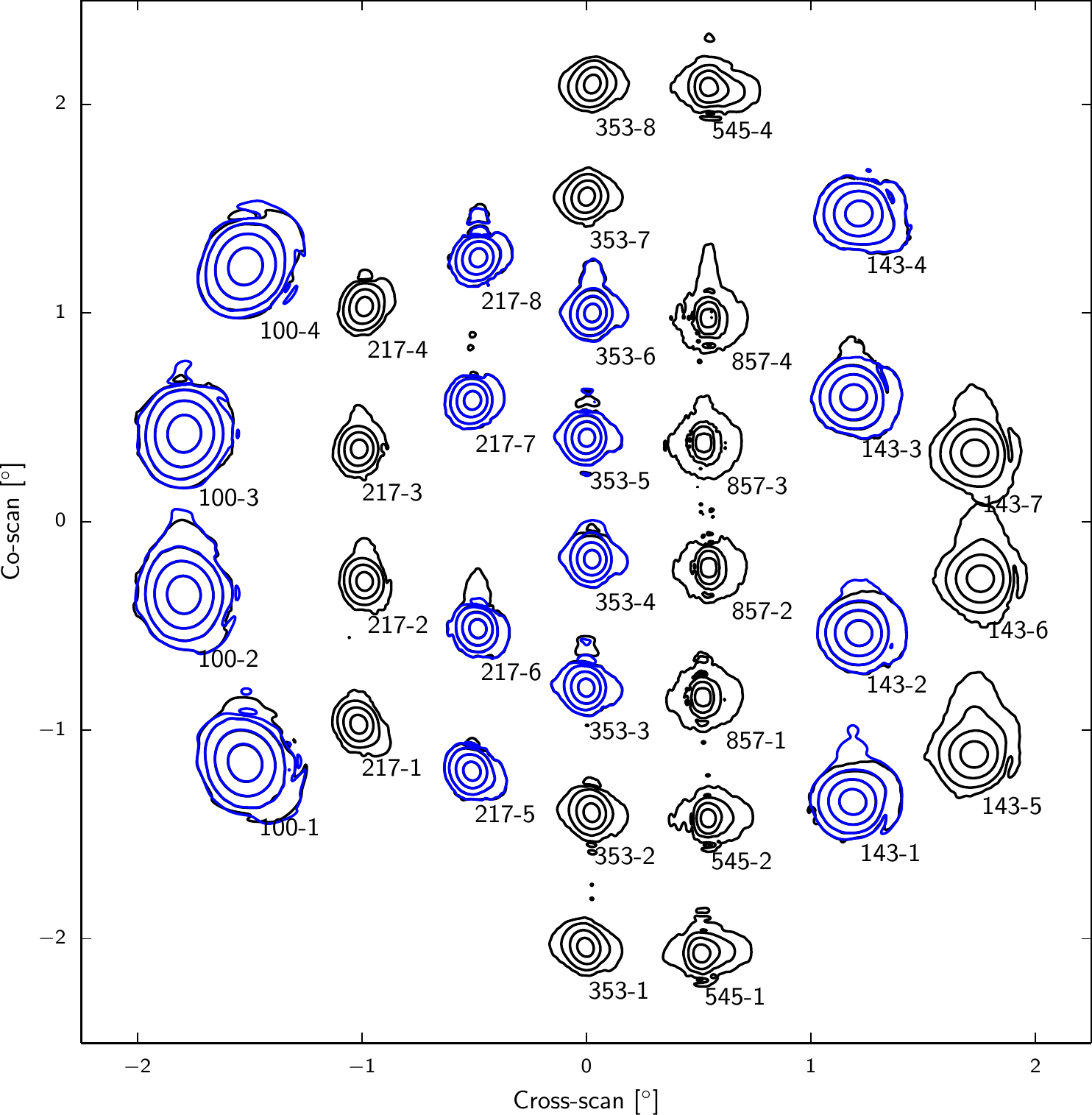}}
\caption{\label{fig:FocalPlanePlot} B-spline hybrid scanning beams
  reconstructed from Mars, Saturn, and Jupiter.  The beams are 
  plotted in logarithmic contours of $-3$, $-10$, $-20$, and $-30$\,dB from the
  peak. PSB pairs are indicated with the $a$ bolometer in black and
  the $b$ bolometer in blue.}
\end{figure*}

A portion of the near sidelobes was not accounted for in the effective beam
window function of the 2013 data~\citep{planck2013-p03c,planck2013-p01a}. To
remedy this, the domain of the main beam reconstruction has been extended to
100\arcm, with no apodization. Saturn data are used where they are 
signal-dominated. Where the signal-to-noise ratio of the Saturn data falls
below 9, azimuthally binned Jupiter data are used.  At larger angles, below the
noise floor of the Jupiter data, we use a power law ($\propto\theta^{-3}$),
whose exponent is derived from {\tt GRASP}, to extend the beam model to
100\arcm.  Figure~\ref{fig:hybridbeamdiagram} shows a diagram of the regions 
handled differently in the hybrid beam model.
A summary of the solid angles of the hybrid beams is shown in
Table~\ref{table:ScanningBeamSolidAngleErrorBudget}.
\begin{table}[tb]                 
\begingroup
\newdimen\tblskip \tblskip=5pt
\caption{ Band-average scanning beam solid angle ($\Omega_{\mathrm{SB}}$) and Monte Carlo-derived errors
  ($\Delta\Omega_{\mathrm{MC}}$)
  including noise, residual glitches, and pointing uncertainty.}
\label{table:ScanningBeamSolidAngleErrorBudget}                            
\nointerlineskip
\vskip -3mm
\footnotesize
\setbox\tablebox=\vbox{
   \newdimen\digitwidth 
   \setbox0=\hbox{\rm 0} 
   \digitwidth=\wd0 
   \catcode`*=\active 
   \def*{\kern\digitwidth}
   \newdimen\signwidth 
   \setbox0=\hbox{-} 
   \signwidth=\wd0 
   \catcode`!=\active 
   \def!{\kern\signwidth}
\halign{\hbox to 2cm{#\leaderfil}\tabskip 1em&\hfil# \tabskip 2em&\hfil# \tabskip 0pt\cr                            
\noalign{\doubleline}
\omit\hfil Band\hfil&\omit\hfil $\Omega_{\mathrm{SB}}$ \hfil&  \omit\hfil
$\Delta\Omega_{\mathrm{MC}}$ \hfil\cr
\omit\hfil [GHz]\hfil&\omit\hfil [arcmin$^2$]\hfil&\cr
\noalign{\vskip 3pt\hrule\vskip 4pt}
100&104.62&0.13\,\%\cr
143&*58.80&0.07\,\%\cr
217&*26.92&0.13\,\%\cr
353&*25.93&0.09\,\%\cr
545&*25.23&0.08\,\%\cr
857&*23.04&0.08\,\%\cr
\noalign{\vskip 3pt\hrule\vskip 5pt}
}
}
\endPlancktablewide                 
\endgroup
\end{table}                        
Figure~\ref{fig:FocalPlanePlot} shows a contour plot of all the
scanning beams referenced to the centre of the
focal plane.

\subsection{Effective beams and window functions}\label{sec:effectivebeams}

As described in \cite{mitra2010} and \cite{planck2013-p03c}, the {\tt FEBeCoP} code is used to
compute the effective beam (the scanning beam averaged over the scanning
history) and the {\tt FEBeCoP} and {\tt Quickbeam} codes are used to compute
the effective beam window functions. 
Statistics of the effective beams are shown in Table~\ref{table:BeamSolidAngle}.

Using Saturn and Jupiter to reconstruct the main beam introduces a small bias
due to the large disc size of the planets as compared to Mars~\citep[see
Fig.~8 of][]{planck2013-p03c}. Additionally, Saturn's ring system introduces a
slight frequency-band dependence of the effective size of the planetary
disc~\citep{planck2014-a33}. Because of Saturn's small size relative to the
HFI beams, for $\ell < 4000$ the symmetric part of Saturn's shape dominates
the window function. 
There is
a small variation with HFI band in the apparent mean size of Saturn due to the
different ring system temperatures, ranging from 9\parcs 25 at 100\GHz\ to
10\parcs 2 at 857\GHz.  Because this variation introduces a bias of less than
2$\times10^{-5}$ in $B^2_\ell$ for $\ell \le 4000$, we ignore it and use a
correction derived for a mean 9\parcs 5 Saturn disc for all bands.

The two effective-beam codes handle temperature-to-polarization leakage
in slightly different ways.  The dominant leakage comes from
differences in the scanning beams of the polarization-sensitive
detectors (See Sect.~\ref{sec:TtoPleakage}).  
The {\tt Quickbeam} code mimics the data model of the mapmaking code and
assumes that every polarization-sensitive detector at a given
frequency has the same beam shape, and thus produces a single
effective beam window function.  Temperature-to-polarization leakage
is handled later as a set of parameters in the likelihood of the angular
power spectra.
The  {\tt FEBeCoP} code produces effective beam window functions
for the polarized power spectra that account for differences in the
main beam.  However, these are computed as the average power leakage of a
given sky signal from temperature to polarization.  Hence, these
polarized window functions are not strictly instrumental 
parameters, since they rely on an assumed fiducial temperature angular power spectrum.

\begin{table*}[tb]                 
\begingroup
\newdimen\tblskip \tblskip=5pt
\caption{ Mean values of effective beam parameters for each HFI 
  frequency. The error in the solid angle $\sigma_\Omega$ comes from
  the scanning beam error budget.  The spatial variation $\Delta \Omega$ is 
  the rms variation of the solid angle across the sky.  The reported FWHM is
  that of the  Gaussian whose solid angle is equivalent to that of the
  mean effective beam. $\Omega_1$ and $\Omega_2$ are the solid angles contained  within circles of radius 1 and 2\,FWHM, respectively
  (used for aperture photometry as described in Appendix A of \citealt{planck2013-p05}).}  
\label{table:BeamSolidAngle}                            
\vskip -3mm
\footnotesize
\setbox\tablebox=\vbox{
   \newdimen\digitwidth 
   \setbox0=\hbox{\rm 0} 
   \digitwidth=\wd0 
   \catcode`*=\active 
   \def*{\kern\digitwidth}
   \newdimen\signwidth 
   \setbox0=\hbox{+} 
   \signwidth=\wd0 
   \catcode`!=\active 
   \def!{\kern\signwidth}
\halign{\hbox to 2cm{#\leaderfil}\tabskip 1em&\hfil#\hfil \tabskip 1em&\hfil#\hfil \tabskip 1em &\hfil#\hfil \tabskip 1em&\hfil#\hfil \tabskip 1em &\hfil#\hfil \tabskip 1em &\hfil#\hfil \tabskip 0pt\cr                            
\noalign{\doubleline}
\omit\hfil Band\hfil&$\Omega$&$\sigma_\Omega$&$\Delta \Omega$&FWHM&$\Omega_1$&$\Omega_2$\cr
\omit\hfil [GHz]\hfil&[arcmin$^2$]&[arcmin$^2$]&[arcmin$^2$]&[arcmin]&[arcmin$^2$]&[arcmin$^2$]\cr
\noalign{\vskip 3pt\hrule\vskip 5pt}
100&106.22&0.14&0.20&9.69&100.78&106.03\cr
143&*60.44&0.04&0.20&7.30&*56.97&*60.21\cr
217&*28.57&0.04&0.19&5.02&*26.46&*28.46\cr
353&*27.69&0.02&0.20&4.94&*25.32&*27.53\cr
545&*26.44&0.02&0.21&4.83&*24.06&*26.09\cr
857&*24.37&0.02&0.12&4.64&*22.58&*23.93\cr
\noalign{\vskip 3pt\hrule\vskip 5pt}
}
}
\endPlancktablewide                 
\endgroup
\end{table*}                        

\subsection{Beam error budget}
As in the 2013 release, the beam error budget is based on an eigenmode
decomposition of the scatter in simulated planet observations.  A
reconstruction bias is estimated from the ensemble average of the simulations.
We generate 100 simulations for each planet observation that include
pointing uncertainty, cosmic ray glitches, and the measured noise
spectrum.  Simulated glitches are injected into the timeline with the correct
energy spectrum and rate~\citep{planck2013-p03e} and are detected and removed
using the deglitch algorithm.  Noise realizations are derived as shown in
Sect.~\ref{sec:noise}.  Pointing uncertainty is simulated by randomizing the
pointing by 1\parcs5 rms in each direction.

The improved signal-to-noise ratio compared to 2013 leads to smaller
error bars; for instance, 
at $\ell=1000$ the uncertainties on $B^2_{\ell}$ are now $(2.2, 0.84,
0.81)\times 10^{-4}$ for 100, 143, and 217\GHz\ frequency-averaged maps
respectively,
reduced from the previous uncertainties  of $(61, 23, 20)\times 10^{-4}$.

A ``Reduced Instrument Model'' (RIMO; see Appendix~\ref{sec:officialHFI}) containing
the effective $B(\ell)$ for temperature and polarization detector sets, for
auto- and cross-spectra at 100 to 217\GHz, is included in the release, for a sky fraction
of 100\,\%. Another RIMO is provided for a sky fraction of 75\,\%.  They both
contain the first five beam error eigenmodes and their covariance
matrix, for the multipole ranges $[0, \ell_{\mathrm{max}}]$ with
$\ell_{\mathrm{max}} =$ 2000, 3000, 3000 at 100, 143, and 217\GHz\
respectively (instead of 
2500, 3000, 4000 previously). These new ranges bracket more closely the ones
expected to be used in the likelihood analyses, and ensure a better
determination of the leading modes on the customized ranges.

As described in Appendix A7 of \cite{planck2013-p08},
these beam window function uncertainty eigenmodes are used to build the
$\tens{C}(\ell)$ covariance matrix used in the
high-$\ell$ angular power spectrum likelihood analysis. It was found that the beam errors are
negligible compared to the other sources of uncertainty and have no noticeable
impact on the values or associated errors of the cosmological parameters.

\subsection{Consistency of beam reconstruction}

\begin{figure*}[ht!]
\begin{center}
\includegraphics[width=\textwidth]{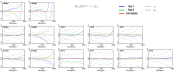}
\caption{\label{fig:Yearly_Beams_Consistency_WF} Comparison of Year~1 and Year~2
  based beams with the reference beam (Full Mission). Window functions are
  calculated using {\tt Quickbeam} in raster scan configuration. Error
  bars are computed using MC simulations for Year~1 and Year~2, taking the
  maximal one for each.}
\end{center}
\end{figure*}

\begin{figure}[ht!]
\begin{center}
\includegraphics[width=1\columnwidth]{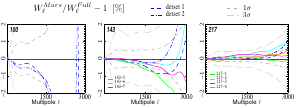}
\caption{\label{fig:Mars_Beams_Consistency_WF} Comparison of Mars-based beams with the reference beam (Full Mission). Mars-based beams are constructed from  Mars data for the main beam ($\lesssim10\arcm$) and Saturn and Jupiter data for the larger scales. Window functions are
  calculated using {\tt Quickbeam} in raster scan configuration. Error
  bars are computed using MC simulations for Mars.}
\end{center}
\end{figure}

To evaluate the accuracy and consistency of the beam reconstruction method,
we have compared beams reconstructed using Mars for
the main beam part instead of Saturn, and using data from Year~1 or Year~2
only. We compared the window functions obtained with
these new beams to the reference ones; new Monte Carlo simulations were
created to evaluate the corresponding error bars, and the results are shown in
Figs.~\ref{fig:Yearly_Beams_Consistency_WF} and \ref{fig:Mars_Beams_Consistency_WF}.        
Note that the effective beam window functions shown in these figures
are not exactly weighted for the scan strategy, rather we plot them in
the raster scan limit, where $B^2_\ell$ is defined as the sum over $m$
the $B^2_{\ell m}$. 
The eigenmodes were evaluated for these different data sets
allowing us to compare them with the reference beam using a $\chi^2$
analysis.  The discrepancy between each data set and the reference beam is
fitted using $N_{\mathrm{dof}}$=5 eigenmodes using
$\ell_{\mathrm{max}}=$1200, 2000, and 2500, for the 100, 143, and 217\GHz\ channels, respectively. The $\chi^2$ is then defined as:
\begin{equation} 
\chi^2 = \sum_{i=1}^{N_{\mathrm{dof}}} (c_i / \lambda_i)^2 
\label{eq:chi2_beam_consistency}
\end{equation}
where $c_i$ is the fit coefficient for eigenvector $i$, with eigenvalue
$\lambda_i$. For each data set, the $p$-value to exceed this $\chi^2$ value for
a $\chi^2$ distribution with $N_{\mathrm{dof}}$ degrees of freedom is indicated in
Table~\ref{tab:Pvalue_Yearly_Beam_Consistency}. A high $p$-value indicates that
the given data set is consistent with the reference beam within the
simulation-determined error eigenvectors, with 100\,\% indicating perfect
agreement.

We find excellent agreement between the yearly and nominal beams for the
SWB bolometer channels.  However, in order to obtain reasonable beam agreement
for the PSB detector sets, we find that the yearly beam errors must be
scaled by a factor of 4.  We conclude that
there is an unknown systematic error in the beam reconstruction that
is not accounted for in the simulations used to estimate the Monte
Carlo error bars. We assign a 
scaling factor of 4 to the error eigenmodes to account for this
systematic uncertainty.  The additional error scaling has a negligible
effect on the cosmological parameters.

\begin{table}[tb]                 
\begingroup
\newdimen\tblskip \tblskip=5pt
\caption{ $P$-values in percent for the $\chi^2$comparison of the  nominal beam with Year~1 and Year~2 data sets, following the definition in Eq.~\eqref{eq:chi2_beam_consistency}. }  
\label{tab:Pvalue_Yearly_Beam_Consistency}                            
\vskip -3mm
\footnotesize
\setbox\tablebox=\vbox{
   \newdimen\digitwidth 
   \setbox0=\hbox{\rm 0} 
   \digitwidth=\wd0 
   \catcode`*=\active 
   \def*{\kern\digitwidth}
   \newdimen\signwidth 
   \setbox0=\hbox{+} 
   \signwidth=\wd0 
   \catcode`!=\active 
   \def!{\kern\signwidth}
\halign{\hbox to 2cm{#\leaderfil}\tabskip 1em&\hfil#\hfil \tabskip 1em &\hfil#\hfil \tabskip 0pt\cr                            
\noalign{\doubleline}
\omit\hfil Band\hfil& Year~1 & Year~2 \cr
\noalign{\vskip 3pt\hrule\vskip 5pt}
100-ds1& 66.95 & 92.44\cr
100-ds2& 85.12 & 75.99\cr
143-ds1& 72.24 & 99.83\cr
143-ds2& *1.14 & 16.81 \cr
143-5& 94.88 & 97.57\cr
143-6& 96.14 & 98.78\cr
143-7& 93.78 & 95.67\cr
217-ds1& 78.12 & 69.74\cr
217-ds2& 27.33 & 30.87\cr
217-4& 95.99 & 68.44\cr
217-1& 99.08 & 97.49\cr
217-2& 97.18 & 98.86\cr
217-3& 97.59 & 99.07\cr
\noalign{\vskip 3pt\hrule\vskip 5pt}
}
}
\endPlancktablewide                 
\endgroup
\end{table}                        

\subsection{Colour correction of the beam shape}

In general, the measured beam shape is a function of the spectral
energy distribution (SED) of the measurement source.  This is of
particular concern, because we measure the beam on a source with a roughly
Rayleigh-Jeans SED, yet we use the effective beam window function
to correct the CMB. 

\cite{planck2013-p03c} described possible levels of this bias derived using {\tt
  GRASP} calculations with the pre-launch telescope
model~\citep{maffei2010,tauber2010b}.  The pre-launch calculations did not
agree well enough with the data to allow a direct application of the colour
correction.  A new telescope model based on flight data, and presented in a
forthcoming paper, predicts beams that agree better with the data at
100--217\GHz, but show worse agreement at 353\GHz.  The new model predicts a
different beam shape colour correction, though at a similar order of magnitude
to that shown in \cite{planck2013-p03c}. This work is ongoing and will be
completed after the 2015 release.

We search for a possible signature of the colour-correction effect in the
data.  First, we check for consistency in the CMB angular power spectra
derived from different detectors and frequency bands using the {\tt SMICA}
algorithm, in order to find discrepancies in beam shape and relative
calibration of different data subsets using the CMB anisotropies~\citep{planck2014-a13}.
There are hints of differences that are orthogonal to the beam error
eigenmodes and appear as changes in relative calibration, but the preferred
changes are at the 0.1\,\% level.  Because this correction is of the same
order of magnitude as the relative calibration uncertainty between detectors,
we treat it as insignificant.

Second, we look at the relative calibration between detectors within a
band for 
compact sources as a function of the source SED.  In this method, any
discrepancies due to solid angle variation with SED are degenerate with an
error in the underlying bandpass.  Additionally, because of the lack of
bright, compact sources with red spectra, this method is limited in its
signal-to-noise ratio.  In the limit that any detected discrepancy is
completely due to solid-angle variation with colour, at the level of 1\,\% in
solid angle we do not detect variations consistent with the physical optics predictions.

Given the lack of measurement of this effect at the current levels of
uncertainty, as well as  uncertainties in the modelling of
the telescope, we note that this effect may be present at a small
level in the data, but 
we do not attempt to correct for it or include it in the error budget.

\subsection{Effective beam window function at large angular scales}
\label{sec:FSLwindowfunction}
The main beam derived from planet observations and {\tt GRASP} modelling extends 100\arcm\
from the beam axis, so the effective beam window function does
not correct the filtering of the sky signal on larger scales (approximately
multipoles $\ell<50$).  In practice, due to reflector and baffle spillover,
the optical response  of HFI extends across the entire
sky~\citep{tauber2010b,planck2013-pip88}.  According to {\tt GRASP}
calculations, the far-sidelobe pattern (FSL) further than $5^\circ$ from the
beam centroid constitutes between 0.05\,\% and 0.3\,\% of the total solid
angle.  To first order, this is entirely described by a correction to the
calibration for angular scales smaller than the
dipole~\citep{planck2013-p01a}.

More correctly, the far-sidelobe beam (defined here as the optical response
more than $5^\circ$ from the beam axis) filters large-angular-scale sky signals
in a way that is coupled with the scan history of the spacecraft, and given a
perfect measurement of the far-sidelobe beam shape, an effective beam window
function could be constructed for the large angular scale with the same
procedures used for the main beam's effective beam window function.

We rely entirely on {\tt GRASP} calculations to determine the large-scale
response of the instrument.  These calculations are fitted to survey
difference maps (see \cite{planck2014-a33}), where the single free parameter is the
amplitude of the {\tt GRASP} far-sidelobe model beam.  Because of a
combination of the low signal-to-noise ratio of the data and uncertainty in
the {\tt GRASP} model, the errors on the fitted amplitude are of the order of
100\,\%. 
While ground-based measurements have given us confidence in our limits on
total spillover, we note that there is a large uncertainty in the exact shape of
the far sidelobes. Concerning a possible evolution of the far sidelobe during
the mission, we note that the background, as measured by the total power on
the bolometers, decreases then stabilizes as a function of time. This proves
that dust deposition on the \Planck\ mirrors is negligible. Hence, the far
sidelobes are likely to be very stable.

\begin{figure}[ht!]
\begin{center}
  \includegraphics[width=1\columnwidth]
{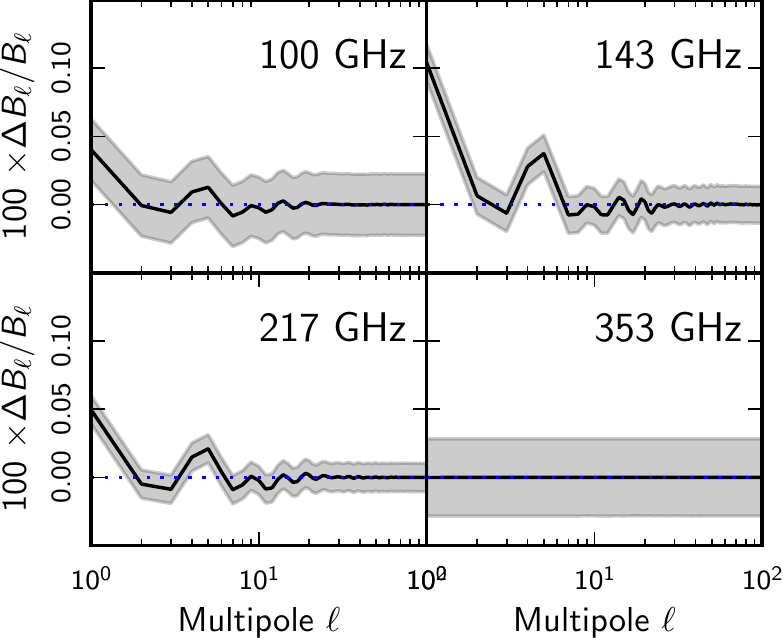}
\caption{\label{fig:FSL_EBWF} Estimate of the effective beam window function
  corrections due to far-sidelobe response (more than $5^\circ$ from the beam
  axis). Shading indicates $\pm1\sigma$ errors. The curves in the 353\GHz\
  panel are divided by a factor of 10; for this frequency the far sidelobe
  response amplitude is very low compared to the estimated
  errors.}
\end{center}
\end{figure}

Fig.~\ref{fig:FSL_EBWF} shows the correction to the effective beam window
function due to the best-fit far-sidelobe model.  The error bars are from the
amplitude error in the fit \citep{planck2014-a33} and they show that the uncertainty is
larger than the predicted beam window function correction except at the dipole
scales ($\ell=1$), aside from a marginally significant, but
negligibly small, bump at $\ell=5$.
The filtering of the CMB signal with the far-sidelobe beam
is thus insignificant and can be ignored as a component of the effective
beam window function.  A map-level correction of Galactic
contamination pickup in the far-sidelobes is more appropriate.
However, given the large uncertainties in the far-sidelobe model
shape, we choose to neglect this correction, and note that residuals from
Galactic pickup may be present in the map.
The only far sidelobe effect that is corrected in the data is an
overall calibration correction, as described in
Paper~2. 

\subsection{Cross-polar response and temperature-to-polarization leakage}
\label{sec:TtoPleakage}
\begin{figure*}[ht!]
\begin{center}
\includegraphics[width=1\textwidth]{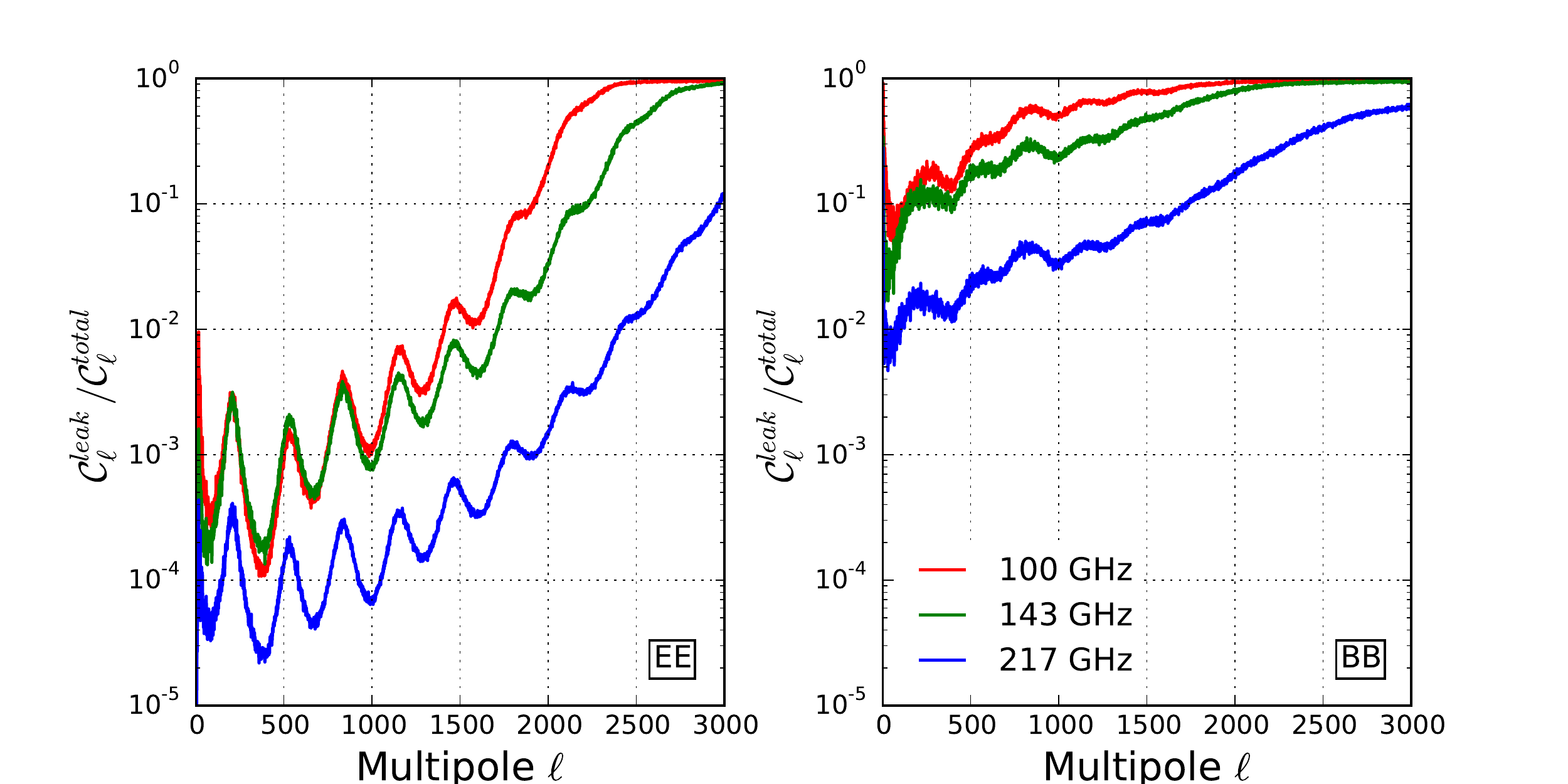}
\caption{\label{fig:mainbeamleak} A simulation of the ratio of
  temperature-to-polarization leakage, given the
  main beam mismatch, compared to $EE$ and $BB$ signal. 
}
\end{center}
\end{figure*}

The data model used for the HFI polarization reconstruction assumes that the entire
cross-polar response of each PSB is due to the detector itself, and so the
beam shape of the cross-polar response is exactly the same as the co-polar
beam shape.
The corrugated feed horns exhibit some internal cross-polar
response~\citep{maffei2010}. {\tt GRASP} simulations show that this
cross-polar response is at the level of 0.1--0.5\,\%, considerably
smaller than the detector cross-polar response (2--5\%).
Simulations using  {\tt FEBeCoP} show that the effect of ignoring the
cross-polar optical beam shape results in a smoothing of polarization power
spectra equivalent to an additional 10--20\arcs\ Gaussian, which we neglect.
Additionally the data model assumes identical beams for every PSB used
to reconstruct the polarization; this creates some
temperature-to-polarization leakage.  Main-beam leakage 
dominates over the differential time response tails.
Figure~\ref{fig:mainbeamleak} shows a simulation of CMB temperature leakage
into polarization power spectra using {\tt FEBeCoP}, given the measured
scanning beams within a band.  

 \section{Validation and consistency tests}\label{sec:consistency}
Here we describe some of the tests that have been done to validate the quality of
the cleaned TOI. Other tests at the map level are described in Paper~2.
First we discuss the impact of the ADC correction. Then we analyse how each
step of the TOI processing pipeline can alter the resulting power
spectra and also study the noise properties. Finally we estimate the filtering function of the
TOI pipeline with end-to-end simulations.

\subsection{ADC residuals}\label{sec:ADCval}
\subsubsection{Gain consistency at ring level}
\begin{figure*}[ht!]
\begin{center}
\includegraphics[width=1\textwidth,angle=180]
{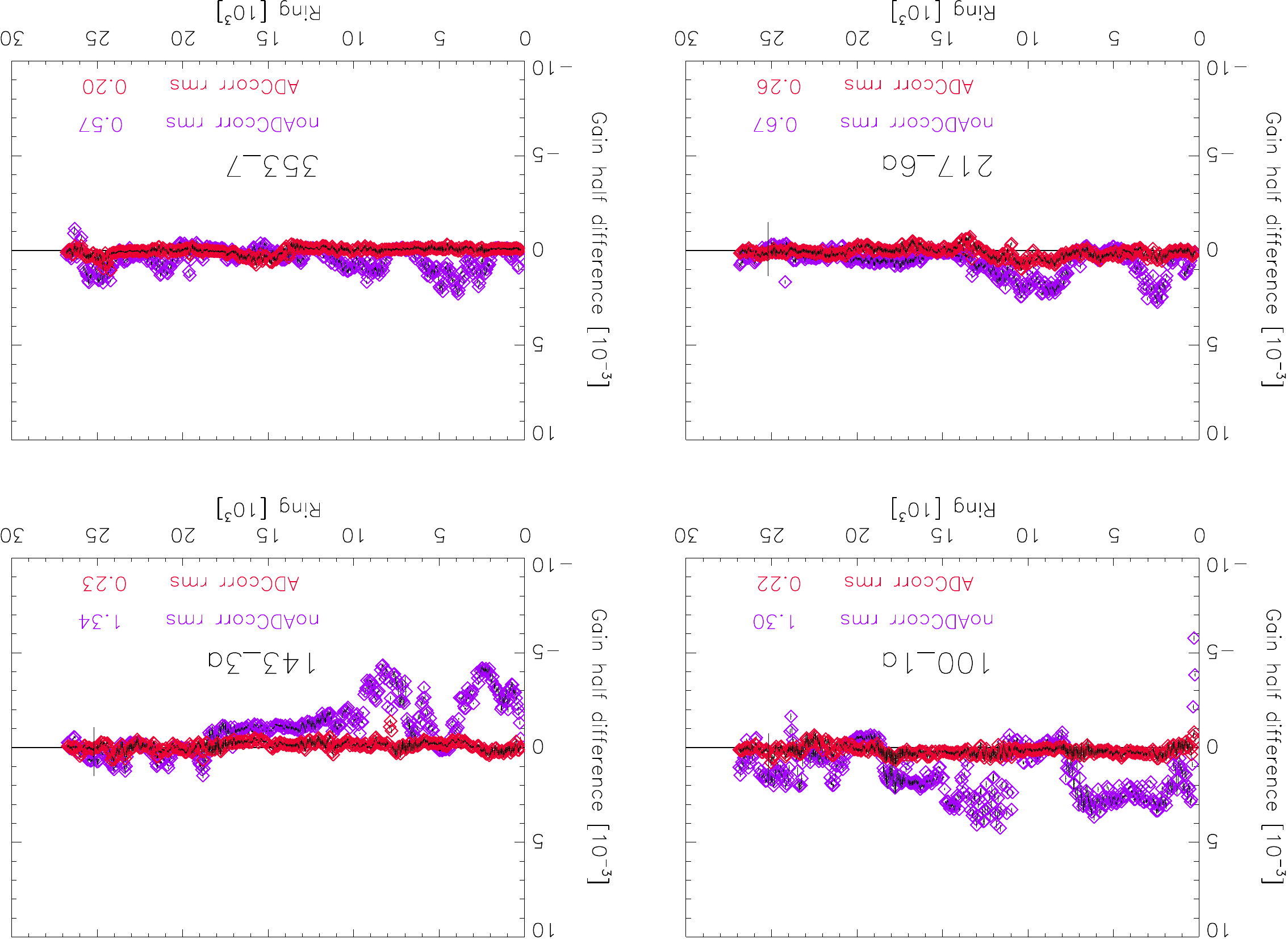}
\caption{\label{fig:ADCgain_test} The relative gain difference between parity
  $+$ and $-$ samples is shown as a function of the ring number. A boxcar
  average of 101 rings has been applied. The purple data points were
  processed with the 2013 pipeline, without ADC correction, while the red points refer to the
  present data release with ADC correction. The median error bars are of the
  order of $10^{-4}$ for the ADC-corrected TOI.}
\end{center}
\end{figure*}

Using the undeconvolved TOI at the ring level, we can monitor the quality of
the ADC correction with respect to the stability of the gain. For that purpose
we can measure the relative gain of parity plus ($g_+$) and parity minus
($g_-$) samples (alternating samples). The two parities sample a very 
different part of the ADC scale, so that a gain mismatch is a diagnostic of
the ADC nonlinearity correction. During a ring, the sky signal from both
parities should be almost identical. We thus correlate a phase-binned ring
(PBR) made of parity plus with the average PBR (made of both parities)
and obtain $g_+$. We similarly obtain $g_-$. The gain half-difference $(g_+-g_-)/2$ is
shown as a function of the ring number in Fig.~\ref{fig:ADCgain_test} for a
representative selection of four bolometers. The improvement obtained with ADC
correction is significant with a root-mean-square dispersion decreasing
by a factor of 2 to 3. Only a handful of bolometers show some discrepancies  at the $10^{-3}$ level after
the ADC correction, namely 143-3b, 217-5b, 217-7a,
217-8a, and 353-3a. This internal consistency test at the ring level is not
sensitive to gain errors common to both parities. However, it agrees 
qualitatively with the overall {\tt Bogopix} gain variations shown in
Paper~2. 

\subsubsection{Parity map spectra}\label{sec:ADCspec}
\begin{figure*}[ht!]
\begin{center}
\includegraphics[width=\columnwidth]{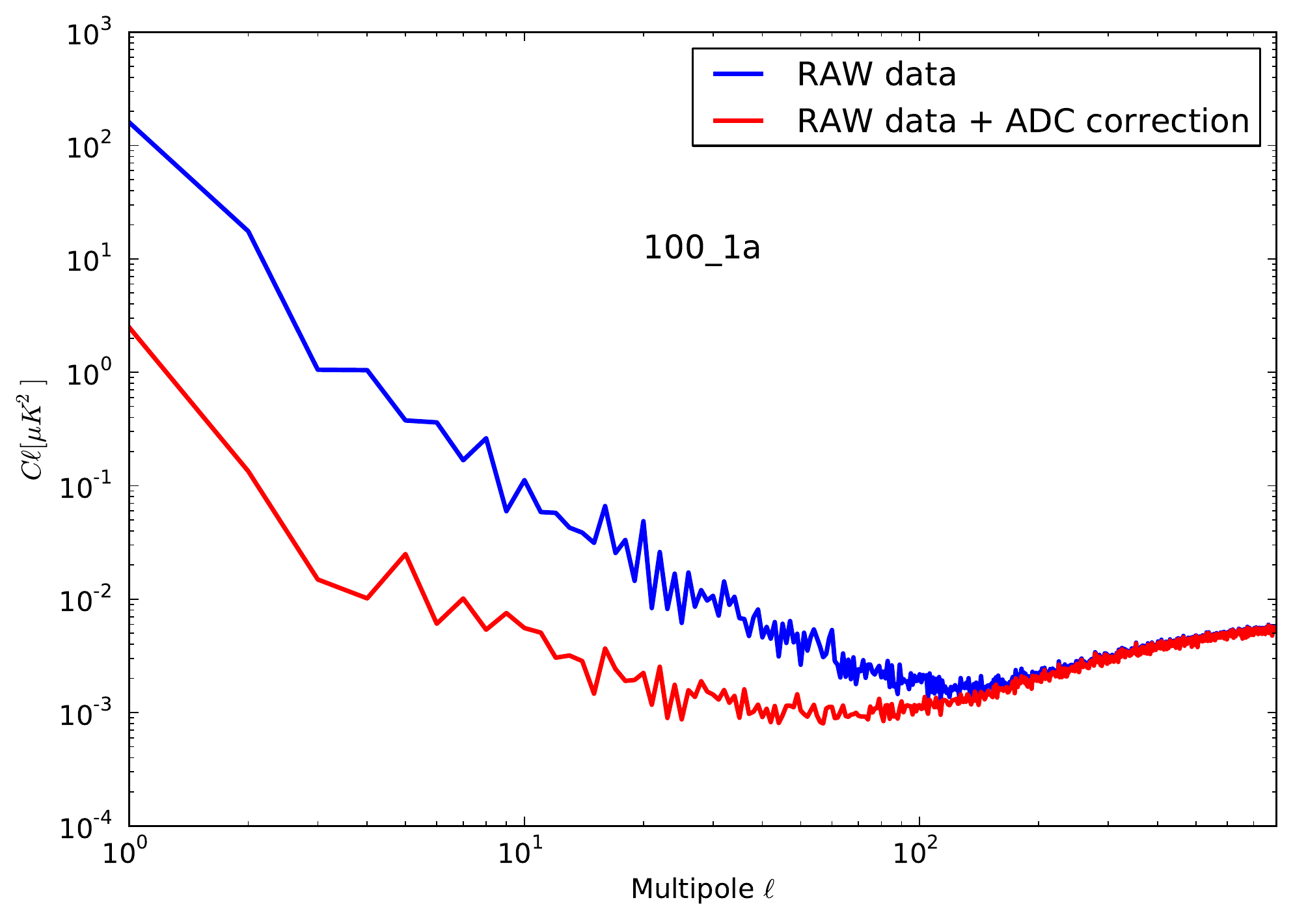}
\includegraphics[width=\columnwidth]{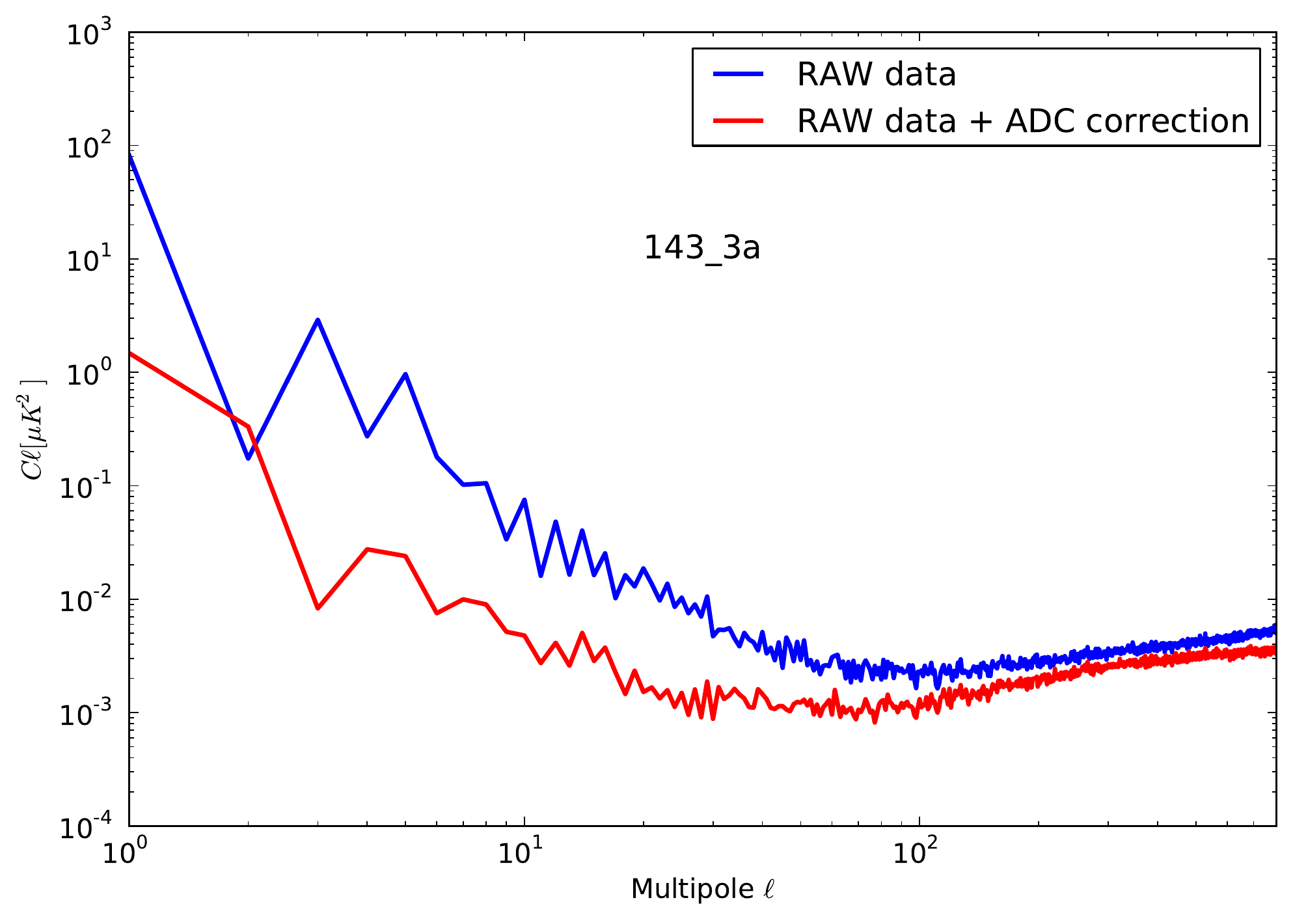}
\includegraphics[width=\columnwidth]{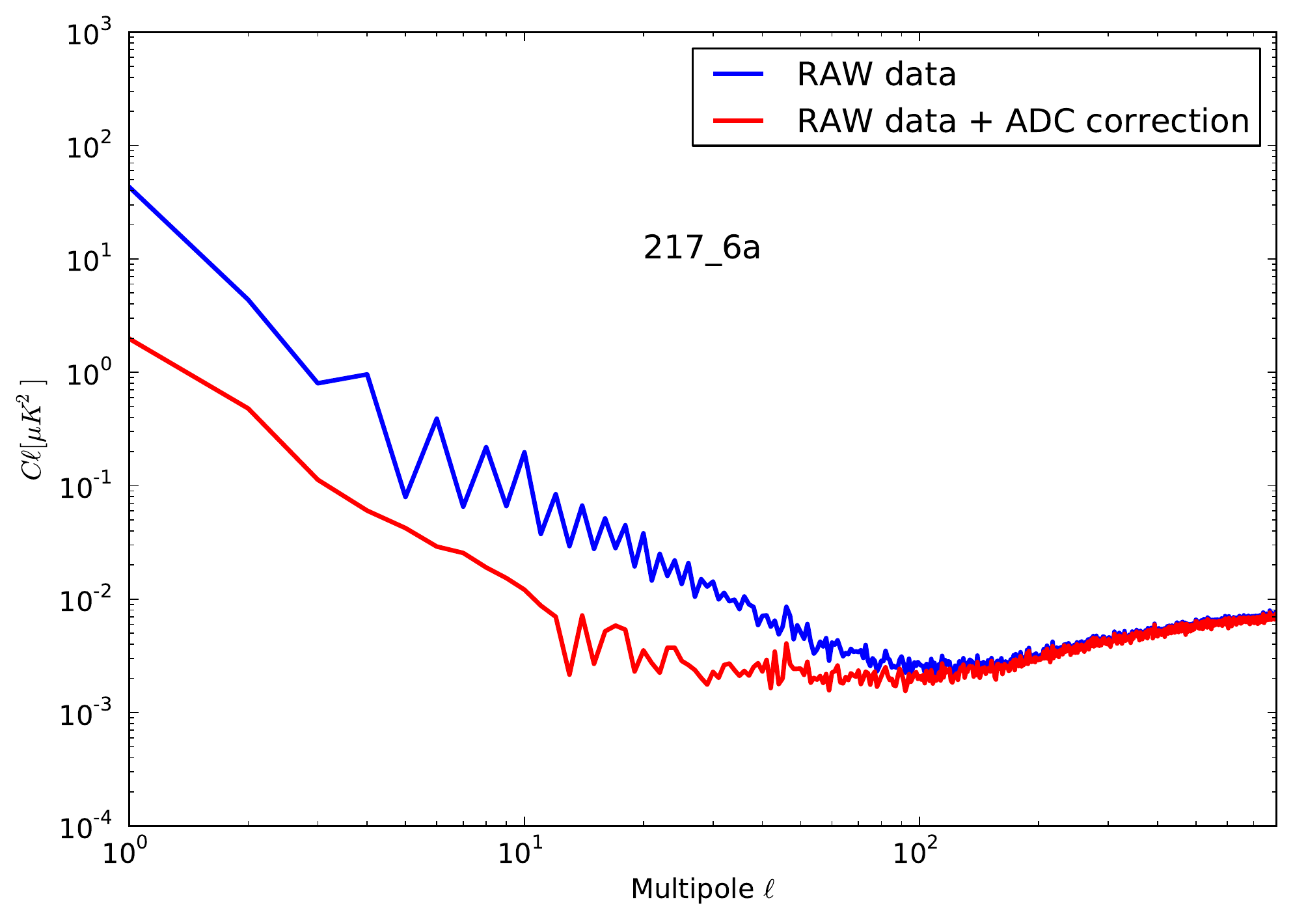}
\includegraphics[width=\columnwidth]{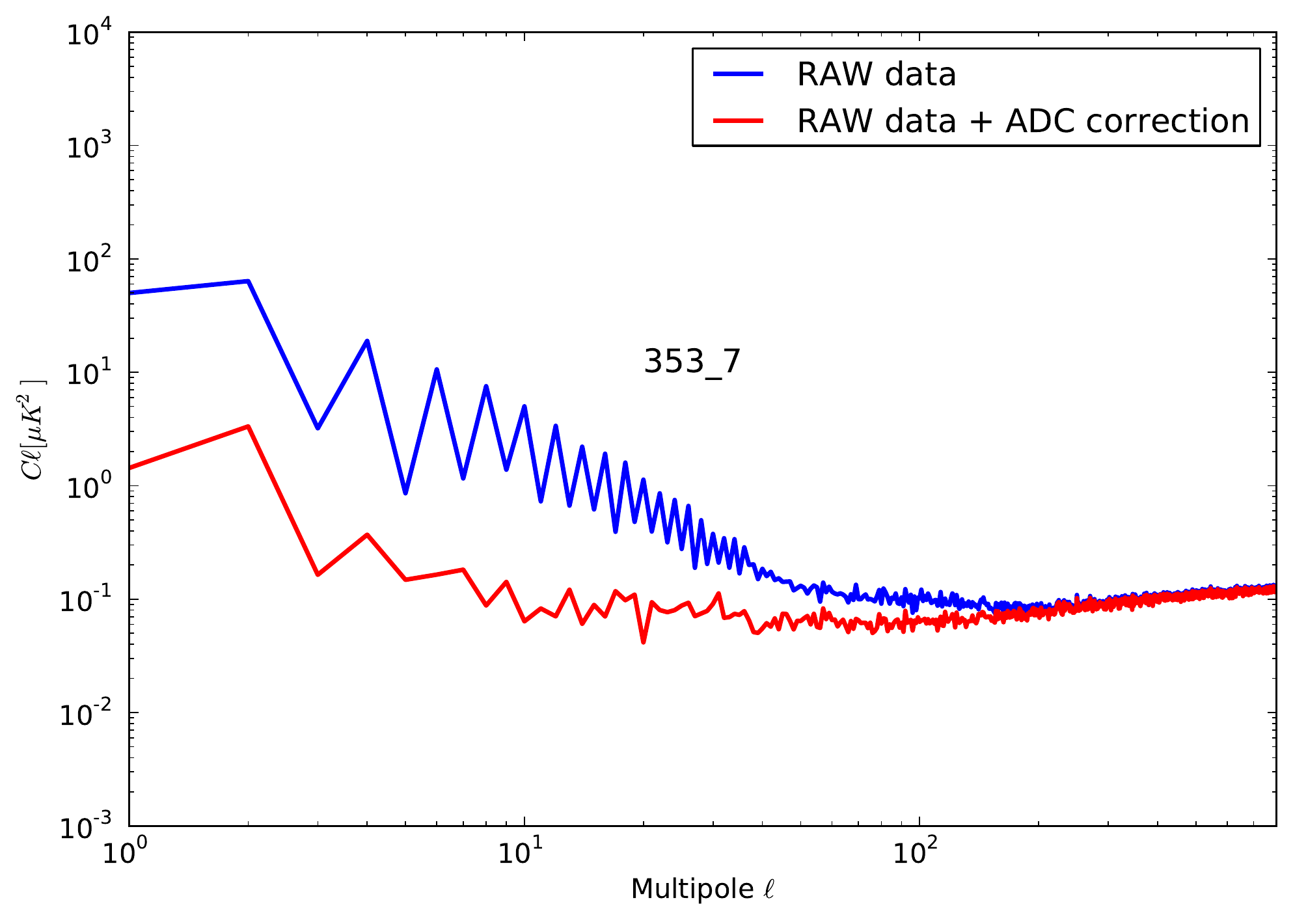}
\caption{\label{fig:ADC_half_parity} Half-parity map difference spectra for          
  the same bolometers as in Fig.~\ref{fig:ADCgain_test}. Blue: 2013
  release without ADC correction. Red: the present
  data release with ADC correction. Below $\ell=100$ the noise should be flat
  when there is no ADC bias.}
\end{center}
\end{figure*}

Two independent sky maps are produced from the two TOI parities; the differences between these
are  again expected to capture the ADC residual effects. Figure~\ref{fig:ADC_half_parity} compares full-sky
spectra derived from maps built on raw and ADC-corrected data. Glitch-flagged samples have been removed. 
The behaviour in $1/\ell^2$ at low $\ell$ is similar to the behaviour of
simulated data as shown in Sect.~\ref{sec:adc_sim}.

These low-$\ell$ residuals, although much improved with respect to the
2013 delivery, are not yet fully under control. At the time of the 2015
delivery, work is continung to estimate how this effect propagates
into the cross-polarization spectra, and to give a coherent
description of the different low-$\ell$ systematics still present in
HFI 2015 data. For this reason, among others, it has been decided to
postpone the release of HFI low $\ell$ polarized data.

\subsection{Filtering effects}\label{sec:filt}

It is customary to evaluate the effects produced by the pipeline on the signal
as a filtering function, depending on the considered angular scale. Here we
evaluate whether or how each pipeline module (see Fig.~\ref{fig:DPCscheme})
filters the signal.

\begin{itemize}
\item ADC correction. The ADC nonlinearities add a spurious component to the
power spectra. Correcting for it removes this component, reducing the $C_\ell$ by an amount smaller than 1\,\%, that depends on the channel, and is very flat in~$\ell$.

\item Glitch removal/flagging. Using simulations, \cite{planck2013-p03e} have
  shown that there is no significant bias ($< 10^{-4}$) induced on the signal
  by the glitch removal procedure, thanks to the joint estimation of glitch
  tail and sky signal. The sky signal is estimated at the ring level using
  spline interpolation in order to correct for signal variations within ring
  pixels. However, the pointing jitter, combined with a strong signal
  gradient, 
  can influence the glitch detection rate, mostly around the Galactic plane and
  planets. We have made simulations in order to evaluate an equivalent
  filtering effect using the glitch module called
  {\tt despike}~\citep{planck2013-p03e}. For CMB channels, the impact is clearly
  negligible. For submillimetre channels, the signal (mostly the interstellar
  dust emission) is changed at the level of $10^{-4}$.

\item AC modulation baseline subtraction. This has no impact on the
  signal, since it affects the TOI at the modulation frequency, which is cut out at the
  end of the TOI processing.

\item Thermal decorrelation. This involves subtracting a TOI that is filtered
  at the minute timescale.  The computation of the offsets per ring (in the
  destriping during the map making process) absorbs the long-term variability
  (typically longer than an hour) but not the mid-term variability (between 1\,min
   and 1\,h). The impact of the thermal decorrelation is to improve
  the accuracy of the destriping offsets. It is therefore not a filtering
  effect. The maximal impact of the thermal decorrelation is estimated by comparing
  $C_\ell$ using data processed with and without the thermal decorrelation.
  In Fig.~\ref{fig:filtering_T0}, we see that the cross-spectrum of
  data that 
  are fully decorrelated in time (as in Year~1 $\times$ Year~2) shows
  differences less than 10\,\% for $\ell<2000$. The very-high-$\ell$ behaviour reflects the noise level of the reference
  spectrum. Consistency tests in the likelihood paper \citep{planck2014-a13}
 are powerful enough to
  assess that the thermal residuals are negligible. 

\begin{figure}[ht!]
  \begin{center}
\includegraphics[width=\columnwidth]{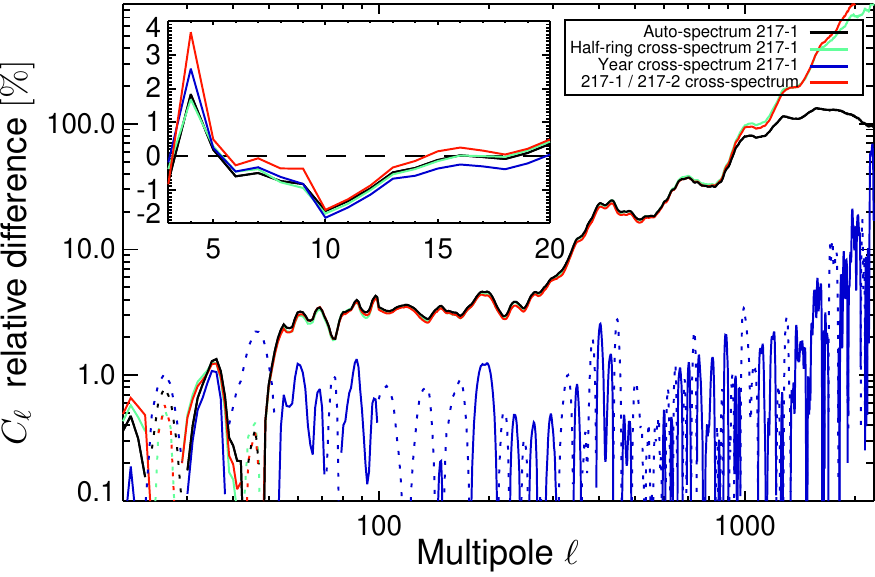}
    \caption{\label{fig:filtering_T0} Relative differences of the
      $C_\ell$, computed on 60\,\% of the sky,
      using data processed with and without the thermal decorrelation. Note
      that the zero-level has been recomputed in all cases to allow the
      comparison to be made. The amplitudes of
      the negative values are shown by dotted lines.}
  \end{center}
\end{figure}

\item 4-K line removal. The maximal impact is estimated by comparing $C_\ell$
  using data analysed with and without the removal of the 4-K line
  frequencies (keeping the same definition of valid rings in both cases). In
  Fig.~\ref{fig:filtering_4K} we see that only two lines have a strong impact:
  the 10\Hz\ line influences the power spectrum at $\ell \approx 600$ and the 30\Hz\
  at $\ell \approx 1800$. For the 10\Hz\ line, the impact on the power spectrum is
  only seen in the auto-spectrum while the 30\Hz\ line remains in cross-spectra
  between two bolometers or between half rings. When the cross-spectrum is
  computed using data that are fully decorrelated in time (as in Year~1 $\times$
  Year~2), no residual is seen. Consistency tests in the likelihood
  paper are powerful for assessing that the 4-K line residuals are negligible,
  thanks to the extra ring discarding scheme described in
  Sect.~\ref{sec:ring_selection}.

\begin{figure}[ht!]
  \begin{center}
    \includegraphics{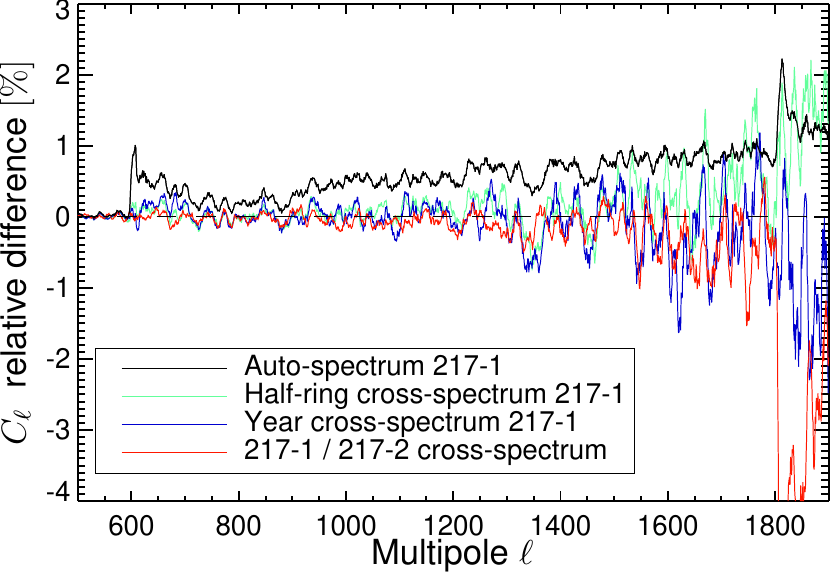}
    \caption{\label{fig:filtering_4K} Relative differences of the
      $C_\ell$, computed on 60\,\% of the sky,
      using data analysed with and without removal of the
      4-K line frequencies.}
  \end{center}
\end{figure}

\item Nonlinearity correction. The impact of this correction on
  measurements of the CMB has been
  propagated through the whole pipeline.  Figure~\ref{fig:filtering_g0} shows
  that the angular power spectrum is underestimated by less than 0.3\,\% between $\ell
  = 20$ and $\ell = 2000$ if no correction is applied, while still keeping the
  same calibration coefficients. We estimate that the nonlinearity correction
  is accurate within a 10\,\% uncertatinty. Thus, the uncertainty of the
  bolometer nonlinearity correction translates into a filtering function
  uncertainty below the $10^{-3}$ level over all angular scales.

\begin{figure}[ht!]
  \begin{center}
    \includegraphics{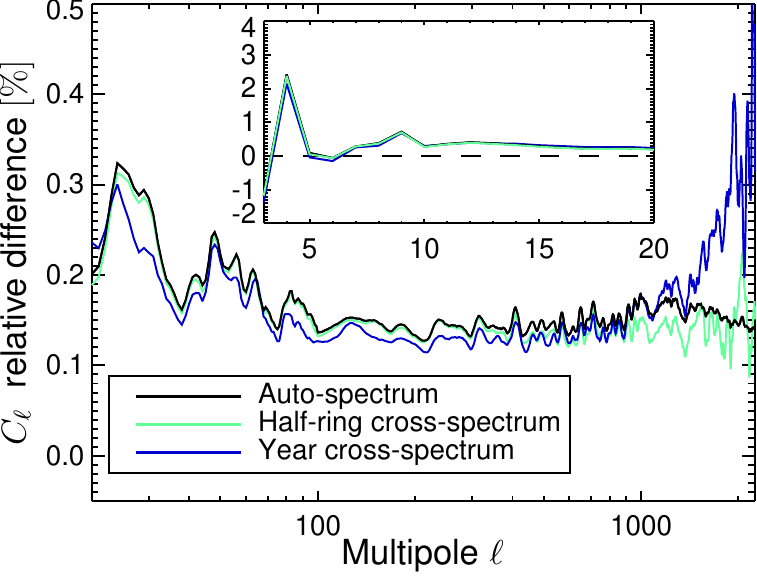}
    \caption{\label{fig:filtering_g0} Relative differences of the
      $C_\ell$, computed on 60\,\% of the sky, using data processed with and without the nonlinearity
      correction of the bolometer response. The example shown here is
      for bolometer 217-1.}
  \end{center}
\end{figure}

\item The Fourier transform module does of course filter the TOI. It
  deconvolves the time transfer function (Sect.~\ref{sec:tau}) but also
  low-pass filters the data~\citep{planck2013-p03c}. It is important for high
  temporal frequencies, but affects all frequencies at the $10^{-4}$ level down
  to the spin frequency where the filter is set to 1. The filtering effect is
  captured in the effective beam for scales of up to $1$--$2^\circ$. However, there could
  remain an effective filtering effect for larger angular scales.

\end{itemize}
In order to test the global pipeline filtering effect that could be induced
by the interaction of several systematic effects, we resort to complete
simulations, described in detail in Sect.~\ref{sec:e2esim}.

\subsection{Noise analyses}\label{sec:noise_analyses}

\subsubsection{Noise Spectrum}\label{sec:noise}

\begin{figure*}[ht!]
\begin{center}
  \includegraphics[width=\textwidth,angle=180]
{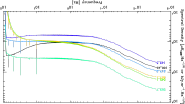}
\caption{\label{fig:noise_spectra} Calibrated noise spectrum for one bolometer
  in each HFI frequency channel. The spectrum is the average of all valid ring
  spectra of the second survey.  The comb-like structure of absorption
  features occurs at multiples of the spacecraft spin frequency (16.7\,mHz),
  and is due to the signal removal procedure. At 10\Hz\ and above there are
  some residuals of the 4-K cooler lines only associated with this
  particular computation. }
\end{center}
\end{figure*}

Noise spectra are best studied at the ring level using the redundancy built into
the scanning strategy. The method is described in \cite{planck2013-p03} and
consists of removing the sky-signal ring average from the TOI for each
ring. Figure~\ref{fig:noise_spectra} shows the amplitude spectral density for
representative HFI bolometers. This is modelled in Sect.~\ref{sec:NoiseModel} for the purpose of providing
accurate inputs for simulations. For bolometers
with relatively long time constants (especially at 100\GHz), the
electronic noise is boosted at high frequency.

\subsubsection{Noise stationarity}\label{sec:noise_statio}

\begin{figure*}[htbp!]
\begin{center}
  \includegraphics[width=0.88\textwidth]{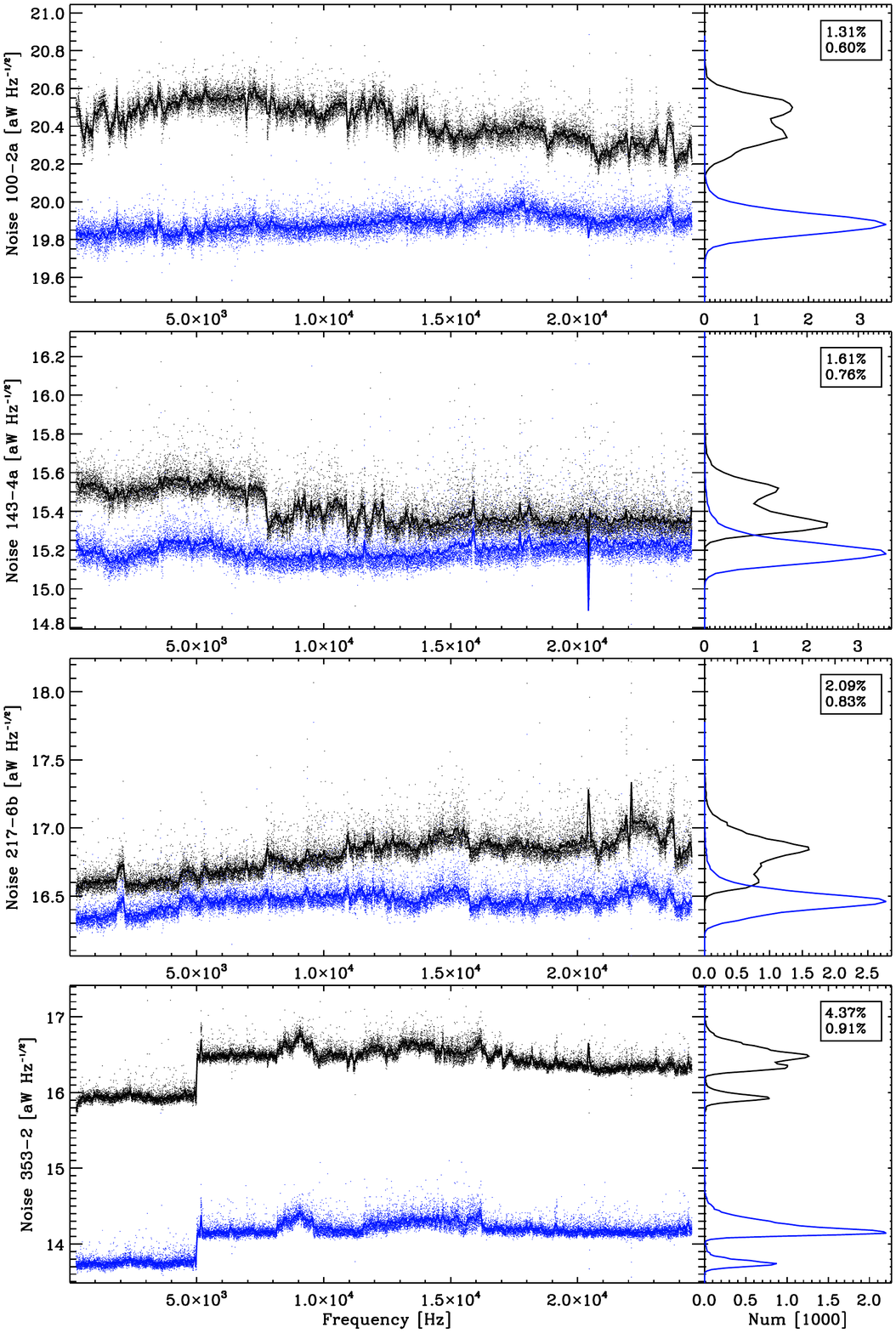}
  \caption{\label{fig:noise_statio} Noise stationarity for a selection of four
    bolometers. The left panels show the total noise for each bolometer (dots)
    as a function of ring number. The total noise is the white noise spectral
    density giving the same rms per sample as the measured one.  The
    solid line  shows a running box average.  The black dots are from
    the 2013 data release 
    and the blue dots from this release. The right panels show histograms
    of the left trends. The box gives the width of the distribution at half
    maximum, as measured on the histogram, normalized to the mean noise
    level. The time response deconvolution has changed between the two data
    releases, so the absolute noise level is different.}
\end{center}
\end{figure*}

The noise stationarity is best studied by measuring the rms deviation of a TOI
sample at the ring level,  at the end of the TOI processing, where the average
ring signal has been subtracted. The statistics are computed for  samples which are
valid, far from the Galactic plane, and far from bright point sources, in order
to avoid strong residual gradients (see \cite{planck2013-p28}).  The
bias induced by the signal subtraction is 
corrected for; the average value (after calibration) over the mission is given in
Table~\ref{tab:Total-noise}. This is compatible with the values quoted in Table~4
of \cite{planck2013-p03}.

\begin{table}[htb]
  \protect\caption{HFI average total noise. The rms noise
    is given for one bolometer. The averaging between bolometers of
    the same band is done with the same weighting scheme as used in the mapmaking
    procedure. The band average (4th column) refers to the sensitivity of the
    collection of $N_\mathrm{b}$ bolometers at the same frequency. 
    \label{tab:Total-noise}}

\centering{}\begingroup
\newdimen\tblskip \tblskip=5pt
\nointerlineskip
\vskip -3mm
\footnotesize
\setbox\tablebox=\vbox{
   \newdimen\digitwidth 
   \setbox0=\hbox{\rm 0} 
   \digitwidth=\wd0 
   \catcode`*=\active 
   \def*{\kern\digitwidth}
   \newdimen\signwidth 
   \setbox0=\hbox{+} 
   \signwidth=\wd0 
   \catcode`!=\active 
   \def!{\kern\signwidth}
\halign{\hfil#\hfil\tabskip=2em&
\hfil#\hfil&
\hfil#\hfil&
\hfil#\hfil&
\hfil#\hfil\/\tabskip=0pt\cr             
\noalign{\doubleline}
Band& $N_\mathrm{b}$& Bol. total noise& Band total noise& Units\cr
\noalign{\vskip 3pt\hrule\vskip 5pt}
100&   8&     113.3&      40.0&   $\mu\mathrm{K_{CMB}\, s}^{1/2}$\cr
143&  11&      57.5&      17.3&   $\mu\mathrm{K_{CMB}\, s}^{1/2}$\cr
217&  12&      83.2&      24.0&   $\mu\mathrm{K_{CMB}\, s}^{1/2}$\cr
353&  12&     282.0&      81.4&   $\mu\mathrm{K_{CMB}\, s}^{1/2}$\cr
\noalign{\vskip 3pt\hrule\vskip 3pt}
545&   3&      45.5&      26.3&   $\mathrm{kJy\, sr^{-1}\, s^{1/2}}$\cr
857&   4&      49.2&      24.6&   $\mathrm{kJy\, sr^{-1}\, s^{1/2}}$\cr
\noalign{\vskip 5pt\hrule\vskip 3pt}}}
\endPlancktable                    
\endgroup
\end{table}

Figure~\ref{fig:noise_statio} shows trends of the 
total noise versus ring number before (2013 release) and after the ADC correction (this release), and histograms of the noise.
In most cases, there is a significant decrease in the relative width of the
histogram when the ADC correction is included, though that decrease is
more significant for the low frequency channels, and since the
distributions are non-Gaussian, very large decreases (or increases)
indicate a large change in the structure of the distribution.  Nevertheless,
there are several cases where a jump in the noise is removed by the ADC
correction (143-2a, 143-4a, 353-2). Other noise jumps (mostly at the
sub-percent level except for bolometer 353-3a) remain unexplained. The
noise of some bolometers shows a linear trend with time, but at the sub-percent
level. For most bolometers, the noise is stationary below the
percent level. 

We have investigated the origin of the jumps in the noise level (e.g., for
bolometer 353-2 at ring 5000 in the lower panel of
Fig.~\ref{fig:noise_statio}). Most of the cases are not due to the
ADC corrections (an exception is the 143-4a bolometer in
Fig.~\ref{fig:noise_statio}). The extra noise component has been analysed by
computing power spectra for selected rings. This falls into two
categories: (1) for some detectors including 353-3a (the strongest case), it
corresponds to an increase of the white noise component; (2) for other
detectors, the extra component in the power spectrum is a bump 
concentrated in the 0.1 to 1\Hz\ range, which could be a very low level of RTS
not detected by the non-Gaussianity tests that are routinely
performed~\citep{planck2013-p03}.
Note that the small changes of noise levels with time are not taken into
account by the mapmaking process, because the loss of optimality can be
neglected.

\subsubsection{Noise modelling}\label{sec:NoiseModel}

Individual ring noise Power Specral Densities (PSDs) suffer from realization noise and signal residuals.
For the purpose of simulating realistic noise from the PSDs, they need to be
regularized by fitting a smooth model to the noisy PSD. To this end, we have
constructed a physically-motivated model of the noise PSD consisting of
\begin{enumerate}
\item a photon/phonon noise component that is subject to the bolometric transfer
  function (suppressed above 10\Hz);
\item an electronic noise component that is only subject to the digital low-pass
  filtering (abrupt suppression close to the Nyquist frequency);
\item a $1/f$ slope comprising the low-frequency thermal fluctuations,
  glitch residuals, and other low frequency fluctuations.
\end{enumerate}
The resulting noise model has four degrees of freedom: three amplitudes and one slope.

\begin{figure}[ht!]
  \begin{center}
    \includegraphics[width=\columnwidth]{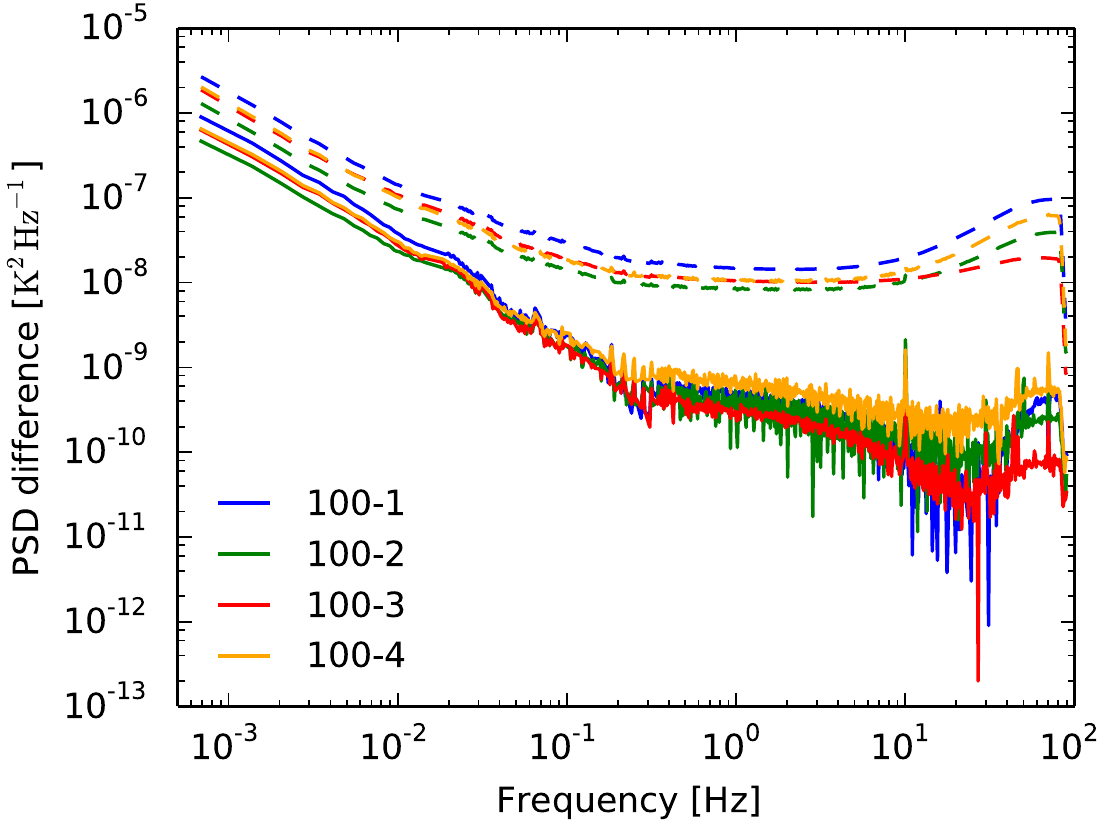}
    \caption{\label{fig:PSDcorr}
      Example of measured summed, $(a+b)$, time stream noise power
      spectral densities (PSDs; dashed lines)
      and the difference in the PSD between $(a+b)$ and $(a-b)$ (solid lines).
      These spectra are bin-by-bin median values across the entire mission and
      were derived by first measuring the sample autocovariance as described
      in \cite{planck2014-a14}.
    }
  \end{center}
\end{figure}

\begin{figure}[ht!]
  \begin{center}
    \includegraphics[width=\columnwidth]{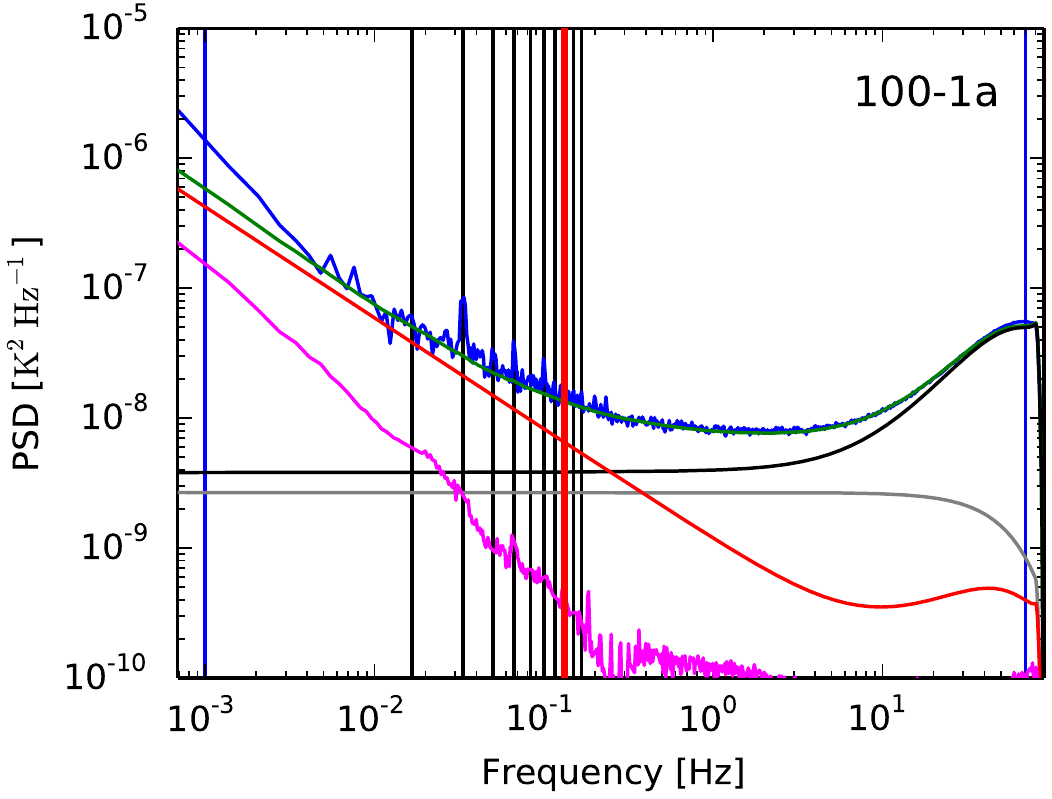} \\
    \includegraphics[width=\columnwidth]{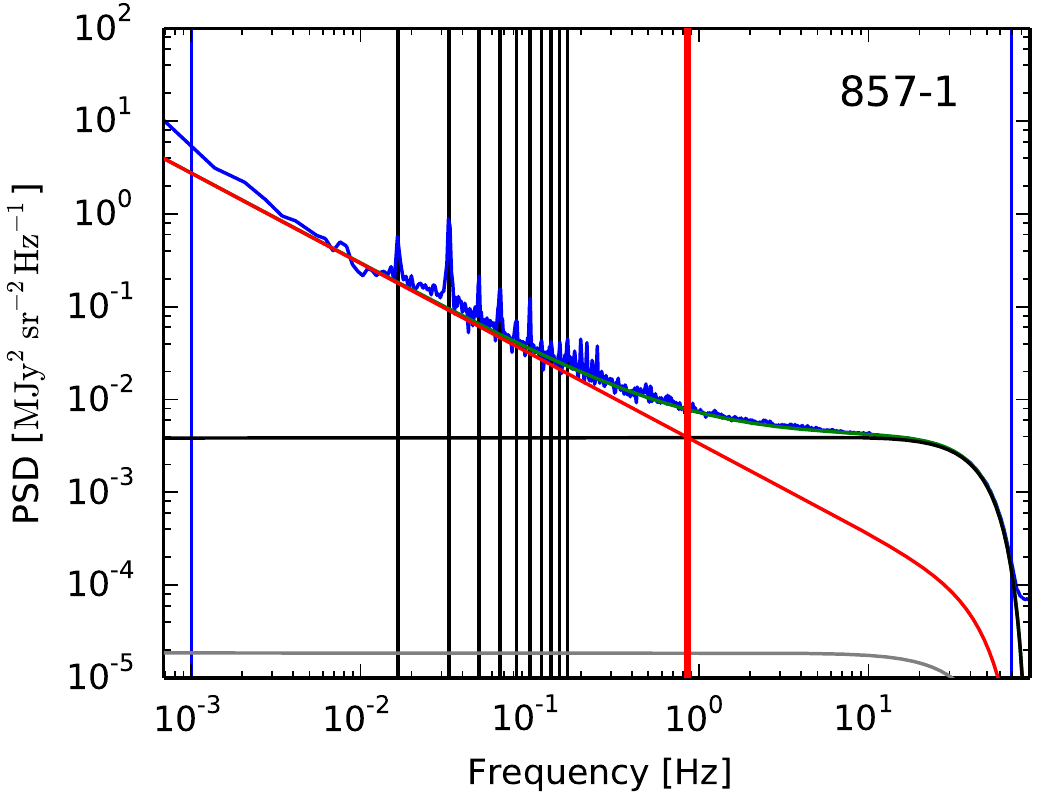}
    \caption{\label{fig:PSDmodel}
      Two examples of the 4-component noise model fits to measured noise PSDs.
      The underlying noise estimation for these fits is described in \cite{planck2014-a14}.
      It measures the sample auto-covariance of signal-subtracted and
      masked data to overcome effects of gaps and signal residuals.
      The blue curve is the measured PSD averaged over 30 pointing periods.
      The red curve is the $1/f$ component, deconvolved with the bolometer
      transfer function. The black curve is the electronic white noise component
      and the grey curve is the photonic white noise component. The magenta
      curve is the measured $a/b$ correlated component. The green curve
      is the sum of all noise model components. Vertical blue lines
      indicate the region where the fit was made and the vertical black lines
      indicate the 10 lowest spin harmonic frequencies.
    }
  \end{center}
\end{figure}

In addition to the noise autospectrum components, we detect the presence of
a correlated noise component, i.e., a common mode between two
polarization-sensitive bolometers in the same
horn (e.g., 100-1a and 100-1b). The common mode was measured by analysing the
noise of sum, $(a+b)$, and difference, $(a-b)$, time streams. If the
bolometers were entirely independent, the two composite streams would have
equivalent noise power. Instead, we systematically find that $(a+b)$ has a higher
noise than $(a-b)$. The measurements for 100\GHz\ are presented in
Fig.~\ref{fig:PSDcorr}.  There are indications that similar common modes exist
across the focal plane, but for the purpose of noise simulation,
the $a/b$ correlated noise that directly affects polarization sensitivity is
the most important one. Examples of noise model fits are presented for two
bolometers in Fig.~\ref{fig:PSDmodel}. Note that a common mode tracing the
global 100\,mK temperature fluctuations at the 1\,min time scale has
already been removed from the processed TOI, by reference to the dark
bolometers. 

\subsection{End-to-end simulations}\label{sec:e2esim}
End-to-end simulations are created by feeding the TOI processing pipeline with
simulated TOI to evaluate and characterize its overall transfer function and
the respective contribution of each individual effect. Simulated TOIs are
produced by applying the real mission scanning strategy to a realistic input
sky specified by the Planck Sky Model~\citep[PSM v1.7.7,][]{delabrouille2012}
containing a lensed CMB realization, Galactic diffuse foregrounds, and the
dipole components (but without point sources).  To this scanning TOI, we then
add a white-noise component, representing the phonon and photon noises. The
very-low-frequency drift seen in the real data is added to the
TOI. The noisy sky TOI is then convolved with the appropriate bolometer
transfer functions. Another white-noise component, representing the
Johnson noise and readout noise with intensities derived in
Fig.~\ref{fig:PSDmodel}, is also added. Simulated cosmic rays using
the measured glitch rates, amplitudes, and
shapes are added to the TOI. This TOI is then 
interpolated to the electronic HFI fast-sampling frequency
($40\times181.3737$\Hz). It is then converted to analogue-to-digital units (ADU)
using a simulated nonlinear analogue-to-digital converter (ADC). Some 4-K
cooler spectral lines are added to the TOI. Both effects (ADC and 4-K lines)
are derived from the measured in-flight behaviour. The TOI finally goes
through the data compression/decompression algorithm used for communications
between the \Planck\ satellite and Earth.

The simulated TOI is then processed in the same way as the real mission data for cleaning
and systematic error removal, calibration, destriping, mapmaking, and power spectrum
computation. The systematic effects added to the signal are simulated, with
the same parameters used by the TOI processing pipeline, but no pointing error
is included  (unlike the simulations of
planet crossings used for beam and focal plane reconstructions).

\begin{figure}[ht!]
\centering
\includegraphics[width=1.02\columnwidth]{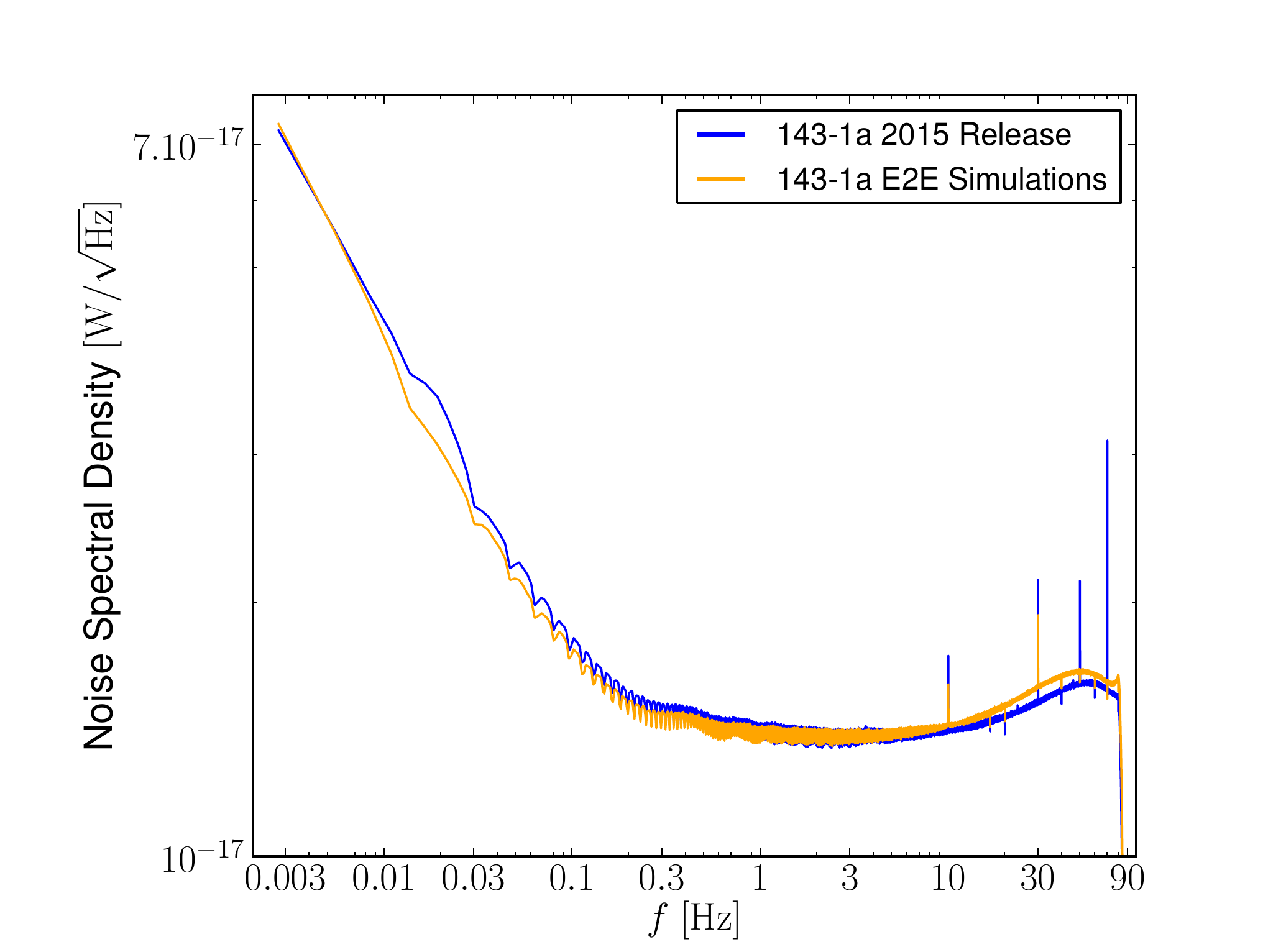}
\caption{The PSD of the true noise for bolometer 143-1a as compared to
  that generated using the end-to-end simulation.}
\label{fig:e2enoise}
\end{figure}

The end-to-end simulations have been used primarily to
characterize our understanding of the properties of the noise  in the final maps, such as
the level and correlations in the noise due to known systematic
effects.   Figure~\ref{fig:e2enoise} compares the
PSD of noise generated by  the end-to-end simulation, showing good agreement
over a wide band: residuals from undetected glitches; long
time-scale thermal drifts; and filtering of detector and electronics
readout noise.  Paper~2 further compares the noise properties of
the end-to-end simulation maps with those of the real data.

In future releases, the end-to-end simulation will be used to characterize
the filtering of the sky signal due to the analysis pipeline as a
function of angular scale.

\section{Conclusions}
\label{sec:conclusion}
 
This paper has described the major improvements in the HFI time-domain processing
and beam analysis in the new 2015 \Planck\ data release. Once the ADC effect
is taken into account, the bolometer gain stability is at the level of one
part in a thousand and the noise is stable within 1\,\% during the whole
mission (29 months of sky survey). The \HeJT\ 4-K cooler lines have been
reduced to an insignificant level at the cost of removing 4\,\% of the
data. The beams are reconstructed out to 100\arcm\ radius at a precision that
gives better than 0.1\,\% errors in the effective beam window
function.  Noise and known systematic errors in the beam measurement
are negligible in the analysis of the CMB data. 
The characteristics that can be best appreciated at the map level are described 
in Paper~2. 
Overall, the instrument CMB sensitivity is
$13.3\,\mu\mathrm{K_{CMB}\, s}^{1/2}$. Systematic effects have been studied
individually via dedicated pipelines and globally via end-to-end
simulations. The main sources of systematic effects are: ADC corrections; glitch impacts; 100-mK base-plate temperature 
fluctuations, 4-K cooler lines, and the bolometer temporal
transfer function. Their impacts have been measured in terms of transfer functions,
noise, and scanning beams.  The systematic errors here are
considered mainly in the context of angular scales with $\ell>10$; as
shown in Paper~2, systematic errors remain in the data and particular
affect the largest scales. A more complete description of the
instrument and systematic errors in the data on the largest scales,
$\ell<10$, will be described in a forthcoming publication. 

Even with these significant changes in the processing of the data
and the understanding of the instrument, there is room for further
improvement.  The ADC correction can be improved with better modelling
of the raw signal and an updated model of the ADC code errors.
The time response reconstruction method which was very effective here
for 353\GHz\ can also be applied to 100--217\GHz.  The cooler lines can be
corrected to greater precision, allowing more samples of the TOI to be retained
for mapmaking.  All of these corrections in the low-level processing
will yield further improvements in the polarization measurements from
HFI. 

\begin{acknowledgements}
The Planck Collaboration acknowledges the support of: ESA; CNES and
CNRS/INSU-IN2P3-INP (France); ASI, CNR, and INAF (Italy); NASA and DoE
(USA); STFC and UKSA (UK); CSIC, MINECO, JA, and RES (Spain); Tekes,
AoF, and CSC (Finland); DLR and MPG (Germany); CSA (Canada); DTU Space
(Denmark); SER/SSO (Switzerland); RCN (Norway); SFI (Ireland);
FCT/MCTES (Portugal); ERC and PRACE (EU). A description of the Planck
Collaboration and a list of its members, indicating which technical or
scientific activities they have been involved in, can be found at
{\url http://www.cosmos.esa.int/web/planck/planck-collaboration}. 

\end{acknowledgements}

 \bibliographystyle{aat}
 \bibliography{Planck_bib,HFIDPC_bib}

\def\eprinttmppp@#1arXiv:@{#1}
\providecommand{\arxivlink[1]}{\href{http://arxiv.org/abs/#1}{arXiv:#1}}
\def\eprinttmp@#1arXiv:#2 [#3]#4@{\ifthenelse{\equal{#3}{x}}{\ifthenelse{
\equal{#1}{}}{\arxivlink{\eprinttmppp@#2@}}{\arxivlink{#1}}}{\arxivlink{#2}
  [#3]}}
\providecommand{\eprintlink}[1]{\eprinttmp@#1arXiv: [x]@}
\providecommand{\eprint}[1]{\eprintlink{#1}}
\providecommand{\adsurl}[1]{\href{#1}{ADS}}
\begin{thebibliography}{31}
\expandafter\ifx\csname natexlab\endcsname\relax\def\natexlab#1{#1}\fi

\bibitem[{Anderson {et~al.}(1999)Anderson, Bai, Bischof, Blackford, Demmel,
  Dongarra, Du~Croz, Greenbaum, Hammarling, McKenney, \& Sorensen}]{lapack}
Anderson, E., Bai, Z., Bischof, C., {et~al.} 1999, {LAPACK} Users' Guide, 3rd
  edn. (Philadelphia, PA: Society for Industrial and Applied Mathematics)

\bibitem[{{Catalano} {et~al.}(2014){Catalano}, {Ade}, {Atik}, {Benoit},
  {Br{\'e}ele}, {Bock}, {Camus}, {Charra}, {Crill}, {Coron}, {Coulais},
  {D{\'e}sert}, {Fauvet}, {Giraud-H{\'e}raud}, {Guillaudin}, {Holmes}, {Jones},
  {Lamarre}, {Mac{\'{\i}}as-P{\'e}rez}, {Martinez}, {Miniussi}, {Monfardini},
  {Pajot}, {Patanchon}, {Pelissier}, {Piat}, {Puget}, {Renault}, {Rosset},
  {Santos}, {Sauv{\'e}}, {Spencer}, \& {Sudiwala}}]{catalano2014}
{Catalano}, A., {Ade}, P., {Atik}, Y., {et~al.}, {Characterization and Physical
  Explanation of Energetic Particles on Planck HFI Instrument}. 2014, Journal
  of Low Temperature Physics

\bibitem[{{Delabrouille} {et~al.}(2013){Delabrouille}, {Betoule}, {Melin},
  {Miville-Desch{\^e}nes}, {Gonzalez-Nuevo}, {Le Jeune}, {Castex}, {de Zotti},
  {Basak}, {Ashdown}, {Aumont}, {Baccigalupi}, {Banday}, {Bernard}, {Bouchet},
  {Clements}, {da Silva}, {Dickinson}, {Dodu}, {Dolag}, {Elsner}, {Fauvet},
  {Fa{\"y}}, {Giardino}, {Leach}, {Lesgourgues}, {Liguori}, {Macias-Perez},
  {Massardi}, {Matarrese}, {Mazzotta}, {Montier}, {Mottet}, {Paladini},
  {Partridge}, {Piffaretti}, {Prezeau}, {Prunet}, {Ricciardi}, {Roman},
  {Schaefer}, \& {Toffolatti}}]{delabrouille2012}
{Delabrouille}, J., {Betoule}, M., {Melin}, J.-B., {et~al.}, {The pre-launch
  Planck Sky Model: a model of sky emission at submillimetre to centimetre
  wavelengths}. 2013, \aap, 553, A96, \eprint{1207.3675}

\bibitem[{{Giorgini} {et~al.}(1996){Giorgini}, {Yeomans}, {Chamberlin},
  {Chodas}, {Jacobson}, {Keesey}, {Lieske}, {Ostro}, {Standish}, \&
  {Wimberly}}]{giorgini1996}
{Giorgini}, J.~D., {Yeomans}, D.~K., {Chamberlin}, A.~B., {et~al.} 1996, in
  Bulletin of the American Astronomical Society, Vol.~28, Bulletin of the
  American Astronomical Society, 1158

\bibitem[{{G{\'o}rski} {et~al.}(2005){G{\'o}rski}, {Hivon}, {Banday},
  {Wandelt}, {Hansen}, {Reinecke}, \& {Bartelmann}}]{gorski2005}
{G{\'o}rski}, K.~M., {Hivon}, E., {Banday}, A.~J., {et~al.}, {HEALPix: A
  Framework for High-Resolution Discretization and Fast Analysis of Data
  Distributed on the Sphere}. 2005, \apj, 622, 759,
  \eprint{arXiv:astro-ph/0409513}

\bibitem[{{Jones} {et~al.}(2003){Jones}, {Bhatia}, {Bock}, \&
  {Lange}}]{Jones2003}
{Jones}, W.~C., {Bhatia}, R., {Bock}, J.~J., \& {Lange}, A.~E. 2003, in Society
  of Photo-Optical Instrumentation Engineers (SPIE) Conference Series, Vol.
  4855, Society of Photo-Optical Instrumentation Engineers (SPIE) Conference
  Series, ed. T.~G. {Phillips} \& J.~{Zmuidzinas}, 227--238

\bibitem[{{Lamarre} {et~al.}(2010){Lamarre}, {Puget}, {Ade}, {Bouchet},
  {Guyot}, {Lange}, {Pajot}, {Arondel}, {Benabed}, {Beney}, {Beno{\^i}t},
  {Bernard}, {Bhatia}, {Blanc}, {Bock}, {Br{\'e}elle}, {Bradshaw}, {Camus},
  {Catalano}, {Charra}, {Charra}, {Church}, {Couchot}, {Coulais}, {Crill},
  {Crook}, {Dassas}, {de Bernardis}, {Delabrouille}, {de Marcillac}, {Delouis},
  {D{\'e}sert}, {Dumesnil}, {Dupac}, {Efstathiou}, {Eng}, {Evesque},
  {Fourmond}, {Ganga}, {Giard}, {Gispert}, {Guglielmi}, {Haissinski},
  {Henrot-Versill{\'e}}, {Hivon}, {Holmes}, {Jones}, {Koch}, {Lagard{\`e}re},
  {Lami}, {Land{\'e}}, {Leriche}, {Leroy}, {Longval},
  {Mac{\'{\i}}as-P{\'e}rez}, {Maciaszek}, {Maffei}, {Mansoux}, {Marty}, {Masi},
  {Mercier}, {Miville-Desch{\^e}nes}, {Moneti}, {Montier}, {Murphy},
  {Narbonne}, {Nexon}, {Paine}, {Pahn}, {Perdereau}, {Piacentini}, {Piat},
  {Plaszczynski}, {Pointecouteau}, {Pons}, {Ponthieu}, {Prunet}, {Rambaud},
  {Recouvreur}, {Renault}, {Ristorcelli}, {Rosset}, {Santos}, {Savini},
  {Serra}, {Stassi}, {Sudiwala}, {Sygnet}, {Tauber}, {Torre}, {Tristram},
  {Vibert}, {Woodcraft}, {Yurchenko}, \& {Yvon}}]{lamarre2010}
{Lamarre}, J., {Puget}, J., {Ade}, P.~A.~R., {et~al.}, {Planck pre-launch
  status: The HFI instrument, from specification to actual performance}. 2010,
  \aap, 520, A9

\bibitem[{{Maffei} {et~al.}(2010){Maffei}, {Noviello}, {Murphy}, {Ade},
  {Lamarre}, {Bouchet}, {Brossard}, {Catalano}, {Colgan}, {Gispert}, {Gleeson},
  {Haynes}, {Jones}, {Lange}, {Longval}, {McAuley}, {Pajot}, {Peacocke},
  {Pisano}, {Puget}, {Ristorcelli}, {Savini}, {Sudiwala}, {Wylde}, \&
  {Yurchenko}}]{maffei2010}
{Maffei}, B., {Noviello}, F., {Murphy}, J.~A., {et~al.}, {Planck pre-launch
  status: HFI beam expectations from the optical optimisation of the focal
  plane}. 2010, \aap, 520, A12

\bibitem[{{Mitra} {et~al.}(2011){Mitra}, {Rocha}, {G{\'o}rski}, {Huffenberger},
  {Eriksen}, {Ashdown}, \& {Lawrence}}]{mitra2010}
{Mitra}, S., {Rocha}, G., {G{\'o}rski}, K.~M., {et~al.}, {Fast Pixel Space
  Convolution for Cosmic Microwave Background Surveys with Asymmetric Beams and
  Complex Scan Strategies: FEBeCoP}. 2011, \apjs, 193, 5, \eprint{1005.1929}

\bibitem[{{Planck Collaboration ES}(2013)}]{planck2013-p28}
{Planck Collaboration ES}. 2013, {The Explanatory Supplement to the Planck 2013
  results, http://pla.esac.esa.int/pla/index.html} ({ESA})

\bibitem[{{Planck Collaboration ES}(2015)}]{planck2014-ES}
{Planck Collaboration ES}. 2015, {The Explanatory Supplement to the Planck 2015
  results, \url{http://wiki.cosmos.esa.int/planckpla/index.php/Main_Page}}
  ({ESA})

\bibitem[{{Planck HFI Core Team}(2011{\natexlab{a}})}]{planck2011-1.5}
{Planck HFI Core Team}, {Planck early results, IV. First assessment of the High
  Frequency Instrument in-flight performance}. 2011{\natexlab{a}}, \aap, 536,
  A4, \eprint{1101.2039}

\bibitem[{{Planck HFI Core Team}(2011{\natexlab{b}})}]{planck2011-1.7}
{Planck HFI Core Team}, {Planck early results. VI. The High Frequency
  Instrument data processing}. 2011{\natexlab{b}}, \aap, 536, A6,
  \eprint{1101.2048}

\bibitem[{{\sorthelp{Planck Collaboration 2014A}}{Planck Collaboration
  I}(2014)}]{planck2013-p01}
{\sorthelp{Planck Collaboration 2014A}}{Planck Collaboration I},
  {\textit{Planck} 2013 results. I. Overview of products and scientific
  results}. 2014, \aap, 571, A1, \eprint{1303.5062}

\bibitem[{{\sorthelp{Planck Collaboration 2014F}}{Planck Collaboration
  VI}(2014)}]{planck2013-p03}
{\sorthelp{Planck Collaboration 2014F}}{Planck Collaboration VI},
  {\textit{Planck} 2013 results. VI. High Frequency Instrument data
  processing}. 2014, \aap, 571, A6, \eprint{1303.5067}

\bibitem[{{\sorthelp{Planck Collaboration 2014G}}{Planck Collaboration
  VII}(2014)}]{planck2013-p03c}
{\sorthelp{Planck Collaboration 2014G}}{Planck Collaboration VII},
  {\textit{Planck} 2013 results. VII. HFI time response and beams}. 2014, \aap,
  571, A7, \eprint{1303.5068}

\bibitem[{{\sorthelp{Planck Collaboration 2014H}}{Planck Collaboration
  VIII}(2014)}]{planck2013-p03f}
{\sorthelp{Planck Collaboration 2014H}}{Planck Collaboration VIII},
  {\textit{Planck} 2013 results. VIII. HFI photometric calibration and
  mapmaking}. 2014, \aap, 571, A8, \eprint{1303.5069}

\bibitem[{{\sorthelp{Planck Collaboration 2014J}}{Planck Collaboration
  X}(2014)}]{planck2013-p03e}
{\sorthelp{Planck Collaboration 2014J}}{Planck Collaboration X},
  {\textit{Planck} 2013 results. X. HFI energetic particle effects:
  characterization, removal, and simulation}. 2014, \aap, 571, A10,
  \eprint{1303.5071}

\bibitem[{{\sorthelp{Planck Collaboration 2014L}}{Planck Collaboration
  XII}(2014)}]{planck2013-p06}
{\sorthelp{Planck Collaboration 2014L}}{Planck Collaboration XII},
  {\textit{Planck} 2013 results. XII. Diffuse component separation}. 2014,
  \aap, 571, A12, \eprint{1303.5072}

\bibitem[{{\sorthelp{Planck Collaboration 2014N}}{Planck Collaboration
  XIV}(2014)}]{planck2013-pip88}
{\sorthelp{Planck Collaboration 2014N}}{Planck Collaboration XIV},
  {\textit{Planck} 2013 results. XIV. Zodiacal emission}. 2014, \aap, 571, A14,
  \eprint{1303.5074}

\bibitem[{{\sorthelp{Planck Collaboration 2014O}}{Planck Collaboration
  XV}(2014)}]{planck2013-p08}
{\sorthelp{Planck Collaboration 2014O}}{Planck Collaboration XV},
  {\textit{Planck} 2013 results. XV. CMB power spectra and likelihood}. 2014,
  \aap, 571, A15, \eprint{1303.5075}

\bibitem[{{\sorthelp{Planck Collaboration 2014P}}{Planck Collaboration
  XVI}(2014)}]{planck2013-p11}
{\sorthelp{Planck Collaboration 2014P}}{Planck Collaboration XVI},
  {\textit{Planck} 2013 results. XVI. Cosmological parameters}. 2014, \aap,
  571, A16, \eprint{1303.5076}

\bibitem[{{\sorthelp{Planck Collaboration 2014ZC}}{Planck Collaboration
  XXVIII}(2014)}]{planck2013-p05}
{\sorthelp{Planck Collaboration 2014ZC}}{Planck Collaboration XXVIII},
  {\textit{Planck} 2013 results. XXVIII. The Planck Catalogue of Compact
  Sources}. 2014, \aap, 571, A28, \eprint{1303.5088}

\bibitem[{{\sorthelp{Planck Collaboration 2014ZF}}{Planck Collaboration
  XXXI}(2014)}]{planck2013-p01a}
{\sorthelp{Planck Collaboration 2014ZF}}{Planck Collaboration XXXI},
  {\textit{Planck} 2013 results. XXXI. Consistency of the \textit{Planck}data}.
  2014, \aap, 571, A31

\bibitem[{{\sorthelp{Planck Collaboration 2015A}}{Planck Collaboration
  I}(2015)}]{planck2014-a01}
{\sorthelp{Planck Collaboration 2015A}}{Planck Collaboration I},
  {\textit{Planck} 2015 results. I. Overview of products and results}. 2015,
  \aap, submitted, \eprint{1502.01582}

\bibitem[{{\sorthelp{Planck Collaboration 2015H}}{Planck Collaboration
  VIII}(2015)}]{planck2014-a09}
{\sorthelp{Planck Collaboration 2015H}}{Planck Collaboration VIII},
  {\textit{Planck} 2015 results. VIII. High Frequency Instrument data
  processing: Calibration and maps}. 2015, \aap, submitted, \eprint{1502.01587}

\bibitem[{{\sorthelp{Planck Collaboration 2015K}}{Planck Collaboration
  XI}(2015)}]{planck2014-a13}
{\sorthelp{Planck Collaboration 2015K}}{Planck Collaboration XI},
  {\textit{Planck} 2015 results. XI. CMB power spectra, likelihood, and
  consistency of cosmological parameters}. 2015, in preparation

\bibitem[{{\sorthelp{Planck Collaboration 2015L}}{Planck Collaboration
  XII}(2015)}]{planck2014-a14}
{\sorthelp{Planck Collaboration 2015L}}{Planck Collaboration XII},
  {\textit{Planck} 2015 results. XII. Simulations}. 2015, in preparation

\bibitem[{{\sorthelp{Planck Collaboration 2015M}}{Planck Collaboration
  XIII}(2015)}]{planck2014-a15}
{\sorthelp{Planck Collaboration 2015M}}{Planck Collaboration XIII},
  {\textit{Planck} 2015 results. XIII. Cosmological parameters}. 2015, \aap,
  submitted, \eprint{1502.01589}

\bibitem[{{\sorthelp{Planck Collaboration 2015ZI}}{Planck Collaboration
  XXXIV}(2015)}]{planck2014-a33}
{\sorthelp{Planck Collaboration 2015ZI}}{Planck Collaboration XXXIV},
  {\textit{Planck} 2015 results. The zodiacal light}. 2015, in preparation

\bibitem[{{Tauber} {et~al.}(2010){Tauber}, {Norgaard-Nielsen}, {Ade}, {Amiri
  Parian}, {Banos}, {Bersanelli}, {Burigana}, {Chamballu}, {de Chambure},
  {Christensen}, {Corre}, {Cozzani}, {Crill}, {Crone}, {D'Arcangelo},
  {Daddato}, {Doyle}, {Dubruel}, {Forma}, {Hills}, {Huffenberger}, {Jaffe},
  {Jessen}, {Kletzkine}, {Lamarre}, {Leahy}, {Longval}, {de Maagt}, {Maffei},
  {Mandolesi}, {Mart{\'{\i}}-Canales}, {Mart{\'{\i}}n-Polegre}, {Martin},
  {Mendes}, {Murphy}, {Nielsen}, {Noviello}, {Paquay}, {Peacocke}, {Ponthieu},
  {Pontoppidan}, {Ristorcelli}, {Riti}, {Rolo}, {Rosset}, {Sandri}, {Savini},
  {Sudiwala}, {Tristram}, {Valenziano}, {van der Vorst}, {van't Klooster},
  {Villa}, \& {Yurchenko}}]{tauber2010b}
{Tauber}, J.~A., {Norgaard-Nielsen}, H.~U., {Ade}, P.~A.~R., {et~al.}, {Planck
  pre-launch status: The optical system}. 2010, \aap, 520, A2

\end{thebibliography}

 \appendix
 \section{HFI TOI and beam product description}
\label{sec:officialHFI}
 
Here we summarize the HFI-specific TOI and  beam productes that are part of the Planck 2015 data release.
A complete description is given in~\cite{planck2014-ES}.

 All the data products are delivered as FITS files containing one or more binary
 table extensions. Non-data products, such as codes, are delivered as
 tar files.

 \subsection{The instrument model} 

 Formally the ``Reduced Instrument Model'' or RIMO (since it is a
 subset of 
 the full model maintained internally by the DPC) this file contains
 instrument parameters and some other product-related information.
\begin{itemize}
 \item{Detector parameters: central frequency; beam size, shape, position,
   and orientation in the focal plane; typical noise level; calibration factor;
   and polarization parameters for the PSBs.}
 \item{Map-level parameters (for the full-mission, full-channel maps only):
   effective frequency; beam characteristics; and noise level.}
 \item{Detector and compound spectral response profiles, for each
   bolometer, for each channel, and for each detector set (``Detset'').}
 \item{Detector noise spectra.}
 \item{Beam window functions and the first five error eigenmodes for the full channel
   maps and their cross-products, and for the Detset and single (unpolarized)
   bolometer maps and their cross-products.}
 \item{Beam correlation matrices, for the beams listed above.}
 \end{itemize}
Note that only the 50 valid bolometers are included.

\subsection{TOI products}

Clean and calibrated signal TOI is provided in six files, one per
frequency  channel, for each operational day. Each file contains the on-board time (OBT) and
associated ``global'' flags (unstable pointing, dark correlation, first
and second half-ring), and the calibrated signal of all bolometers of that
channel (in K$_\mathrm{CMB}$ for the 100 to 353\GHz\ channels and MJy\,sr$^{-1}$ for
the 545 and 857\GHz\ channels) each with a ``local'' flag (not-valid
data, glitch, galactic plane, strong source, etc.).

The term ``clean'' means that all known systematic effects, including the
solar and the orbital dipoles, have been removed, and the
data, if destriped and projected onto a map using the data processing centre
(DPC) tools, would yield the DPC maps.
The flagged regions of the TOI (glitches, big planets, decorrelated
dark bolometers) have
been replaced by an estimate of the signal determined by reading the maps at
the pointing position, thus they are easily recognized as regions of lower
noise.  This was done to preserve the continuity of the flagged
timelines, a necessary step for processing them with Fourier techniques.

In addition to the signal TOI, a pointing TOI is also delivered.  It has the
same basic structure as the signal TOI, i.e., one file per frequency channel
per operational day, but there are three columns per detector, corresponding
to Galactic coordinates $\phi$, $\theta$, $\psi$ for each sample, i.e., the
longitude, the colatitude (= latitude + 90$^{\circ}$), and the roll angle.

The total volume of these data is around 30\,TB, of which 90\,\% is the
pointing data.


\subsection{ROI products}

Ring-ordered information (ROI) products contain a
unique data point for each stable pointing period, or ring. 
In order to reproduce the maps delivered by the HFI, some baseline or
destriping offsets must be applied. These offsets are computed for each
detector and for each ring using the full mission data, as described in
Paper~2.  Three sets of offsets are provided: one for the full rings, and two
for each half-ring. These offsets are given in a large table which is
$N_\mathrm{rings}$ rows $\times N_\mathrm{dets}$ columns $= 26766
\times N_\mathrm{b}$, 
in which each cell contains a vector with the three offsets, where $N_\mathrm{b}$ is
the number of bolometers with a delivered TOI.  These values are
given in the second extension. The first extension contains the indices of the
first sample of each ring (and also the ESA pointingID and the ring start
time) to indicate where each offset value has to be applied.    

 \subsection{Software}

\begin{itemize}
 \item{Colour-correction / unit-conversion code: this is a suite of IDL scripts
to manipulate the bandpass profiles contained in the RIMO to determine colour
corrections and unit conversions.}
 \item{Likelihood code: see~\cite{planck2014-a13}.}
\end{itemize}
 
\section{Scanning beam pipeline}
\label{sec:scanningbeampipeline}
This section describes the special processing applied to planet data, leading
to the scanning beam maps, emphasizing changes from \cite{planck2013-p03c}.
We use four observations of Saturn and four to five observations of Jupiter
for the 100--353\GHz\ channels.\footnote{HFI's dilution cooler began to
  warm during the fifth observation of Jupiter, making only four observations
  available for the 100\GHz\ bolometers and the 217\GHz\ SWBs.}  For
these channels, we chose to trade the excellent cross-scan coverage of the
first Mars observation for a higher signal-to-noise ratio.  This choice
introduced complications into the pipeline in order to handle data from
observations separated in time by months, as well as vastly different planet
brightnesses.  

The submillimetre channels (545\GHz\ and 857\GHz) are driven nonlinear by
Jupiter and Saturn, so we use two observations of Mars to build the beam
near the peak of the main beam, supplemented by Saturn and Jupiter data to
measure the near sidelobes.

The representation of the main beam's peak and its closest sidelobes (to
approximately 30\arcm) uses B-spline functions~\citep{planck2013-p03c}, while the
diffraction pattern beyond this radius follows an analytic function
$A\theta^{-3}$, where $A$ is an amplitude factor derived from 
the mirror edge taper~\citep{tauber2010b} and $\theta$ is the angular distance
from the main beam axis. For beams using Saturn data in the peak, we apply the
correction for the planet's finite size at the effective beam window function
level (Sect.~\ref{sec:effectivebeams}). Since the scanning beam maps
are evaluated in the detector frame,  the
scanning direction is always the same, and because the spacecraft rotation rate
is nearly constant, temporal residuals can be spatially constrained in those maps. We apply a
specific treatment for the time-response residual region, to improve 
sensitivity to the residual; specifically, B-splines are used instead of the diffraction 
model. See Sect.~\ref{sec:time_response_uncertainties} for
estimates of transfer-function residuals that cannot be accounted for in the
scanning beam, due to the finite size of the reconstructed maps used 
for window function estimation. 

\subsection{Planet data processing}
\begin{figure}[ht!]
  \begin{center}
    \includegraphics[width=\columnwidth]{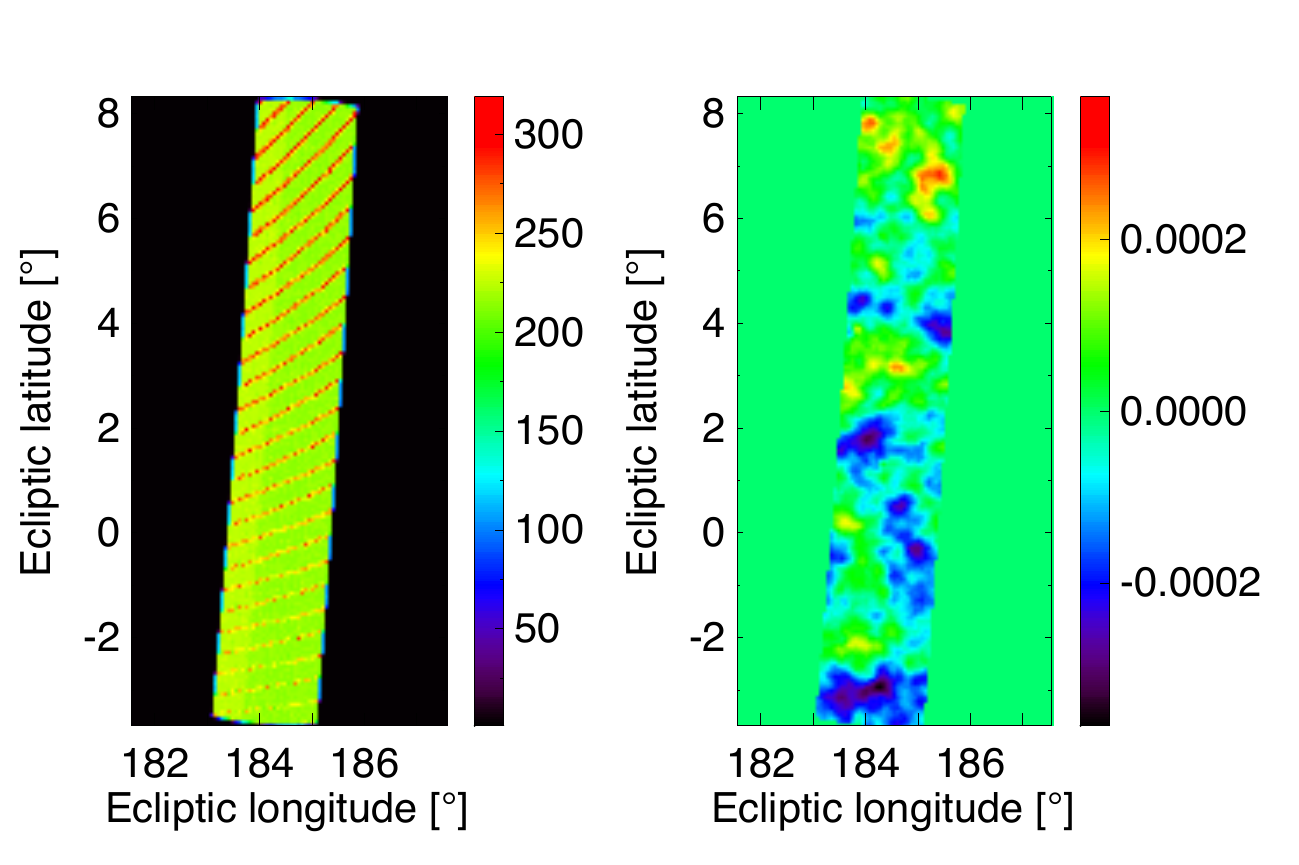} \\
    \caption{\label{fig:backgroundremoval}Hit count map (left
      panel) of the corresponding background map (right panel) of the
      first Saturn scan, for bolometer 143-5, in $\mathrm{K}_\mathrm{CMB}$. 
      }
  \end{center}
\end{figure}

Planet data from the standard TOI processing pipeline cannot be used
to recover the scanning beam to sufficient accuracy ($<1\%$ at the window
function level). We apply a suite of post-processing operations to the
planet data in order to reduce the contribution of identified
systematic errors to the bias in the scanning beam representation and  the window
function estimate. Simulations show that the
additional processing applied to planetary data reduces the effective
window function bias to less than the measurement errors. The following
operations are applied to the planet data and are specific to the beam
reconstruction pipeline:
\begin{itemize}
 \item{background subtraction;}
 \item{planet proper motion pointing correction;}
 \item{destriping;}
 \item{pointing offset correction per planet crossing;}
 \item{merging of different calibrators (renormalization);}
 \item{additional glitch residuals flagging.}
\end{itemize}
The background subtraction uses the full-mission, full-frequency science maps
(the planet crossings are masked). A cubic-spline interpolation is performed
from the pixel centres to the planet crossing pointing
(Fig.~\ref{fig:backgroundremoval}). 

The proper motion of the planets is corrected using ephemerides from the JPL
Horizons
package\footnote{\url{http://ssd.jpl.nasa.gov/?horizons}}~\citep{giorgini1996}
to rotate the pointing into the \Planck\ spacecraft rest frame.

\begin{figure}[ht!]
  \begin{center}
    \includegraphics[width=\columnwidth]{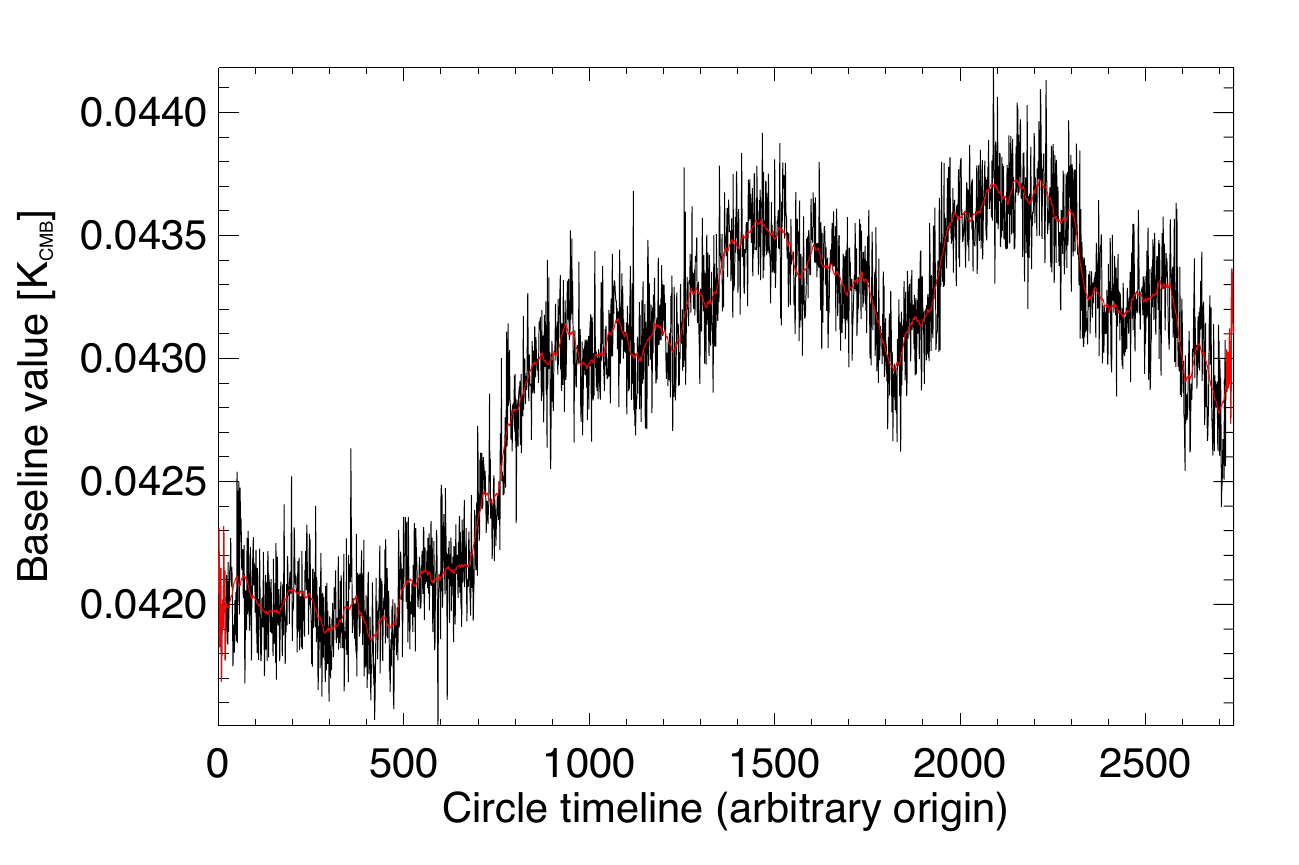} \\
    \caption{\label{fig:destriper1}The baseline for each circle for
      bolometer 100-3b and the first scan of Mars (black line), and
      smoothed with a 40-circle sliding window (red line).
	}
  \end{center}
\end{figure}

\begin{figure}[ht!]
  \begin{center}
    \includegraphics[width=\columnwidth]{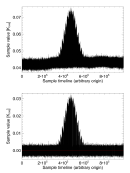} \\
    \caption{\label{fig:destriper2}TOI for bolometer 100-3b during 
      the first scan of Mars, shown both before (upper 
      panel) and after (lower panel) subtraction of the baseline (Fig.~\ref{fig:destriper1}).
	}
  \end{center}
\end{figure}

The destriping removes long-timescale baseline drifts.  Here we subtract a
baseline offset for each circle (60\,s timescales) rather than for each
ring (30--60\,min time scales) as in the mapmaking. We estimate the baseline
offset in a range from $3^\circ$ to $6^\circ$ away from the beam peak as the median of
the samples prior to the planet crossing, to minimize the bias from time-response
residuals. We then smooth the overall trend per circle using a
sliding window with a 40-circle width, as seen in Fig.~\ref{fig:destriper1}.
Figure~\ref{fig:destriper2} shows an example of the destriping procedure.

During the destriping process, we flag and remove the entire circle if one of
the following criteria is met \emph{in the baseline timeline}:
\begin{itemize}
\item the number of flagged samples exceeds 25\,\% of the total samples
  in the circle;
\item there is an event exceeding 3.3 times the estimated local rms
  (residual glitch biasing the estimate of the baseline);
\item the number of positive and negative samples is not balanced,
  such that their
  relative contribution in the total number of used samples differs by more
  than 20\,\%.
\end{itemize}
An additional 5\,\% of the planet data are flagged during this procedure,
largely in the main-beam area.

We expect an error of a few arc seconds in the pointing centroid of each
planetary observation, given the current pointing
model~\citep{planck2014-a01}. We apply a re-centreing procedure to each planet
crossing by fitting a beam template to each observation, where the free
parameters include the pointing shifts in both directions (co-scan and
cross-scan). The pointing of each planet crossing is corrected with these
offsets. The planet data processing is an iterative process, with a loop over the
beam template used. For the first iteration, the template is a prior version
of the beam.

\begin{figure}[ht!]
  \begin{center}
    \includegraphics[width=\columnwidth]{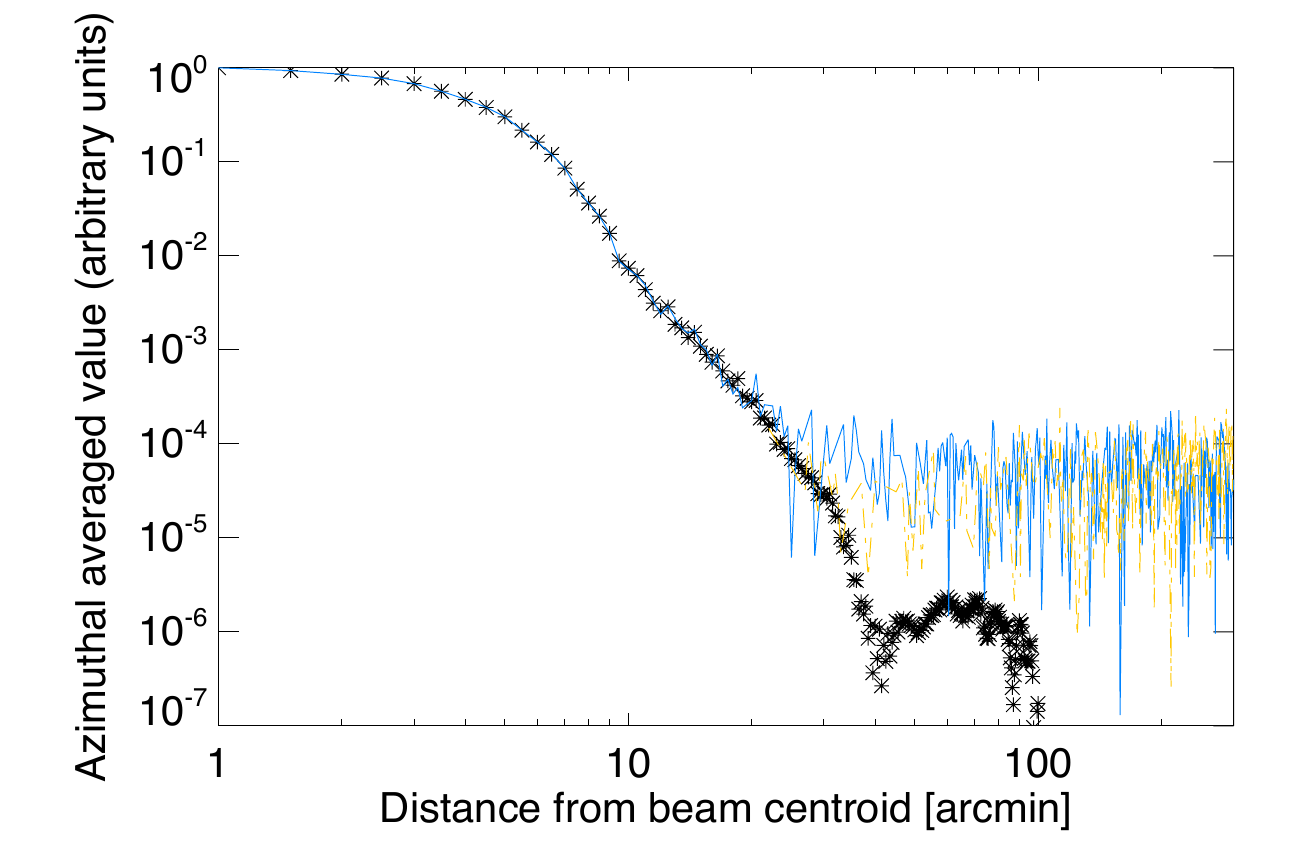} \\
    \caption{\label{fig:renorm1}Renormalization of the beam. The black stars show the beam azimuthal profile
      template used to renormalize the data. The pointing information of the
      scan is reflected through the values of the planet scan's
      template, 
      since the two-dimensional scanning reference beam map is interpolated to
      the transit pointing before the azimuthal averaging. The
      renormalization convention is arbitrary, and is chosen so that the template
      resulting from uniform azimuthal coverage is set to the value 1 at 1\arcm\
      from the beam centroid. The blue curve traces the positive
      data from the planet scan azimuthal average after the
      renormalization procedure; the yellow dashed  curve shows  the negative data
      converted to positive values for plotting purposes. For Jupiter, only
      data below the conservative threshold are used in the renormalization
      procedure (see table \ref{tab:nljup}) to avoid biases from the
      finite planetary
      disc size and detector's nonlinearity response effects. 
	}
  \end{center}
\end{figure}

\begin{figure}[ht!]
  \begin{center}
    \includegraphics[width=\columnwidth]{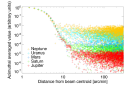} \\
    \caption{\label{fig:renorm2}Renormalized planet data (positive
      values). Planet scans are prepared for the merging into a single data
      timeline using the azimuthal profile template process. For 
      illustration purposes, we also show Neptune and Uranus scans. 
	}
  \end{center}
\end{figure}

Each planet crossing has a different peak brightness, so we renormalize the
planet signals in order to merge them into a common beam map. We interpolate
the beam template to the pointing of each crossing to take into account the
non-uniform coverage of the calibrator observation and perform an azimuth
angle average (see Fig.~\ref{fig:renorm1}) of the samples. We then recover an
amplitude factor by fitting the azimuthally-averaged planet data to an 
azimuthally averaged interpolated template. For Jupiter data, only the portion
of the profile not affected by nonlinearity is used in the fitting 
procedure. Since the same template is common to all calibrators, this
procedure, by construction, forces agreement between the planet data.
Figure~\ref{fig:renorm2} shows the renormalized profiles.  We apply an
interpolation on the two-dimensional beam map template to take into
account the change in the coverage per scan, distorting the uniformly
distributed azimuthal template, since beams are asymmetric and the
sampling is sparse during a planet observation due to the \Planck\ scan. 
The spread at low radius between different scans and planets is
dominated by this effect, while the spread at high radius is mainly
attributed to the different signal-to-noise properties of each
calibrator. 

\begin{figure*}[ht!]
  \begin{center}
    \includegraphics[width=\textwidth]{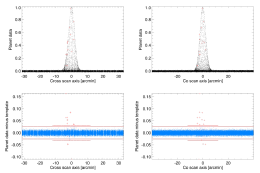} \\
    \caption{\label{fig:xtraflag}Additional flagging process in the beam
      pipeline. Data for samples near the peak
      response (marked with red crosses in the upper panels) are rejected
      when the residual timeline (planet data minus the
      template, blue points in the  lower panels) is either beyond the threshold
      set by 3 times the rms noise (inner red lines), or beyond the pointing
      error threshold (outer red lines).}
  \end{center}
\end{figure*}

The performance of the standard TOI deglitching is degraded by the high signal
gradients of planet scans, relative to CMB data; with no
additional flagging, glitch residuals create a bias of several percent in
the effective beam window function. We therefore add an 
extra stage of glitch flagging to the planet data processing that excludes
samples exceeding 3 times the estimated local noise rms in the low gradients
region of the main beam, and 3 times the expected rms of high-frequency
pointing errors in the strong gradient regions of the main beam
(see Fig.~\ref{fig:xtraflag}).  The fact that the excluded samples are
vertically aligned when projected on the cross-scan axis indicates
that they are likely to be within the same scanning circle,
correlating them in time. The co- and cross-scan projection indicates
they are localized in the beam peak region, where the performance of
the standard glitch removal process is seriously affected.  

\begin{figure}[ht!]
  \begin{center}
    \includegraphics[width=\columnwidth]{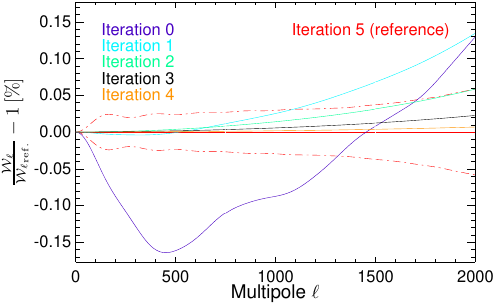} \\
    \caption{\label{fig:convergence}Convergence process of the planet data
      post-treatment for a 143\GHz\ channel. The first guess (iteration 0) is a
      beam built using a different (obsolete) recovery of the temporal
      transfer function. The figure illustrates how sensitive the
      reconstruction is to any change in the time response and in this case
      how it readjusts through the iterations. The dash-dot lines are the
      standard deviation of the spread from the associated Monte Carlo
      ensembles. The number of iterations was chosen so that the iteration
      variation within the last loop (iterations 4 and 5) is conservatively
      within the error bars for all channels from 100\GHz\ to 353\GHz.
	}
  \end{center}
\end{figure}

The planetary data processing converges in five iterations to within
the Monte Carlo error bars (see
Fig.~\ref{fig:convergence}), and is not dependent on the template used in the
first iteration. 

\subsection{Hybrid representation for main beams}

We use Saturn data to measure the main beam and Jupiter to measure the
closest sidelobes for all channels, except 545 and 857\GHz\ where the peak is
described with Mars data. The transition regions consists of a linear
weighting scheme between the data sets. The transition region is
defined as the span of signal samples that are lower
than 9 times the estimated noise rms of the peak planet (Saturn or
Mars) and greater than the Jupiter threshold. The Jupiter threshold is set such
that we keep only data where nonlinearities of the detectors and the
finite size of Jupiter's disc have negligible effects compared to the noise uncertainties.
Table~\ref{tab:nljup} summarizes the Jupiter threshold levels in peak maximum amplitude. They
have been set to conservative values.
 
\begin{table}
	\protect\caption{Jupiter threshold values as a percentage of peak maximum amplitude.
          \label{tab:nljup}}
	\centering{}\begingroup
	\newdimen\tblskip \tblskip=5pt
	\nointerlineskip
	\vskip -3mm
	\footnotesize
	\setbox\tablebox=\vbox{
   	\newdimen\digitwidth 
   	\setbox0=\hbox{\rm 0} 
   	\digitwidth=\wd0 
   	\catcode`*=\active 
   	\def*{\kern\digitwidth}
	\newdimen\signwidth 
	\setbox0=\hbox{+} 
	\signwidth=\wd0 
	\catcode`!=\active 
	\def!{\kern\signwidth}
\halign{\hfil#\hfil\tabskip=2em&
	\hfil#\hfil&
	\hfil#\hfil&
	\hfil#\hfil&
	\hfil#\hfil&
	\hfil#\hfil&
	\hfil#\hfil\/\tabskip=0pt\cr
\noalign{\doubleline}
Band [GHz]& 100& 143& 217& 353& 545& 857\cr
\noalign{\vskip 3pt\hrule\vskip 3pt}
Threshold [\%]& 50& 33&  20& 10& 1& 1\cr
\noalign{\vskip 5pt\hrule\vskip 3pt}}}
\endPlancktable                    
\endgroup
\end{table}

We further spatially average the combined Jupiter data in regions where the signal
is lower than 3 times the estimated Jupiter rms noise.  Physical optics
simulations show that we should not expect high-spatial-frequency variations in the
beam in those regions. The
regularization filter is a square boxcar average, 0\parcm5 on a side
for 100--353\GHz, and 10\arcs\ on a side for 545--857\GHz.

For the final beam model, we fit the B-spline coefficients to the
renormalized planet timelines and the filtered Jupiter data, using a
least squares estimator. As opposed to \cite{planck2013-p03c} (in
particular, their equation~C.4), no additional smoothing criterion is
added to the scoring function in the inversion process, since the
signal-to-noise ratio of four Saturn crossings is high enough to
prevent bias in the B-spline functions coming from high-frequency noise. The
inversion process uses the Cholesky method from the {\tt LAPACK}
module \citep{lapack}. Table~\ref{tab:bsknots} summarizes the control
point (knot) parameters.  The knot locations are tuned so that the
final bias at the window function level is included in the error
budget.  The geometrical shape of the B-spline functions defines the
properties of the intrinsic low-pass filter acting on the data, while
preventing the reconstruction being too sensitive to high frequency
errors. The parameter choice relies on the data quality and the
coverage of the combined planet scan.

\begin{table*}[t]
  \protect\caption{B-spline knot spacing for the scanning beam model at each frequency band.
    \label{tab:bsknots}}
	\centering{}\begingroup
	\newdimen\tblskip \tblskip=5pt
	\nointerlineskip
	\vskip -3mm
	\footnotesize
	\setbox\tablebox=\vbox{
   	\newdimen\digitwidth 
   	\setbox0=\hbox{\rm 0} 
   	\digitwidth=\wd0 
   	\catcode`*=\active 
   	\def*{\kern\digitwidth}
	\newdimen\signwidth 
	\setbox0=\hbox{+} 
	\signwidth=\wd0 
	\catcode`!=\active 
	\def!{\kern\signwidth}
\halign{\hfil#\hfil\tabskip=2em&
	\hfil#\hfil&
	\hfil#\hfil&
	\hfil#\hfil&
	\hfil#\hfil&
	\hfil#\hfil&
	\hfil#\hfil\/\tabskip=0pt\cr
\noalign{\doubleline}
Band [GHz]& 100& 143& 217& 353& 545& 857\cr
\noalign{\vskip 3pt\hrule\vskip 3pt}
Knot spacing [arcmin]& 2& 1.5&  1.25& 1.25& 1.25& 0.8\cr
\noalign{\vskip 3pt\hrule\vskip 3pt}}}
\endPlancktablewide                    
\endgroup
\end{table*}

In the transition area describing the near sidelobes, we further process 
the B-spline representation in order to capture the spatial variations of this
intermediate regime while allowing a smooth transition into the analytic
diffraction pattern. To do this, we build azimuthally-averaged
profiles in a number of ranges of azimuth angle, 
and propagate these templates with a spline interpolation to the final
map. The number of azimuthal templates is constrained by the expected
asymmetries and secondary lobes derived from optical predictions in this
regime of the main beam. Table~\ref{tab:angularchunks} summarizes the number
of angular regions at each frequency band.  Since the azimuthal
templates are equally distributed in angle, the number of templates
sets the separation between each angularly averaged template, and hence
the fidelity of the reconstruction to the azimuthal asymmetries of 
the main beam in this regime. 

\begin{figure}[ht!]
  \begin{center}
    \includegraphics[width=\columnwidth]{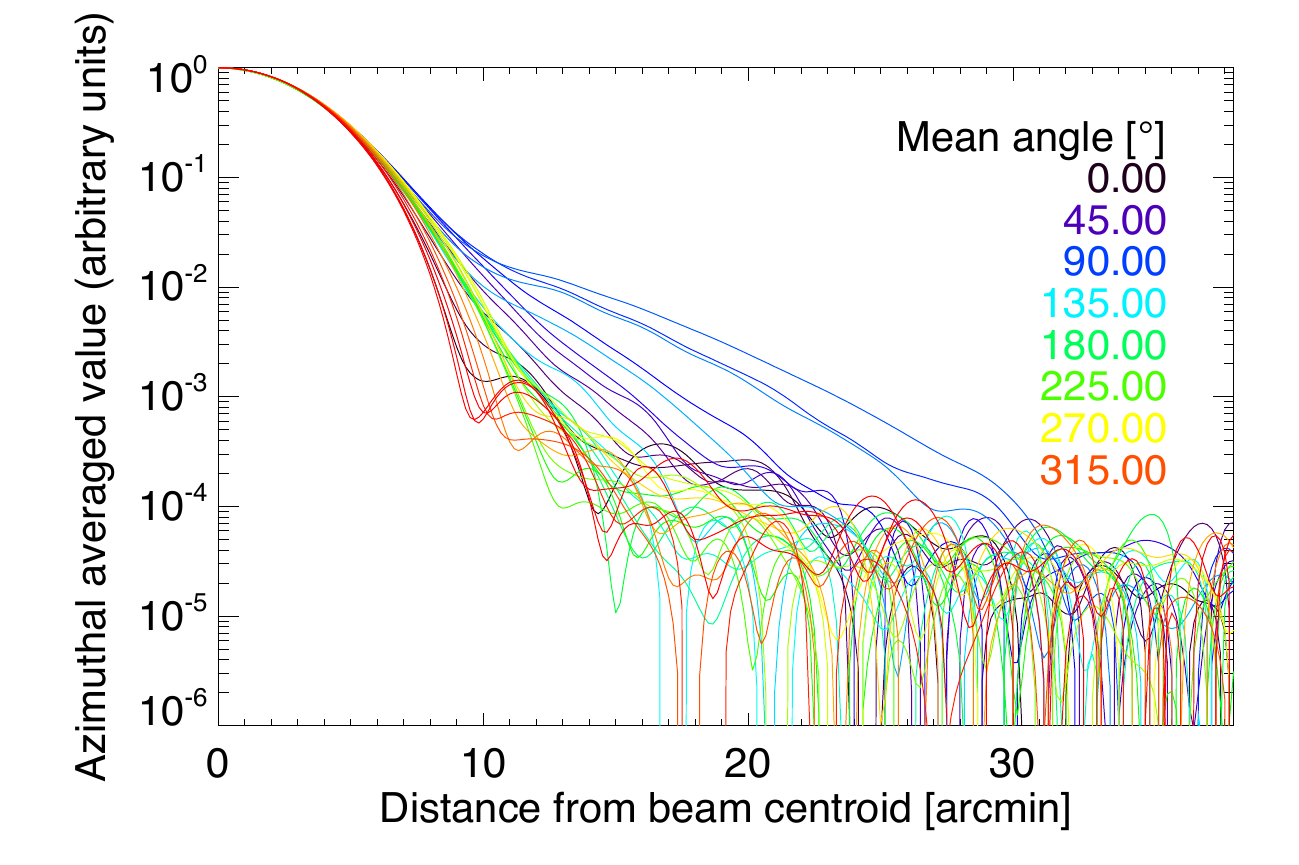} \\
    \caption{\label{fig:htemplates}The 32 radial profiles (averaged
      over a range of azimuthal angles) for the
      143-5 beam. The  azimuthal angles are defined so that 90$^\circ$ 
      corresponds to the profile  closest to the residual time
      response feature (dark blue). 
}
  \end{center}
\end{figure}

Figure~\ref{fig:htemplates} shows
the azimuthally-averaged profiles as a function of the radius from the beam
centroid. The differences in the templates beyond a 5\arcm\ radius
show the optical asymmetry of the beam.  In addition, the 
templates close to $90^\circ$ azimuth angle, shown in blue,
show time response residual effects. In
this region, the initial B-spline evaluations are kept with no
additional azimuthal averaging, in order to better reproduce the
uncertainties in the time response of the detector and readout electronics.
The templates are only used in the regions where Jupiter's rms noise is
not negligible, down to the assumed symmetric diffraction pattern
dominant regime. 
\begin{table*}[t] 
  \protect\caption{Azimuth angle averaging parameters for linking the closest sidelobes to the diffraction pattern, around 25\arcm\ from the beam centroid.
          \label{tab:angularchunks}}
	\centering{}\begingroup
	\newdimen\tblskip \tblskip=5pt
	\nointerlineskip
	\vskip -3mm
	\footnotesize
	\setbox\tablebox=\vbox{
   	\newdimen\digitwidth 
   	\setbox0=\hbox{\rm 0} 
   	\digitwidth=\wd0 
   	\catcode`*=\active 
   	\def*{\kern\digitwidth}
	\newdimen\signwidth 
	\setbox0=\hbox{+} 
	\signwidth=\wd0 
	\catcode`!=\active 
	\def!{\kern\signwidth}
\halign{\hbox to 10.8cm{#\leaderfil}\tabskip=2em&
	\hfil#\hfil&
	\hfil#\hfil&
	\hfil#\hfil&
	\hfil#\hfil&
	\hfil#\hfil&
	\hfil#\hfil\/\tabskip=0pt\cr
\noalign{\doubleline}
\omit&\multispan6\hfil Band [GHz]\hfil\cr
\noalign{\vskip -3pt}
\omit&\multispan6\hrulefill\cr
\noalign{\vskip 3pt}
\omit& 100& 143& 217& 353& 545& 857\cr
\noalign{\vskip 3pt\hrule\vskip 5pt}
Number of azimuthal templates& 16& 32&  32& 32& 64& 64\cr
Angular separation& 22\pdeg5& 11\pdeg2&  11\pdeg2& 11\pdeg2& 5\pdeg6& 5\pdeg6\cr
Azimuthal averaging upper threshold [Jupiter noise rms]& 9& 9&  7& 7& 5& 5\cr
Diffraction pattern dominant regime [Azimuthal averaged Jupiter profile rms]& 3.3& 3.3& 3.3& 3.3& 3.3& 3.3\cr
\noalign{\vskip 5pt\hrule\vskip 3pt}}}
\endPlancktablewide                    
\endgroup
\end{table*}

The intermediate sidelobes beyond 30\arcm\ from the beam centroid are
dominated by a diffraction pattern and are modelled from optical
predictions.  The predictions are compatible with Jupiter azimuthally-averaged profiles (see, e.g., \citealt{planck2013-p01a}). We make the
assumption that the intermediate sidelobe behaviour is azimuthally
symmetric and we keep the B-spline representation in the transfer
function residual region (temporal causal part of the beam) to
properly account for temporal transfer function residuals.

\section{Time Response }\label{sec:appendix_time_response}

\subsection{Measurement of 353~GHz bolometer time constants}
As for the other HFI detectors, the 353\GHz\ bolometer thermal time
response is described in the Fourier
domain as the sum of single-pole low-pass filters, each parameterized by a
time constant and an amplitude. However, the measurement procedure followed a different
process than for the 100--217\GHz\ bolometers (described in
Sect.~\ref{sec:tau}), jointly solving for the optical beam and time
response.  
When applied to 353\GHz\ bolometers, the nominal time response
procedure created scanning beams with strong (several percent of the
peak amplitude) non-optical asymmetries in the scanning (also causal in time) direction, 
extending beyond 30\arcm\ from the beam centroid. This indicated that the
recovered time constants from the previous method were far from the
true ones, distorting the corresponding scanning beams. Therefore, a new
method was developed. 

The procedure described here, created for the 353\GHz\ time response,
will potentially be used in the future at all frequencies,
due to the improved recovery of the very low frequency portion of the time
response, as well as better reconstruction of the scanning beam close
to the centroid.

\subsubsection{Normalization of the glitch slow component}
The stacked glitch template used for glitch removal (Sect.~\ref{sec:deglitching} and Fig.~\ref{fig:glitchvsjupiter}; see also
\citealt{planck2013-p03e}) is built using data filtered with a 3-point finite
impulse response filter.\footnote{The data are convolved with the kernel
  $[0.25,0.5,0.25]$ in the time domain.}  Hence, the time response inferred
from glitches, and in particular the relative amplitude of slow response to
fast response, is strongly biased and must be renormalized.

\subsubsection{Starting parameters and dipole correction}
The time constants and associated amplitudes where $\tau >$ 300\,ms
are set to the glitch values, but renormalized in overall amplitude
using planet data with the scanning beam method described in
Appendix~\ref{sec:scanningbeampipeline}. The results of the fitting procedure
for the time response is dependent on the initial guess because of the strong
correlation in parameters and because planet data (considered only
within 100\arcmin\ of the centroid) only weakly constrain the very slow portion of
the time response ($\tau >$ 1\,s). The initial guess of the glitch
renormalization amplitude is set such that the created power compared to the
previous release matches the correction needed to remove half of the shifted
dipole effect~\citep{planck2014-a09}, assuming that the other half of the
change will be provided by the change of the fast components of the time
response.

Because of the convention we adopted for the global normalization of the time
response model, any additional response at very low frequencies added
to match the long timescale response of the glitches changes the amplitudes of the  
existing time response. Thus, for all the other parameters, the starting
values are the previous-release parameters, with amplitudes adjusted to include
the power introduced by the renormalized glitch template in the time
response.

\subsubsection{Beam-forming while recovering the parameters}
The simultaneous fitting of an optical beam shape and time-response
parameters for the 353\GHz\ detectors follows the procedure
outlined in \cite{planck2013-p03c}, apart from slight differences in
the distance from the planet centroid where the fitting is
performed. 
A symmetrized beam profile along the scanning direction is derived from initial-guess
time response parameters, then updated after each iteration.

\begin{figure}[ht!]
  \begin{center}
    \includegraphics[width=\columnwidth]{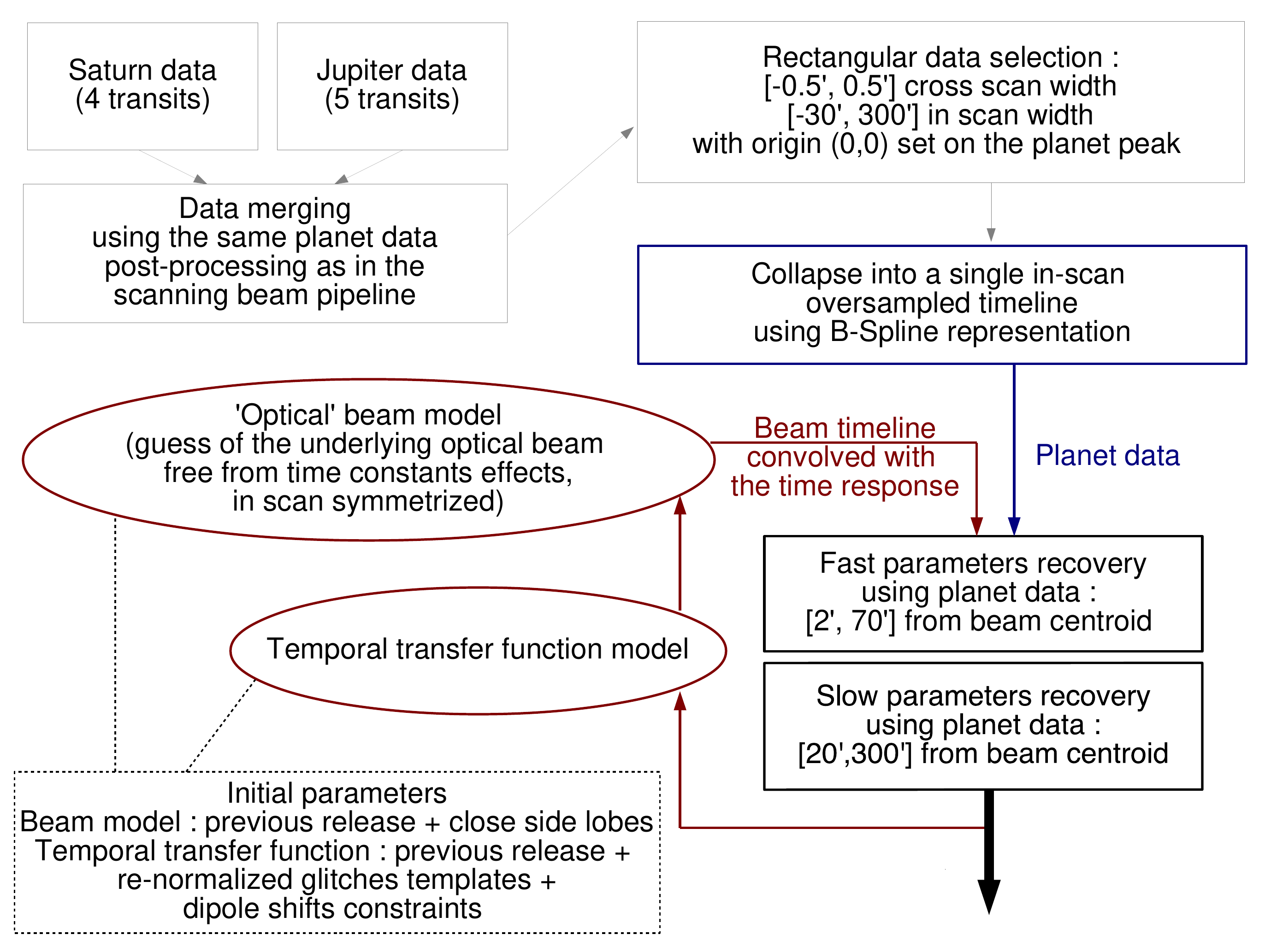} \\
    \caption{\label{fig:BS_TTT}Schematic diagram of the time-constant fitting
      process for the 353\GHz\ channels. The main difference from the CMB
      channels is the fact that the longest time constants from the
      glitch 
      templates that cannot be constrained by the planet signal are 
      renormalized so that their relative contribution is injected properly in the
      temporal model, and the fast part of the time response is re-evaluated
      to reduce the co-scan distortion and induce a beam closer to
      optical. This is particularly significant for PSB pairs, where
      incorrectly deconvolved fast time constants can artificially
      induce beam mismatch and therefore spurious polarization.}
  \end{center}
\end{figure}

The fit is performed as a loop over two steps, alternating fast (10\,ms $<\tau <$ 100\,ms)
and slow (100\,ms $< \tau <$ 300\,ms) parameter adjustments. For fast
parameters, the recovery is performed over the range $3$\arcmin\ to $70$\arcmin\ from the
beam peak centroid, while for slow parameters the range starts at
$20$\arcmin\ 
and extends up to $300$\arcmin. In the first step, the fast parameters are
fitted to combined Saturn and Jupiter data, with other parameters fixed
(intermediate ones and glitch template amplitudes). The second step adjusts
intermediate parameters and the glitch template amplitude to match Jupiter
data with other parameters fixed (fast parameters). We stop the loop when both
of the following requirements are met: (1) the symmetrized beam template is
stable within 1\,\% accuracy in the peak domain; and (2) the relative
reduced $\chi^2$ is stable within $10^{-4}$ in each
iteration. Figure~\ref{fig:BS_TTT} illustrates the overall process for the
353\GHz\ channels. The final parameters are listed in
Table~\ref{table:time_response_pars}.
Notably, this solution requires the inclusion of two extra time-constants
and associated amplitudes. 

\begin{table*}[tb]                 
\begingroup
\newdimen\tblskip \tblskip=5pt
\caption{Parameters for time response models that are deconvolved from the data.}                          
\label{table:time_response_pars}                            
\nointerlineskip
\vskip -3mm
\footnotesize
\setbox\tablebox=\vbox{
   \newdimen\digitwidth 
   \setbox0=\hbox{\rm 0} 
   \digitwidth=\wd0 
   \catcode`*=\active 
   \def*{\kern\digitwidth}
   \newdimen\signwidth 
   \setbox0=\hbox{+} 
   \signwidth=\wd0 
   \catcode`!=\active 
   \def!{\kern\signwidth}
\halign{\hbox to 1.6cm{#\leaderfil}\tabskip 1em&\hfil#\hfil \tabskip 1em&\hfil#\hfil \tabskip 1em &\hfil#\hfil \tabskip 1em&\hfil#\hfil \tabskip 1em&\hfil#\hfil \tabskip 1em&\hfil#\hfil \tabskip 1em&\hfil#\hfil \tabskip 1em&\hfil#\hfil \tabskip 1em&\hfil#\hfil \tabskip 1em&\hfil#\hfil \tabskip 1em&\hfil#\hfil \tabskip 1em&\hfil#\hfil \tabskip 1em&\hfil#\hfil \tabskip 1em&\hfil#\hfil \tabskip 1em&\hfil#\hfil \tabskip 1em&\hfil#\hfil \tabskip 0pt\cr                            
\noalign{\doubleline}
\omit Bolometer & $a_1$ & $\tau_1$ & $a_2$ & $\tau_2$ & $a_3$ & $\tau_3$ & $a_4$ & $\tau_4$ & $a_5$ & $\tau_5$ & $a_6$ & $\tau_6$ & $a_7$ & $\tau_7$ & $\tau_{\mathrm{stray}}$ & $S_{\mathrm{phase}}$ \cr
\omit & [$10^{-3}$] & [ms] & [$10^{-3}$] & [ms] & [$10^{-3}$] & [ms] & [$10^{-3}$] & [ms] & [$10^{-3}$] & [ms] & [$10^{-3}$] & [ms] & [$10^{-3}$] & [ms] & [ms] & [ms] \cr
\noalign{\vskip 3pt\hrule\vskip 5pt}
100-1a & 394 & 10.0 & 537 & 20.9 & 54.6 & 52.3 & 11.6 & 246 & 2.09 & 1375 & \dots & \dots & \dots & \dots & 1.593 &1.386 \cr
100-1b & 478 & 10.3 & 457 & 19.2 & 45.9 & 56.3 & 13.4 & 305 & 5.74 & 2187 & \dots & \dots & \dots & \dots & 1.489 &1.386 \cr
100-2a & 483 & 6.84 & 429 & 13.6 & 72.8 & 43.1 & 13.3 & 298 & 1.28 & 1910 & \dots & \dots & \dots & \dots & 1.321 &1.247 \cr
100-2b & 132 & 5.84 & 749 & 15.1 & 104 & 43.4 & 13.4 & 409 & 2.26 & 2925 & \dots & \dots & \dots & \dots & 1.378 &1.247 \cr
100-3a & 741 & 5.39 & 222 & 14.7 & 29.4 & 47.7 & 7.88 & 327 & 0.09 & 2113 & \dots & \dots & \dots & \dots & 1.421 &1.247 \cr
100-3b & 601 & 5.48 & 348 & 15.5 & 38.8 & 43.5 & 9.83 & 249 & 2.37 & 1058 & \dots & \dots & \dots & \dots & 1.663 &1.247 \cr
100-4a & 410 & 8.20 & 513 & 17.8 & 60.6 & 54.4 & 11.8 & 325 & 4.41 & 1839 & \dots & \dots & \dots & \dots & 1.251 &1.247 \cr
100-4b & 690 & 11.3 & 283 & 24.3 & 20.3 & 90.0 & 2.26 & 295 & 4.57 & 1924 & \dots & \dots & \dots & \dots & 1.381 &1.386 \cr
\omit\cr
143-1a & 816 & 4.47 & 143 & 12.0 & 27.7 & 39.4 & 11.7 & 281 & 1.03 & 1250 & \dots & \dots & \dots & \dots & 1.418 & 1.247 \cr
143-1b & 498 & 4.72 & 338 & 15.6 & 129 & 55.7 & 32.1 & 352 & 3.44 & 1656 & \dots & \dots & \dots & \dots & 1.485 & 1.247 \cr
143-2a & 910 & 4.70 & 76.4 & 17.0 & 8.47 & 100 & 4.36 & 290 & 1.19 & 1370 & \dots & \dots & \dots & \dots & 1.481 & 1.247 \cr
143-2b & 925 & 5.24 & 51.6 & 16.7 & 15.2 & 49.0 & 7.37 & 279 & 0.63 & 1269 & \dots & \dots & \dots & \dots & 1.461 & 1.247 \cr
143-3a & 687 & 4.19 & 276 & 9.56 & 27.3 & 43.6 & 8.34 & 308 & 1.40 & 1499 & \dots & \dots & \dots & \dots & 1.451 & 1.247 \cr
143-3b & 820 & 4.47 & 131 & 13.2 & 36.9 & 39.9 & 10.0 & 324 & 1.65 & 1735 & \dots & \dots & \dots & \dots & 1.609 & 0.832 \cr
143-4a & 911 & 5.69 & 71.7 & 18.9 & 10.6 & 49.4 & 5.38 & 318 & 1.78 & 1786 & \dots & \dots & \dots & \dots & 1.592 & 1.247 \cr
143-4b & 429 & 6.06 & 508 & 6.06 & 55.6 & 26.7 & 7.05 & 363 & 0.46 & 2208 & \dots & \dots & \dots & \dots & 1.820 & 1.247 \cr
143-5 & 523 & 6.64 & 423  & 6.64 & 38.9 & 42.5 & 11.4 & 322 & 3.76 & 1425 & \dots & \dots & \dots & \dots & 2.024 & 1.386 \cr
143-6 & 535 & 5.51 & 423  & 5.51 & 31.5 & 38.7 & 9.39 & 338 & 1.53 & 2227 & \dots & \dots & \dots & \dots & 1.529 & 1.109 \cr
143-7 & 415 & 5.43 & 565  & 5.43 & 14.8 & 43.3 & 4.40 & 366 & 0.80 & 2708 & \dots & \dots & \dots & \dots & 1.860 & 1.386 \cr
\omit\cr
217-1 & 13.6 &3.46  &  955  &3.46  &   27.5 &25.0   &    2.95 &384  &    0.96  &3284 &\dots&\dots&\dots&\dots&1.590  &1.109 \cr
217-2 & 979 &3.52  &  14.0  &26.1 &   2.91 &41.7   &    2.73 &362  &    1.28  &2748 &\dots&\dots&\dots&\dots&1.602  &1.247 \cr
217-3 & 934 &3.55  &  33.7  &3.55  &   27.8  &31.7   &    4.20 &321  &    0.38  &1754 &\dots&\dots&\dots&\dots&1.742  &1.247 \cr
217-4 & 658 &1.35  &  320  &5.55  &   17.9  &27.9   &    2.74 &402  &    0.80  &2920 &\dots&\dots&\dots&\dots&1.710  &1.109 \cr
217-5a & 905 &6.69  &  79.7 &21.6  &    7.68 &66.2   &    6.34 &252  &    1.79  &1160 &\dots&\dots&\dots&\dots&1.573  &1.109 \cr
217-5b & 924 &5.76  &  60.9 &18.0  &    7.49 &65.8   &    6.19 &343  &    1.42  &2742 &\dots&\dots&\dots&\dots&1.867  &1.247 \cr
217-6a & 872 &6.45  &  69.7 &19.7  &   46.8  &44.8   &   11.3  &270  &    0.46  &1226 &\dots&\dots&\dots&\dots&1.545  &1.247 \cr
217-6b & 285 &6.23  &  667  &6.23  &   38.9  &26.8   &    8.65 &267  &    0.22  &1266 &\dots&\dots&\dots&\dots&1.455  &1.109 \cr
217-7a & 344 &5.48  &  576  &5.48  &   71.6  &25.1   &    6.82 &282  &    1.77  &1279 &\dots&\dots&\dots&\dots&1.515  &1.386 \cr
217-7b & 847 &5.07  &  127  &14.4  &   17.3  &49.9   &    6.19 &348  &    2.36  &1787 &\dots&\dots&\dots&\dots&1.505  &1.386 \cr
217-8a & 498 &7.22  &  441  &7.22  &   50.7  &30.2   &    9.05 &266  &    1.04  &1260 &\dots&\dots&\dots&\dots&1.789  &1.109 \cr
217-8b & 512 &7.03  &  411  &7.03  &   63.5  &27.7   &    9.65 &312  &    1.22  &2212 &\dots&\dots&\dots&\dots&1.731  &1.247 \cr
\omit\cr
353-1 & 740   &1.23  &  187  &6.07  &   59.3  &16.1   &   106   &71.9 &    1.92 &1378 &0.57 &1725 & 0.90 &4034 &1.211   &0.911 \cr
353-2 & 944   &5.32  &  44.9 &19.8  &    6.63 &103    &\dots    &\dots  &    0.86 &879  &2.99 &2611 & 0.17 &7084 &3.445   &0.970 \cr
353-7 & 686   &0.55 &  283  &3.60  &   24.5  &14.6   &    3.98 &99.9 &    2.69 &682  &0.05 &1180 & 0.22 &3498 &1.089   &1.247 \cr
353-8 & 716   &1.97  &  261  &5.95  &   17.3  &31.7   &\dots    &\dots  &    4.25 &169  &0.59 &1052 & 1.51 &2943 &1.987   &1.109 \cr
353-3a & 22.7 &3.03  &  869  &6.90  &   81.2  &22.5   &   16.2 &80.5 &   10.3 &490  &0.11 &1310 & 0.17 &3611 &1.782   &1.247 \cr
353-3b & 218  &2.67  &  667  &6.95  &   88.9  &19.5   &   19.4 &70.0 &    5.73 &597  &0.08 &1402 & 0.19 &3770 &1.574   &1.109 \cr
353-4a & 463  &1.89  &  464  &6.26  &   55.5  &22.1   &\dots    &\dots&   10.8 &119  &4.94 &1507 & 1.69 &3311 &1.479   &1.247 \cr
353-4b & 809  &4.50  &  89.0 &15.5  &   19.9  &28.9   &   73.5 &96.2 &    8.68 &326  &0.24 &2181 & 0.08 &4611 &1.726   &1.109 \cr
353-5a & 778  &5.94  &  160  &12.4  &   41.3  &32.4   &   14.2 &104  &    0.21 &559  &6.59 &1334 & 0.08 &3356 &1.571   &1.109 \cr
353-5b & 782  &5.89  &  115  &10.8  &   27.3  &39.9   &   67.3 &9.57 &    6.15 &272  &0.65 &1429 & 1.51 &3510 &1.893   &1.109 \cr
353-6a & 93.7  &1.27  &  834  &5.99  &   58.0  &24.2   &   11.1 &112  &    4.06 &701  &0.20 &1565 & 0.15 &3666 &1.998   &1.247 \cr
353-6b & 53.7 &1.24  &  911  &5.54  &   23.1  &28.8   &    6.29 &118  &    5.84 &763  &0.07 &1201 & 0.18 &3444 &2.895   &1.109 \cr
\omit\cr
545-1 & 991 & 2.93 & 7.43 & 26 & 1.39 & 2600 & \dots & \dots & \dots & \dots & \dots & \dots & \dots & \dots & 2.160 & 1.109 \cr
545-2 & 985 & 2.77 & 12.8 & 24 & 2.46 & 2800 & \dots & \dots & \dots & \dots & \dots & \dots & \dots & \dots & 1.868 & 0.970 \cr
545-4 & 972 & 3.00 & 27.7 & 25 & 0.78 & 2500 & \dots & \dots & \dots & \dots & \dots & \dots & \dots & \dots & 2.222 & 1.109 \cr
\omit\cr
857-1 & 974 & 3.38 & 22.9 & 25.0 & 3.49 & 2200 & \dots & \dots & \dots & \dots & \dots & \dots & \dots & \dots & 1.765 &1.109 \cr
857-2 & 840 & 1.48 & 158 & 6.56 & 2.49 & 3200 & \dots & \dots & \dots & \dots & \dots & \dots & \dots & \dots & 2.202 &1.247 \cr
857-3 & 360 & 0.04 & 627 & 2.40 & 11.1 & 17 & 2.00&1900 & \dots & \dots & \dots & \dots & \dots & \dots & 1.524 &1.263 \cr
857-4 & 278 & 0.40 & 719 & 3.92 & 1.62 & 90 & 1.52&800 & \dots & \dots & \dots & \dots & \dots & \dots & 1.490 &0.558 \cr
\noalign{\vskip 3pt\hrule\vskip 5pt}
}
}
\endPlancktablewide                 
\endgroup
\end{table*}                        

\raggedright 
\end{document}